\documentclass[11pt]{article}
\pdfoutput=1    % This is to ensure ArXiv compiles it correctly
\usepackage{setspace}
\usepackage[T1]{fontenc}                        % Use 8-bit encoding that has 256 glyphs
\usepackage{microtype} 				            % Slightly tweak font spacing for aesthetics
\usepackage{anyfontsize}                        % Allow for generic font sizes, only needed to suppress annoying warnings
\usepackage{amssymb,amsmath,amscd}
\usepackage[UKenglish]{babel}
\usepackage{titlesec}
\usepackage{graphicx,xcolor,xparse}
\usepackage{enumitem}
\usepackage{multirow}
\usepackage[absolute]{textpos}
\usepackage[margin=1.5em,labelfont=bf]{caption}
\usepackage{cite}
\usepackage{bm}                                 % Bold greek letters
\usepackage{mathrsfs}                           % Provides the \mathscr font
\usepackage{placeins}                           % This allows me to put floatbarriers
\usepackage[hang,flushmargin]{footmisc}         % Remove footnote indents
\usepackage{orcidlink}                          % ORCID links
\usepackage{hyperref}                           % hyperref settings
\hypersetup{
	colorlinks=true,
	linkcolor=midnight,
	citecolor=cardinal,
	urlcolor=darker-green,
	linktoc=all
}

\usepackage{geometry} 					% Document margins
\numberwithin{equation}{section} 

\titleformat{\section}[block]{\Large\bfseries\centering}{\thesection}{1em}{} 
\titleformat{\subsection}[block]{\bfseries}{\thesubsection}{1em}{} 

\titlespacing*{\section}{0pt}{1em}{1em}
\titlespacing*{\subsection}{0pt}{0.75em}{0.75em}

\addto\captionsUKenglish{
  \renewcommand{\contentsname}%
    {Table of Contents}%
}

\definecolor{gray}{gray}{0.30}
\definecolor{dark-gray}{gray}{0.20}
\definecolor{light-gray}{gray}{0.80}
\definecolor{dark-red}{rgb}{0.7,0,0}
\definecolor{forestgreen}{rgb}{0.13, 0.55, 0.13}
\definecolor{dark-green}{rgb}{0,0.5,0}
\definecolor{darker-green}{rgb}{0.1,0.4,0}
\definecolor{light-blue}{rgb}{0.8,0.8,1}
\definecolor{dark-blue}{rgb}{0.3,0.3,0.7}
\definecolor{darker-blue}{rgb}{0.2, 0, 0.8}
\definecolor{midnight}{rgb}{0, 0, 0.5}
\definecolor{golden}{rgb}{0.92, 0.7, 0}
\definecolor{dark-red}{rgb}{0.7,0,0}
\definecolor{cardinal}{rgb}{0.6,0,0}

\graphicspath{{./figures/}}

\usepackage[vcentermath]{youngtab}
\usepackage{tikz} 
\usepackage{tikz-cd} 
\usetikzlibrary{matrix,calc} 
\usetikzlibrary{decorations.markings, shapes.geometric, shapes.misc} 
\usetikzlibrary{decorations.pathreplacing,positioning} 
\usetikzlibrary{arrows}

\tikzset{middlearrow/.style={
		decoration={markings,
			mark= at position 0.5 with {\arrow{#1}} ,
		},
		postaction={decorate}
	}
}

\def\mop#1{\mathop{\rm #1}\nolimits}

\def\Re{\mop{Re}}
\def\Im{\mop{Im}}
\def\diag{\mop{diag}}
\def\ii{{\rm i}}
\def\tr{\mop{tr}}

\newcommand{\dd}{{\rm d}}

\newcommand{\e}{\mathrm{e}}
\newcommand{\bz}{\bar{z}}

\newcommand\Tr{\mathrm{Tr}\,}

\renewcommand\Re{{\rm Re}} 
\renewcommand\Im{{\rm Im}}

\newcommand{\noo}{\raisebox{-0.5pt}{\tiny$\circ$} \hspace{-3.85pt}\raisebox{+3.5pt}{\tiny$\circ$}\hspace{1pt}}

\newcommand{\f}[2]{\frac{#1}{#2}}

\newcommand{\acomm}[2]{\left\{{#1},{#2}\right\}}
\newcommand{\llbrak}[2]{\left\{{#1}\,{}_\lambda\,{#2}\right\}}
\newcommand{\lllbrak}[3]{\left\{{#1}\,{}_{\lambda_1}\,{#2}\,{}_{\lambda_2}\,{#3}\right\}}

\newcommand{\nn}{\nonumber}
\newcommand{\wti}[1]{\widetilde{#1}}

\def\overleftrightarrow#1{\vbox{\ialign{##\crcr
	$\leftrightarrow$\crcr\noalign{\kern-0pt\nointerlineskip}
	$\hfil\displaystyle{#1}\hfil$\crcr}}}

\newcommand{\del}{\partial}

\newcommand\bA{\mathbf{A}}
\newcommand\bB{\mathbf{B}}
\newcommand\bC{\mathbf{C}}

\newcommand\bG{\mathbf{G}}

\newcommand\bJ{\mathbf{J}}

\newcommand\bL{\mathbf{L}}
\newcommand\bM{\mathbf{M}}
\newcommand\bN{\mathbf{N}}
\newcommand\bO{\mathbf{O}}
\newcommand\bP{\mathbf{P}}
\newcommand\bQ{\mathbf{Q}}
\newcommand\bR{\mathbf{R}}
\newcommand\bS{\mathbf{S}}

\newcommand\bZ{\mathbf{Z}}

\newcommand\cA{\mathcal{A}}

\newcommand\cF{\mathcal{F}}

\newcommand\cH{\mathcal{H}}
\newcommand\cI{\mathcal{I}}
\newcommand\cJ{\mathcal{J}}
\newcommand\cK{\mathcal{K}}
\newcommand\cL{\mathcal{L}}
\newcommand\cM{\mathcal{M}}
\newcommand\cN{\mathcal{N}}
\newcommand\cO{\mathcal{O}}
\newcommand\cP{\mathcal{P}}
\newcommand\cQ{\mathcal{Q}}
\newcommand\cR{\mathcal{R}}
\newcommand\cS{\mathcal{S}}

\newcommand\fg{\mathfrak{g}}
\newcommand\ff{\mathfrak{f}}

\newcommand\bbC{\mathbb{C}}

\newcommand\bbR{\mathbb{R}}

\newcommand{\bfbeta}{{\text{\boldmath$\beta$}}}
\newcommand{\bfgamma}{{\text{\boldmath$\gamma$}}}

\newcommand{\bfphi}{{\text{\boldmath$\phi$}}}

\newcommand{\bflambda}{{\text{\boldmath$\lambda$}}}

\newcommand\dota{{\dot{\alpha}}}
\newcommand\dotb{{\dot{\beta}}}

\newcommand\SO{\mathrm{SO}}

\newcommand\UU{\mathrm{U}}
\newcommand\SU{\mathrm{SU}}

\newcommand\Spin{\mathrm{Spin}}

\newcommand{\lie}{\mathfrak}
\renewcommand\sl{\mathfrak{sl}}
\newcommand\so{\mathfrak{so}}
\newcommand\su{\mathfrak{su}}

\newcommand\psl{\mathfrak{psl}}
\newcommand\gl{\mathfrak{gl}}

\newcommand\uu{\mathfrak{u}}

\NewDocumentCommand{\llb}{s m m m}{%
  \IfBooleanTF{#1}{%
    \left\{\!\left\{ #2 {}_{#3} #4 \right\}\!\right\}%
  }{%
    \{\!\{ #2 {}_{#3} #4 \}\!\}%
  }%
}

\NewDocumentCommand{\lb}{s m m m}{%
  \IfBooleanTF{#1}{%
    \left\{  #2  {}_{#3}  #4  \right\}%
  }{%
    \{  #2 \, {}_{#3} \, #4 \}%
  }%
}

\usepackage{tikz,tikz-3dplot}
\usepackage{pgfplots}
\pgfplotsset{compat=1.17}
\usetikzlibrary{shapes,arrows,cd,chains,decorations.markings,decorations.pathmorphing,calc, patterns}

\tikzset{
    ->-/.style args={#1rotate#2}{
        decoration={
            markings, 
            mark=at position #1 with {\arrow[scale=1.5, rotate=#2]{stealth}}
        }, 
        postaction={decorate}
    }
}

\tikzstyle{GraphNode}=[circle, draw=black, fill=black, inner sep=2pt, minimum size=5pt]
\tikzstyle{GraphNodeH}=[circle, draw=black, fill=none, inner sep=2pt, minimum size=5pt]
\tikzstyle{GraphEdge}=[black]

\tikzstyle{vertex}=[circle, draw, minimum size=0.2cm, inner sep = 1pt, fill=black] 
\tikzstyle{line}=[line width=0.3mm,->- = {0.5rotate0}]
\tikzstyle{lineShimmy}=[line width=0.3mm,->- = {0.53rotate0}]
    
\pgfmathsetmacro{\gS}{1}

\usepackage{tikzit}
\tikzstyle{red dot}=[fill=red, draw=none, shape=circle, minimum size=5pt, inner sep=2pt]
\tikzstyle{black dot}=[fill=black, draw=none, shape=circle, minimum size=5pt, inner sep=2pt]

\tikzstyle{red edge}=[-, draw=red]
\tikzstyle{black edge}=[-]
\tikzstyle{new edge style 0}=[fill={rgb,255: red,191; green,0; blue,64}, draw=black, dots, =]

\title{
\fontsize{20pt}{23pt}\selectfont\textbf{Unravelling the Holomorphic Twist II:} \\[2mm] 
\fontsize{20pt}{23pt}\selectfont\textbf{Anomalies and Extended Supersymmetry}
}

\author{
\Large Pieter Bomans \footnote{~\href{mailto:pieter.bomans@maths.ox.ac.uk}{pieter.bomans@maths.ox.ac.uk}\, \orcidlink{0000-0002-0907-9830}} , \hspace{2mm} Niklas Garner \footnote{~\href{mailto:niklas.garner@maths.ox.ac.uk}{niklas.garner@maths.ox.ac.uk}\, \orcidlink{0000-0002-4646-2599}} ,\\
\Large Brian R. Williams \footnote{~\href{mailto:bwill22@bu.edu}{bwill22@bu.edu}\, \orcidlink{0000-0001-9641-861X}} , \hspace{2mm} and \hspace{2mm} Jingxiang Wu \footnote{~\href{mailto:jingxiang.wu@maths.ox.ac.uk}{jingxiang.wu@maths.ox.ac.uk}\, \orcidlink{0000-0001-6867-1407}}\\[8mm]
    \normalsize \textit{${}^{*, \dagger, \S}$ Mathematical Institute, University of Oxford}\\ 
    \normalsize \textit{Andrew Wiles Building, Radcliffe Observatory Quarter}\\ 
    \normalsize \textit{Woodstock Road, Oxford, OX2 6GG, U.K.}\\[3mm]
    \normalsize \textit{${}^\ddagger$Department of Mathematics \& Statistics, Boston University}\\
    \normalsize \textit{665 Commonwealth Ave, Boston, MA, 02215}
}

\date{}

\begin{document}
\pagestyle{empty}

\maketitle
\thispagestyle{empty}
\vspace{\stretch{1}}

\begin{abstract}
\noindent Twists of four-dimensional supersymmetric quantum field theories (SQFTs) isolate protected sectors with rich algebraic structures. We develop a unified framework for analyzing symmetries and anomalies in four-dimensional holomorphically twisted SQFTs, combining complex-geometric and algebraic perspectives. This approach clarifies the connections between existing formulations in the literature and resolves several open questions left unanswered in the first installment of this series.

\noindent We place particular emphasis on theories with extended supersymmetry, where the holomorphic twist gives rise to enhanced algebraic and geometric structures. We explain how these features emerge and govern the organization of the twisted theory. Furthermore, we demonstrate how a superconformal deformation of the twisted theory naturally leads to the associated vertex operator algebra, clarifying how the higher algebraic structures of the holomorphically twisted theory give rise to vertex algebras structures.
\end{abstract}

\vspace{\stretch{4}}
 
\newpage

{ 
\setstretch{1.1}                % Locally adjust line spacing to fit toc on one page
\hypersetup{linkcolor=black}
\setcounter{tocdepth}{2}
\tableofcontents
}

\setcounter{page}{0}

\clearpage
\pagestyle{plain}

%%%%%%%%%%%%%%%%%%%%%%%%%%%%%%%%%%%%%%%%%	
\section{Introduction}                  %
\label{sec:intro}                       %
%%%%%%%%%%%%%%%%%%%%%%%%%%%%%%%%%%%%%%%%%

Twists of four-dimensional supersymmetric quantum field theories (SQFTs) have emerged as powerful tools to isolate protected sectors whose algebraic structure often enables exact computations.\footnote{By a twist of a superconformal theory we refer to the process of taking (derived) invariants with respect to an arbitrary square-zero supercharge that may or may not be translation invariant, i.e. we view this square-zero supercharge as a differential and observables as a chain complex with respect it; ``physical'' observables of the twisted theory thus correspond to cohomology classes.}
A prominent example for $\cN \geq 2$ superconformal quantum field theories (SCFTs) is the VOA twist of \cite{Beem:2013sza}, in which the Schur sector is encoded in a two-dimensional holomorphic theory, a vertex operator algebra (VOA).
Among all twists, the richest class are minimal, or holomorphic twists \cite{Costello:2011np,Costello:2013zra,Budzik:2023xbr,Johansen:1994aw,Johansen:2003hw}.
A square-zero supercharge $\bQ$ corresponding to such a twist is characterized by the property that it is translation invariant and that the subspace of translations spanned by super-brackets of the form $\{\bQ,X\}$ is minimal (in four dimensions the minimal subspace is two-dimensional).
The twist of a four-dimensional theory with respect to such a supercharge yields a \textit{holomorphic field theory} endowed with a rich (derived) operator product algebra and symmetry structure.
This construction captures the protected data of all multiplets in the parent theory satisfying a chiral shortening condition.
In other words, it receives contributions from all operators that contribute to the full superconformal or supersymmetric index!

One of the primary advantages of holomorphic field theories, such as those arising from twisting four-dimensional SQFTs, is that they remain amenable to powerful algebraic and geometric techniques, many of which are familiar from two-dimensional chiral conformal field theory (CFT).
Two-dimensional CFTs are distinguished by their $\infty$-dimensional symmetry algebras, most notably the Virasoro algebra and affine Kac–Moody algebra. 
The operator product expansions (OPEs) of conserved currents with local operators encode these symmetries, while central extensions of the algebras capture their quantum anomalies. 
A familiar example is the central extension of the Virasoro algebra, whose coefficient $c$ governs the conformal, or Weyl, anomaly of the chiral CFT.
An analogous structure arises in higher-dimensional holomorphic field theories \cite{gwilliamHigherKacMoodyAlgebras2018, faonteHigherKacMoodyAlgebras2019, Saberi:2019fkq, Budzik:2023xbr, Williams:2024mcc}.
The analogs of the Virasoro and Kac--Moody algebras play familiar geometric roles: the first as local conformal transformations/holomorphic coordinate transformations, and the latter as holomorphic flavor symmetries.

The OPE, being holomorphic, is controlled by the Dolbeault, or equivalently coherent, cohomology of the configuration space of points.
In contrast to the case of $\bC = \bR^2$, the configuration space of points in $\bC^d$, with $d > 1$, is no longer affine as an algebraic variety, and hence it has non-vanishing higher cohomology.
As a result, the $\infty$-dimensional symmetry algebras present in higher-dimensional holomorphic QFT are manifestly \textit{derived}, and the resulting algebras are homotopy Lie, or $L_\infty$, algebras.%
\footnote{An $L_\infty$ algebra is a weakening of the concept of a dg Lie algebra where the Jacobi identity holds only up to coherent homotopy.
Explicitly, an $L_\infty$ algebras are vector spaces equipped with $n$-ary operations for $n \geq 1$ which satisfy higher Jacobi-like identities \cite{Lada_1993}.} %
Just as in two-dimensions, anomalies are captured by central extensions of these $L_\infty$ algebras; for example, the four-dimensional analogue of the Witt algebra has a two-dimensional space of such central extensions, capturing the conformal $a$ and $c$ anomalies of the untwisted theory \cite{Bomans:2023mkd, Williams:2024xrq}.%
\footnote{In six dimensions the corresponding Virasoro algebra has a four-dimensional space of central extensions \cite{Williams:2024mcc} and these correspond to the unique type A anomaly, proportional to the Euler density, and the three remaining type B anomalies corresponding to all the Weyl-invariant scalar polynomials formed from the Ricci tensor including covariant Laplacians \cite{Deser:1993yx}.} %
One motivation for characterizing these infinite-dimensional symmetry algebras echoes their use in chiral CFT and the theory of vertex algebras: such algebras can be effective in constraining the operator product expansion (or its derived versions) of local operators in a given theory.

While significant progress has been made in understanding holomorphic twists, particularly for Lagrangian theories and $\cN=1$ SCFTs, most existing approaches are purely algebraic or purely geometric.
On the algebraic side \cite{gwilliamHigherKacMoodyAlgebras2018, faonteHigherKacMoodyAlgebras2019, costelloFactorizationAlgebrasQuantum2021, Budzik:2022mpd, Budzik:2023xbr, Bomans:2023mkd, Gaiotto:2024gii}, one is often interested in structural properties of local observables, such as secondary products capturing properties of integrated correlators.
The complementary complex-geometric perspective \cite{Costello:2011np, Costello:2013zra, Saberi:2019fkq, Saberi:2019ghy, Williams:2024mcc, Williams:2024xrq}, seeks to understand how the holomorphic field theory responds, perhaps anomalously, to placing the theory on a curved spacetime or turning on a background holomorphic flavor symmetry bundles.
Of course, these two approaches to holomorphic field theory should speak to one another through the conserved currents coupling to these backgrounds, just as it does in more familiar QFT.
A more holistic approach that incorporates the anomalous couplings to background fields through the algebraic properties of local operators in a systematic way has been lacking.

In this work we bridge this gap by making explicit the way complex-geometric backgrounds impact the algebraic structures possessed by local operators.
Our approach applies uniformly to theories with $\cN \geq 1$ supersymmetry.
We explain in detail the derived $\infty$-dimensional nature of these symmetry algebras, with a particular emphasis on extended supersymmetry, and explain how anomalies are systematically encoded through higher homotopical operations of the corresponding currents.
In addition, we clearly connect these structures to more conventional constructions in four-dimensional untwisted QFT.
As a concrete application of our analysis, we show how turning on a suitable background for twisted $\cN=2$ superconformal symmetry can be used to recover the VOA twist of \cite{Beem:2013sza}, cf. \cite{Saberi:2019fkq}, making explicit how the higher algebraic structures of the twisted theory give rise to the vertex algebra structure; see e.g. \cite{Costello:2016mgj, Elliott:2020uwn, Borsten:2025hrn} for discussions on the connection between background fields and twisting.

%-------------------
\subsection{Summary of results}
%-------------------

As mentioned above, four-dimensional holomorphic field theories, i.e. those living in two complex dimensions, admit $\infty$-dimensional symmetry algebras that arise in much the same way as in two dimensions, although there are some additional ingredients that are not present in the latter case.
In the context of minimally twisted $\cN \geq 1$ SQFT, we fix a holomorphic supercharge $\bQ$, which always exists in this case and is unique up to equivalence; more generally $\bQ$ should be viewed as the BV/BRST differential capturing the gauge redundancies and equations of motion of the holomorphic field theory.
We start our discussion in Section \ref{sec:holo-twist} by reviewing the origin of these four-dimensional symmetry algebras and setting up a useful organizational framework using a notion of superfield suitable to holomorphic field theories.
For twisted SQFTs, these \emph{reduced superfields} can be thought of as organizing fields into a multiplet including its holomorphic descendants. 
Each $\cN=1$ multiplet obeying an anti-chiral shortening condition contributes a single $\bQ$-closed operator $O$, naturally identified as the bottom component of a \emph{semi-chiral} superfield $\bO$, which is a special type of reduced superfield.

When the untwisted theory possesses a flavor symmetry, the spectrum includes a conserved current multiplet, which in the twisted theory gives rise to a semi-chiral superfield $\bJ$ satisfying the conservation law $\bar\del \bJ = 0$.
Just as in ordinary QFTs, the action of this flavor symmetry on a given local operator can be implemented by integrating the current $\bJ$ around it.
The $\infty$-dimensional enhancement is then realized by witnessing an enhancement of global flavor transformations to local, holomorphic flavor transformations, which are implemented by integrals of the current $\bJ$ weighted by holomorphic functions of two variables.
The true novelty of working in complex dimension $d > 1$ is that a holomorphic function on $\bC^d \backslash \{0\}$ necessarily extends over $0$, a result due to Hartogs; this is in stark contrast with the case in $d = 1$ where the are genuinely meromorphic functions.
Instead, these singular ``functions'' appear in higher form degree: punctured affine space $\bC^d \backslash \{0\}$ has higher Dolbeault cohomology.
Correspondingly, there are yet more symmetries we can use that arise from integrals of $\bJ$ weighted by a general holomorphic, i.e. $\bar \partial$-closed, $(0, \bullet)$-forms.
Thus, the $\infty$-dimensional symmetry algebras of four-dimensional holomorphic field theories have a much more homotopical flavor than those familiar to two dimensions.
A convenient way to describe a portion of this structure is through $\lambda$-brackets, which serve as generating functions for the higher modes of the currents and organize the algebra into an $L_\infty$ analogue of a Lie conformal algebra \cite{Budzik:2022mpd, Budzik:2023xbr, Gaiotto:2024gii}.

We focus on two of the most important examples of these symmetries in Section \ref{sec:symmetries}: flavor symmetries, which enhance to four-dimensional analogues of affine Kac-Moody symmetry algebras, and spacetime symmetries, which enhance to an analogue of the Virasoro algebra.
In the case of a flavor symmetry, this resulting symmetry algebra is a higher-dimensional analogue of the Kac-Moody symmetry of a two-dimensional CFT first described in \cite{faonteHigherKacMoodyAlgebras2019}, see also \cite{gwilliamHigherKacMoodyAlgebras2018}.
Similarly, when the parent theory possesses an unbroken $\UU(1)_r$ $R$-symmetry, the twisted theory inherits a pair of semi-chiral superfields $\bS_i$, $i = 1,2$, satisfying the conservation law $\bar \del \bS_i = 0$ and generate an algebra we denote $\lie{vir}(2)_{a,c}$ which is a higher-dimensional analogue of the Virasoro symmetry of a two-dimensional CFT, also first described in \cite{faonteHigherKacMoodyAlgebras2019} but see also \cite{williamsHolomorphicSmodelIts, Saberi:2019fkq}.
In each of these cases we describe how to couple a theory with these symmetries to background fields; in the former case this simply amounts to introducing a background principal bundle with connection whereas in the latter such a background is a deformation of the complex structure of spacetime.
We can treat these backgrounds in exactly the same way as in ordinary QFT and, expanding on \cite{gwilliamHigherKacMoodyAlgebras2018, Bomans:2023mkd, Williams:2024xrq}, show that these are directly related to the holomorphic anomaly polynomial, which in turn can be related to the standard anomaly polynomial.
Moreover, through a descent construction, we demonstrate explicitly how these central extensions are intimately tied to anomalies.

As an illustrative example, consider a $\UU(1)$ flavor symmetry.
The cubic flavor anomaly leads to the relation
\begin{equation}
    \bQ \bJ = \bar\del\bJ -\f{\hbar k}{2\pi^2} \del \bA 
    \del \bA\,,
\end{equation}
where $k$ is the corresponding ’t Hooft anomaly coefficient.
This equation shows that the non-conservation of the current in the holomorphic twist corresponds to the failure of the superfield $\bJ$ to be semi-chiral.
If the right-hand side involves only background fields, such as the background connection $\bA$, the theory has an 't Hooft anomaly.
By contrast, if it involves dynamical gauge fields, denoted by $\mathbf{c}$ in the main text, the quantum theory suffers from a gauge or ABJ anomaly depending on whether $\mathbf{c}$ is the gauge field coupling to $\bJ$ or not; we will not consider situations with gauge anomalies as they do not lead to consistent quantum-mechanical theories.
In the more interesting case of an ABJ anomaly, one often simply says that the symmetry is broken, although there are ways to recover some portion of it using a non-invertible modification thereof \cite{Choi:2022jqy, Cordova:2022ieu}. Here we advocate for a novel perspective: the above non-conservation equation should instead be viewed as the conservation equation for a different symmetry. This new symmetry comes equipped with an internal differential, mapping the symmetry generated by $\bJ$ to that generated by $\del \mathbf{c} \del \mathbf{c}$. This differential captures in a natural way the ABJ anomaly of the physical symmetry.

For simplicity we considered a $\UU(1)$ flavor symmetry in the above.
However, the same considerations apply equally to non-abelian flavor symmetries as well as to spacetime symmetries.
A careful treatment of these cases will be given in the main text.
With this careful understanding of anomalies in place, we can adapt the $a$-maximization algorithm \cite{Intriligator:2003jj} to the holomorphically twisted setting.
In particular, the same principles that determine the exact superconformal R-symmetry in the untwisted theory now guide us in identifying the correct twisting homomorphism allowing us to extract the IR central charges $a$ and $c$.

Having established a general framework for $\cN=1$ theories, we turned to the case of extended supersymmetry in Section \ref{sec:extended-SUSY}.
The presence of additional supercurrents enlarges the structure of the holomorphic symmetry algebra, leading to supersymmetric generalizations of the higher Virasoro algebra \cite{Saberi:2019fkq}.
For instance, in $\cN=2$ theories the stress tensor multiplet contains several semi-chiral superfields $(\bS_i\,, \bR\,, \bG_i\,, \widetilde{\bG})$, whose brackets generate a super-analogue of $\lie{vir}(2)_{a,c}$ that we call $\lie{svir}(2|1)_{a,c}$.
We provide a detailed account of these algebras for $\cN=2,3,4$, including super-geometric realizations thereof as dg Lie superalgebras of derived holomorphic vector fields on suitable complex super-manifolds.
Anomalies again appear as ternary brackets among these generators, and in cases with $\cN\geq 3$ supersymmetry we recover the well-known equality $a=c$ \cite{Aharony:2015oyb, Sohnius:1978wk, Sohnius:1981sn} from the uniqueness of the associated 3-cocycle \cite{Fuks1986}.
While the non-centrally extended algebra can be expressed invariantly as currents on some (punctured) superspace, we have not been able to derive expressions for the three-cocycles as invariant expressions on superspace in all examples.
For instance, we know that the twisted $\mathcal{N}=2$ superconformal symmetry is holomorphic vector fields on a superspace with two complex bosonic directions and one complex odd direction. 
In this example, we do not know super-geometric expressions for the three-cocycles which correspond to $a,c$---but, we do know that the space of three-cocycles modulo equivalence is nevertheless two-dimensional \cite{pimenov}.
On the other hand, we do have component-level expressions for the two independent three-cocycles at the level of the twist.

In the presence of extended supersymmetry, a variety of inequivalent twists emerge, leading to holomorphic–topological, topological, or even lower-dimensional theories \cite{Eager:2018dsx, Elliott:2020ecf, Elliott:2024jvw}. 
These additional twists can often be realized as deformations of the minimal holomorphic twist by specific semi-chiral superfields \cite{Elliott:2015rja,Saberi:2019fkq}.
We focus on a particularly important example of this in Section \ref{sec:deformations}: the {\it superconformal deformation} of \cite{Saberi:2019fkq} implemented by turning on a background for the twisted $\cN=2$ superconformal symmetry and realizes the SCFT/VOA correspondence of \cite{Beem:2013sza} as a deformation of the holomorphic twist.
This deformation localizes the four-dimensional holomorphic theory to the plane $z^2=0$, giving rise to a genuine two-dimensional holomorphic field theory.
We illustrate this explicitly in the case of the free hypermultiplet, where the resulting algebra of local operators is precisely the symplectic boson VOA.
More generally, we develop an explicit and systematic formalism that allows one to extract the vertex algebra structure directly from the four-dimensional holomorphic $\lambda$-brackets.

%-------------------
\subsection{Outlook}
%-------------------

The holomorphic twist offers a natural categorification of the superconformal index: it upgrades the index from a protected numerical invariant to a cohomologically graded operator algebra. 
This refined structure retains detailed algebraic and geometric information about the theory, providing access to the operator product expansion (OPE), symmetry algebra, and deformation patterns that are invisible to the index alone. In particular, the holomorphic twist yields a sensitive diagnostic for detecting (super)symmetry enhancement along the renormalization group (RG) flow, allowing one to probe subtle IR phenomena in a controlled, computable framework.

This work opens many interesting future directions. We list a few notable ones:

\begin{itemize}
    \item An application of these ideas appears in our upcoming study of twisted Argyres–Douglas theories \cite{upcoming1}. Building on the symmetry structures developed here, we will analyze $\cN=1$ Lagrangian constructions of Argyres–Douglas theories in the spirit of \cite{Maruyoshi:2016tqk,Maruyoshi:2016aim,Agarwal:2016pjo,Maruyoshi:2018nod,Cho:2024civ}, focusing on the emergence of an $\cN=2$ structure in the holomorphic twist. The holomorphic perspective provides a sharper and more robust test of their proposed IR supersymmetry enhancement than previously available. In addition, this study will clarify various aspects of RG flows in holomorphic theories.
    \item As emphasized in Section~\ref{sec:aext}, our formulation of holomorphic $a$ maximization is not fully intrinsic to the holomorphic setting, but rather imports certain ingredients from the untwisted physical theory. In particular, the role of the exact superconformal R-symmetry remains somewhat opaque from a purely holomorphic perspective. A natural next step would be to identify additional structures in the twisted theory that justify, from first principles, the emergence of a single tensor structure in the ternary $\lllbrak{\bS_i}{\bS_j}{\bJ_f}$ bracket as in Eq.~\eqref{eq:ssj-superconformal}. Such an understanding would complete an intrinsically holomorphic proof of $a$-extremization. To further establish $a$-maximization on holomorphic grounds, one would also need a sharper notion of unitarity in the twisted setting. Recent progress in the VOA sector of $\cN=2$ SCFTs \cite{ArabiArdehali:2025fad} has introduced the idea of graded unitarity as the appropriate positivity criterion in VOAs, corresponding to four-dimensional unitarity in the parent SCFT. It is a compelling open question whether this concept, or some generalization thereof, can be extended to the holomorphic twist. 
    \item We have shown that in the presence of extended supersymmetry, additional twists can often be realized as deformations of the holomorphically twisted theory itself. A key example is the superconformal twist, obtained by deforming the minimal twist by $z^2 \wti{\bG}$. Motivated by this construction, one may also consider more exotic deformations by $(z^2)^m \wti{\bG}$, which are reminiscent of the so-called R-twists studied in \cite{Cecotti:2010fi,Cecotti:2011iy,Cecotti:2015lab} and revisited more recently in \cite{Dedushenko:2023cvd,Gaiotto:2024ioj,ArabiArdehali:2024vli,Kim:2024dxu}. Clarifying the precise relationship between these two perspectives would be valuable, and could perhaps provide a concrete framework for analyzing such generalized twists directly within the holomorphic twist. 
    \item For twists of $\cN=1$ superconformal field theories, we have a geometric description of the central extensions of the higher Witt algebra in terms of the degree four universal characteristic classes $(c_1^3, c_1 c_2)$. 
    From these classes, we can associate the explicit $3$-cocycles which define the higher Virasoro algebras \cite{Williams:2024mcc}.
    To better understand the twist of more supersymmetric examples, it would be beneficial to have a supergeometric interpretation of the central extensions.
    For example, when $\cN=2$, we know that the holomorphic twist of the superconformal algebra admits a two-parameter family of extensions corresponding to the \textit{degree four} combinations of Chern classes $(c_1^2, c_2)$ \cite{pimenov}.
    It would be useful to understand the correspondence of these classes with explicit formulas for the corresponding $3$-cocycles appearing in $\mathfrak{svir}(2|1)$.
    \item In our discussion of ABJ anomalies we commented on the appearance of discrete (non-invertible) symmetries, replacing the original anomalous symmetry \cite{Cordova:2022ieu,Choi:2022jqy}. We demonstrated how the same mechanism arises in holomorphic field theories. Unlike in the physical parent, the spectrum of line operators is trivial in $\bQ$-cohomology, so the non-invertible character of these symmetries is not immediately manifest. A more detailed study of their action on holomorphic surface operators may clarify how this non-invertible structure is realized in the holomorphic setup. More interesting is that this construction applies equally well to holomorphic spacetime symmetries. When conformal symmetry is broken by an ABJ anomaly, a discrete non-anomalous dilation symmetry survives. It would be interesting to study this in more detail, and more generally discuss how spacetime symmetries and twists can be embedded into the symTFT framework, cf. \cite{Apruzzi:2025hvs}.
\end{itemize}

%-------------------
\subsection{Structure of the paper}
%-------------------

Section~\ref{sec:holo-twist} introduces the holomorphic twist of four-dimensional supersymmetric field theories and the algebraic structures underlying the twisted theories. In Section~\ref{sec:symmetries}, we investigate the structure of symmetries and anomalies in the twisted theory, develop a descent formalism and anomaly inflow for holomorphic theories, and establish explicit connections to the corresponding constructions in the untwisted setting. Section~\ref{sec:extended-SUSY} focuses on theories with extended supersymmetry, where we analyze the enhanced algebraic structures that arise in the holomorphic twist. In Section~\ref{sec:deformations}, we explore superconformal deformations of the twisted theory and demonstrate how the associated vertex operator algebra can be recovered. Finally, Appendices~\ref{app:conventions} and~\ref{app:4dSCA} summarize our conventions and provide an overview of the relevant superconformal algebras.

%%%%%%%%%%%%%%%%%%%%%%%%%%%%%%%%%%%%%%%%%
\section{The Holomorphic Twist}         %
\label{sec:holo-twist}                  %
%%%%%%%%%%%%%%%%%%%%%%%%%%%%%%%%%%%%%%%%%

Consider a four-dimensional Euclidean SQFT with $\mathcal{N} = 1$ supersymmetry. Such a theory preserves (at least) four supercharges, denoted $\cQ_\alpha$ and $\wti{\cQ}_{\dota}$, where $\alpha = \pm$ and $\dota = \dot{\pm}$. These supercharges satisfy the following anti-commutation relations:\footnote{When $\cN>1$ the supersymmetry algebra can be centrally extended. We will not consider such central extensions in this work.}
\begin{equation}
    \acomm{\cQ_\alpha}{\wti\cQ_{\dota}} = \, \cP_{\alpha\dota}\,,\qquad\qquad \acomm{\cQ_\alpha}{\cQ_\beta} = \acomm{\wti\cQ_{\dota}}{\wti\cQ_{\dotb}} = 0\,,
\end{equation}
where $\cP_{\alpha\dota}$ are the translation generators. The supercharges are rotated by the R-symmetry $G_R$. For a detailed review of the four-dimensional superconformal algebra, we refer the reader to Appendix~\ref{app:4dSCA}.

For the purposes of this paper, performing a twist (minimal or otherwise) of the theory consists of the following steps: 
\renewcommand\labelenumi{(\theenumi)}
\begin{enumerate}
    \item Choose a nilpotent supercharge $\bQ$ in the supersymmetry algebra. The space of such supercharges is called the \emph{nilpotence variety} \cite{Eager:2018dsx}, see also \cite{Elliott:2020ecf}. The nilpotence variety is acted upon by rotations, parity transformations, and $R$-symmetries and is stratified by the number of momenta that are homotopically trivial in the twist; a \emph{minimal} twist is one where the fewest number of momenta are trivial.
    \item Add $\bQ$ to the BV/BRST differential of the theory and treat the total differential $\dd_{BRST}+ \bQ$ as the BV/BRST differential of the twisted theory. The space of observables in the twisted theory is the cohomology with respect to this differential. 
\end{enumerate}
It is often advantageous to extend this procedure by two additional, albeit optional, steps. These steps provide the necessary data to, first, place the twisted theory on curved spacetimes endowed with appropriate tangential structures  (such as a complex structure or, more generally, a transverse holomorphic foliation) and second, equip the theory with a cohomological grading.
\begin{enumerate}[resume]
    \item Choose a homomorphism $\phi: G \rightarrow {\rm Spin}(d) \times G_R$ under which the preferred supercharge $\bQ$ transforms trivially. This gives rise to a new action of the (reduced) Lorentz group on the fields of the twisted theory. Using this new action we have changed the spins of the fields as well as the supercharges. For the minimal twist of four-dimensional theories, such a twisting homomorphism is often taken to be of the form
    \begin{equation}
        \phi: {\rm MU}(2,\bC) \rightarrow {\rm Spin}(4)\times G_R\,,
    \end{equation}
    where ${\rm MU(2,\bC)}$ is the metaplectic unitary group, a double cover of $\UU(2)$.
    \item Choose a homomorphism $\alpha: \UU(1) \to G_R$ under which the preferred supercharge $\bQ$ transforms non-trivially. Such a choice allows us to define a cohomological grading by modifying ghost number by suitable $R$-charges so that $\bQ$ has a well-defined charge.
\end{enumerate}
\renewcommand\labelenumi{\theenumi.}
We note that we have not placed the requirement that the homomorphism $\alpha$ gives $\bQ$ unit charge as this need not be possible. 
A situation where this is the case is in theories with fractional $R$-charges; this is simply saying the true $\UU(1)$, i.e. the one which gives all observables integral charges and has no subgroups acting trivially, does not give the supercharges charges $\pm 1$.
We also note that it is possible to extend the allowed twisting homomorphisms and gradings by replacing $G_R$ by its product with the flavor symmetries of the untwisted theory; this can often allow for less restrictive backgrounds, e.g. by foregoing or relaxing the need for spin structures, and can be used to correlate the Grassmann parity of observables with the cohomological grading $\!\!\mod 2$. 

The four-dimensional $\cN = 1$ supersymmetry algebra contains a unique nilpotent supercharge.\footnote{More precisely, any other supercharge is in the same orbit under (complexified) rotations, $R$-symmetries, and parity transformations.} Without loss of generality, we choose $\bQ = \cQ_-$ as the twisting supercharge. This supercharge is holomorphic, i.e. the image $\acomm{\bQ}{\bullet}$ is half the dimension of spacetime \cite{Costello:2011np,Saberi:2019ghy,Elliott:2020ecf,Elliott:2024jvw}. More explicitly, the choice of supercharge singles out a complex structure of $\bbR^4 \cong \bbC^2$ in which $(z^{\dota},\bz^{\dota}) = (x^{+\dota},x^{-\dota})$ are holomorphic and anti-holomorphic coordinates, respectively. A direct computation shows that the anti-holomorphic translations are $\bQ$-exact,
\begin{equation}
    \acomm{\bQ}{\wti\cQ_{\dota}} = \cP_{-\dota} = -\ii \,\del_{-\dota} = -\ii\,\del_{\bz^{\dot{\alpha}}}\,.
\end{equation}
Consequently, after passing to $\bQ$-cohomology, the twisted theory becomes holomorphic in a cohomological sense: the anti-holomorphic translations act trivially, and the operator product expansion (OPE) is meromorphic. However, by Hartogs' lemma -- which forbids the existence of meromorphic functions on $\bbC^2$ with isolated singularities -- the OPE in $\bQ$-cohomology must be non-singular. Thus, the operators surviving in cohomology form a ring structure, referred to as the semi-chiral ring \cite{Budzik:2023xbr}, by analogy with the chiral ring. Beyond this ring structure, the twisted theory carries a rich complex geometric and algebraic structure which we will discuss in the remainder of this and the next section.

\textit{Comment on notation:} In the twisted theory, the dependence on the un-dotted spinor indices $\alpha$ is cohomologically trivialized. 
To streamline the notation, we rename the remaining dotted indices $\dota$ as $i = 1,2$. Concretely, we identify the holomorphic coordinates as $z^1 = z^{\dot{+}}$ and $z^2 = z^{\dot{-}}$, and similarly for the anti-holomorphic coordinates.

%-------------------
\subsection{Semi-chiral superfields}
%-------------------

It is often useful to formulate $\cN = 1$ SQFTs in a manifestly supersymmetric way using $\cN=1$ superfields. In this formalism, spacetime is promoted to a supermanifold by adding odd coordinates $\theta^\alpha$ and $\bar\theta^{i}$. The components of a superfield $\cO(z,\bz,\theta,\bar\theta)$ are related through successive actions of the supersymmetry generators, thereby encoding an entire multiplet within a single superfield.\footnote{Unless otherwise specified, all superfields are understood to be $\cN=1$ superfields, allowing us to uniformly treat theories with $\cN \geq 1$ supersymmetry.} Further details on our conventions for superfields and supercovariant derivatives can be found in Appendix \ref{app:4dSCA}.

In the twisted theory, 
the coordinates $\theta^\alpha$ and $\bar\theta^{i}$ transform respectively as scalars and anti-holomorphic one-forms. For this reason we can identify $\bar\theta^{i} = \ii\,\dd\bz^{i}$ and introduce the reduced superfield
\begin{equation}
    \bO \equiv \e^{-\ii \dd\bz^{i} \wti\cQ_{i}}\bO^{(0)} = \bO^{(0)} + \bO^{(1)} + \bO^{(2)}\,,
\end{equation}
obtained by setting $\theta^\alpha = 0$ in the conventional superfield. We then define a semi-chiral superfield as a reduced superfield that is annihilated by the supercovariant derivative $D_-$,
\begin{equation}
    D_- \bO = (\bQ - \bar\del)\bO = 0\,,
\end{equation}
where we introduced $\bar\del = \dd \bz^i \, \del_{\bz^i}$. Similarly, we define a semi-chiral field to be a field annihilated by $\bQ$. Note that from the definitions above it follows that the bottom component of a semi-chiral superfield is a semi-chiral field, i.e. $\bQ \bO^{(0)}=0$, while its higher form components, often called descendants, satisfy the holomorphic descent equations,
\begin{equation}
    \bQ \bO^{(k)} = \bar\del \bO^{(k-1)}\,.
\end{equation}
Moreover, given a semi-chiral field $O$ we can build a semi-chiral superfield $\bO$ with $O = \bO^{(0)}$ by applying $\e^{-\ii \dd \bz^i \wti\cQ_i}$.
In other words, semi-chiral superfields are in one-to-one correspondence with $\bQ$-closed local operators. Moreover, since BPS observables do not depend on $\bQ$-exact deformations, the algebraic structure defined by descent is sensitive to the cohomology classes of semi-chiral operators. 

%-------------------
\subsection{Higher operations}
\label{sec:higheroperations}
%-------------------

As already noted, Hartogs’ lemma prohibits the appearance of singular terms in the OPE of $\bQ$-closed local operators in a holomorphic theory in two complex dimensions: the correlation functions thereof are holomorphic away from the diagonals where insertion points are coincident and hence must extend over these diagonals. Nevertheless, there \emph{can} be singularities in OPEs of operators that are not $\bQ$-closed, e.g. the descendants appearing in a semi-chiral superfield. Holomorphic theories are equipped with a collection of $n$-ary secondary operations that allow us to access singular (physical) OPE data by probing simultaneous multi-operator configurations. These operations arise naturally from the derived structure of the space of local operators: instead of forming a strict associative algebra, local operators in the twisted theory organize into a homotopy algebra, typically a dg Lie algebra or more generally $L_\infty$-algebra, where higher brackets encode obstructions to strict associativity and Jacobi identities. Concretely, these secondary operations are defined by integrating suitable descendant operators over compact cycles in configuration space, thereby capturing the residual singularities associated with multi-point collisions. We now turn to a more explicit construction of these secondary products. For a detailed perturbative analysis of these brackets and their interrelations, see \cite{Gaiotto:2024gii}.

Consider a patch of spacetime with semi-chiral operator $O_1$ placed at a point $z^i$ and a second $O_2$ at the origin. Just as in the more familiar setting of complex dimension 1, we can extract the operators appearing in their OPE of these operators and their descendants via $S^3$ integrals of the semi-chiral superfields $\bO_1$ around $\bO_2$ of the form
\begin{equation}
    \oint_{S^3} \f{\dd^2 z}{(2\pi \ii)^2} \rho(z) [\bO_1(z) \bO_2(0)]\,,
\end{equation}
where the (binary) product of operators, $\left[\bO_1(z)\bO_2(0)\right]$, is understood as the OPE and $\rho(z) \in \Omega^{0,\bullet}(\bC^2 \backslash \{0\})$. This integral is again a semi-chiral superfield when $\rho$ is holomorphic $\overline{\del} \rho = 0$ and is $\bQ$-exact if $\rho = \overline\del \wti\rho$, i.e. the $\bQ$-cohomology class of the resulting semi-chiral operator only depends on the class $[\rho] \in H^{0,\bullet}(\bC^2\backslash\{0\})$.
The Dolbeault cohomology, $H^{0,\bullet}(\bC^2 \backslash \{0\})$, is concentrated in degrees zero and one and we define a basis of representatives as follows,
\begin{equation}\label{eq:dolbeault-representatives}
    \rho_{m,n} = \left\{\begin{array}{lll}
        \left(z^1\right)^m \left(z^2\right)^n & \qquad\in\Omega^{0,0}(\bC^2 \backslash \{0\})\qquad & m,n\geq 0\,,\\[2mm]
        \del_{z^1}^{-m-1}\del_{z^2}^{-n-1}\omega_{\rm BM} & \qquad\in\Omega^{0,1}(\bC^2 \backslash \{0\})\qquad & m,n < 0\,,\\[2mm]
        0 & ~ & \text{otherwise}\,,
    \end{array}\right.
\end{equation}
where the Bochner-Martinelli $\omega_{\rm BM}$ is the Green's function for $\bar\del$ on $\bC^2$ and can be written explicitly as
\begin{equation}
        \omega_{\rm BM} = \frac{1}{(2\pi \ii)^2} %\f{1}{(2\pi \ii)^2}
        \f{\bz^1\dd\bz^2 - \bz^2 \dd \bz^1}{|z|^4}\in \Omega^{0,1}\left( \bC^2 \backslash \{0\} \right)\,.
\end{equation}
This allows us to define an infinite collection of binary secondary brackets as
\begin{equation}
    \acomm{\bO_1}{\bO_2}_{m,n} =\, \oint_{S^3} \f{\dd^2 z}{(2\pi \ii)^2} \,\rho_{m,n}\,\left[\bO_1(z)\bO_2(0)\right]\,.
\end{equation}
Hence we see that, analogous as in one(-complex)-dimensional VOAs, the positive modes encode the singular part of the OPE, while the negative modes capture the regularized products $\del_{z^1}^m\del_{z^2}^n \bO_1(z)\bO_2(0)$.

The binary $\lambda$-bracket is then defined as the generating function collecting all positive modes\footnote{The generating function of negative modes will be reviewed later in section \ref{sec:2dbracketfrom4d} and appendix \ref{app:bracketaxiom}.},
\begin{equation}
    \llbrak{\bO_1}{\bO_2} = \oint_{S^3}\f{\dd^2 z}{(2\pi\ii)^2}\e^{\lambda\cdot z}[\bO_1(z)\bO_2(0)]\,,
\end{equation}
where $\lambda\cdot z = \lambda_i z^i$
As such, the $\lambda$-bracket can be understood as the Fourier transform of the singular part of the OPE.

It is useful to consider the space $\bC^2 \backslash \{0\}$ as a configuration space of two points with one point fixed at the origin, 
\begin{equation}
    \bC^2 \backslash \{0\} = {\rm Conf}^0_2(\bC^2)\,.
\end{equation}
With this in mind, we can rewrite the above $S^3$ integral as an integral over this configuration space:
\begin{equation}
\begin{aligned}
    \acomm{\bO_1}{\bO_2}_{m,n} =&\, \oint_{S^3} \f{\dd^2 z}{(2\pi \ii)^2} \,\rho_{m,n}\,\left[\bO_1(z)\bO_2(0)\right]\\
    =&\, \int_{{\rm Conf}^0_2} \f{\dd^2 z}{(2\pi\ii)^2} \rho_{m,n} \bar\del \left[\bO_1(z)\bO_2(0)\right]\\
    =&\, \int_{{\rm Conf}^0_2} \f{\dd^2 z}{(2\pi\ii)^2} \rho_{m,n} \bQ \left[\bO_1(z)\bO_2(0)\right]\,.
\end{aligned}
\end{equation}
The definitions above can then be formally generalized to arbitrary $(n+1)$-ary secondary brackets defined as\footnote{These brackets are defined at the level of cohomology, where $\rho$ is $\bar\del$-closed and the integration domain is a cycle. More generally, when working at the chain level, one can consider brackets defined by integrating non-closed forms over arbitrary chains (not necessarily cycles), which may yield homotopically meaningful but cohomologically trivial contributions.} 
\begin{align}
    \left\{\bO_1\,,\,\bO_2\,,\,\cdots\,,\,\bO_{n+1}\right\}_{\Sigma,\rho} =
    \int_{\Sigma} \prod_{k=1}^n\f{\dd^2 z_k}{(2\pi\ii)^2} \rho \, \bQ \left[\bO_1(z_1)\bO_2(z_2)\cdots \bO_{n+1}(0)\right]\\
    =
    \int_{\Sigma} \prod_{k=1}^n\f{\dd^2 z_k}{(2\pi\ii)^2} \rho \, \bar\del \left[\bO_1(z_1)\bO_2(z_2)\cdots \bO_{n+1}(0)\right]\\
    =
    \int_{\del\Sigma} \prod_{k=1}^n\f{\dd^2 z_k}{(2\pi\ii)^2} \rho \, \left[\bO_1(z_1)\bO_2(z_2)\cdots \bO_{n+1}(0)\right]\,,
\end{align}
where $\Sigma$ denotes a (possibly unbounded) chain in ${\rm Conf}^0_{n+1}(\bC^2)$, the configuration space of $n+1$ points with one point at the origin, and $\rho$ represent a cohomology class $\rho \in H^{0,\bullet}({\rm Conf}^0_{n+1}(\bC^2))$. 
The product of operators $\left[ \bO_1\cdots \bO_{n+1} \right]$ appearing in this integral can be thought of as a $(k+1)$-ary generalization of the OPE, depending on the integration cycle in configuration space.%
\footnote{In the physical theory, the strict associativity of the OPE ensures that all multi-operator collision limits are fully determined by successive applications of the binary OPE; no genuinely new data arises from simultaneous multi-point collisions. In the holomorphic twist, however, the projection onto $\bQ$-cohomology removes unprotected operators from the spectrum. This projection disrupts the strict associativity, resulting in an OPE that is associative only up to homotopy. Consequently, multi-point collision limits in the twisted theory cannot in general be reconstructed solely from the binary OPE data of protected operators. Instead, the higher brackets encode additional, cohomologically protected information about the multi-operator dynamics which is not accessible from the binary OPE alone, though it ultimately originates from the underlying physical OPE structure.}
Although the Dolbeault cohomology of such configuration spaces currently lacks a concrete algebraic model or explicit basis of representatives, it is straightforward to see that the zeroth degree cohomology $H^{0,0}({\rm Conf}^0_{n+1}(\bC^2))$ is generated by monomials in $n$ complex variables $z_1^{k_1}\cdots z_n^{k_n}$.

Hence, generalizing the definition above, we define the $(n+1)$-ary $\lambda$-bracket as
\begin{equation}\label{eq:n-bracket-np}
    \left\{\bO_1\,{}_{\lambda_1}\,\bO_2\,{}_{\lambda_2}\,\cdots\,{}_{\lambda_n}\,\bO_{n+1}\right\} = \int_{{\rm Conf}^0_{n+1}} \prod_{k=1}^n\f{\dd^2 z_k}{(2\pi\ii)^2} \e^{\lambda_k\cdot z_k} \bQ \left[\bO_1(z_1)\bO_2(z_2)\cdots \bO_{n+1}(0)\right]\,,
\end{equation}
The ($n>2$)-ary brackets introduced above should be understood as formal expressions, since at the time of writing there is no fully satisfactory non-perturbative definition of the generalized OPEs that appear in these constructions. In perturbation theory we can be more precise. Define the free $\lambda$-brackets with a subscript $0$
\begin{equation}
    \left\{\bO_1\,{}_{\lambda_1}\,\cdots \bO_{k}\,{}_{\lambda_k}\bO_{k+1}\right\}_0 = \sum_{\Gamma}\left( \int_{\bC^{2n}}\prod_{k=1}^n\f{\dd^2 z_k}{(2\pi \ii)^2} \e^{\lambda_k\cdot z_k} W_{\Gamma}\left[\bQ\big(\bO_1(z_1)\cdots\bO_n(z_n)\bO_{n+1}(0)\big)\right] \right)\,,
\end{equation}
where $W_\Gamma[\cdots]$ corresponds to performing Wick contractions in the free theory according to the graph $\Gamma$ and we sum over all 2-Laman graphs with $n+1$ nodes \cite{Budzik:2022mpd,Gaiotto:2024gii}. The interacting bracket can then be recovered in perturbation theory by including higher-loop diagrams with additional insertions of the interactions $\cI$:
\begin{equation}\label{eq:pert-int}
    \left\{\bO_1\,{}_{\lambda_1}\,\cdots \bO_{n}\,{}_{\lambda_n}\bO_{n+1}\right\} = \sum_{l=0}^\infty \f{\hbar^l}{l!} \big\{ \underbrace{\cI\,{}_0\,\cdots\,\cI\,{}_0\,}_{l \text{ times}} \bO_1\,{}_{\lambda_1}\,\cdots \bO_{n}\,{}_{\lambda_n}\bO_{n+1}\big\}_0 \,,
\end{equation}
where we added a formal loop counting parameter $\hbar$ and the $l$ additional $\lambda$ parameters are set to zero. 

Similarly, until now we always considered the differential $\bQ$ abstractly as a fully quantum corrected object. However, in practice we are always forced to use the perturbative quantum corrected operator $\bQ$ which can be expressed as
\begin{equation}
    \bQ = \sum_{l=0}^\infty \hbar^l \bQ_l \,,
\end{equation}
where $\bQ_l$ is the $l$-loop supercharge whose action on a field is defined as the free $l$-loop $\lambda$-bracket,
\begin{equation}
    \bQ_l\cdot \bO = \f{1}{(l+1)!}\big\{ \underbrace{\cI\,{}_0\,\cdots\,\cI\,{}_0\,}_{l+1 \text{ times}} \bO \big\}_0\,.
\end{equation}
%

%-------------------
\subsection{Lagrangian \texorpdfstring{$\cN \geq 1$}{N >= 1} theories}
\label{sec:N=1example}
%-------------------

To make these constructions more concrete, we now present a selection of examples that will serve as running illustrations throughout the text. 
All our examples consist of Lagrangian $\cN\geq 1$ SQFTs, which admit an explicit free field realization in terms of coupled $\mathbf{b}\mathbf{c}$-$\bfbeta\bfgamma$ systems, as formulated in \cite{Costello:2011np,Costello:2013zra,Saberi:2019ghy,Elliott:2020ecf,Budzik:2023xbr}. 

Consider a gauge theory with $\fg$ vector multiplets and a collection of chiral multiplets in representation $R$, possibly with a non-trivial holomorphic superpotential $W$. 
As shown in \cite{Costello:2013zra}, the holomorphically twisted theory can be obtained by adding the supercharge $\bQ$ to the BRST differential. 
Doing so carefully one obtains a holomorphically twisted theory equivalent to the following holomorphic field theory with BV action
\begin{equation}
    S_{\rm BV} = \int_{\bbC^2} \dd^2z\ \Tr \mathbf{b}\left(\bar{\del} \mathbf{c} + \frac12  [\mathbf{c},\mathbf{c}]\right) +  \bfbeta \left(\bar \del \bfgamma + \mathbf{c} \bfgamma \right) + W(\bfgamma), \label{eq:generalBVaction}
\end{equation}
where the fields appearing in this action are reduced superfields defined as
\begin{equation}
    \begin{aligned}
    \mathbf{b}& =\mathbf{b}^{(0)}+\mathbf{b}^{(1)}+\mathbf{b}^{(2)} \in \Omega^{0,\bullet}(\bbC^2,\fg^\vee), &
    \mathbf{c}&=\mathbf{c}^{(0)}+\mathbf{c}^{(1)}+\mathbf{c}^{(2)} \in \Omega^{0,\bullet}(\bbC^2,\fg)[1] \, ,\\
    \bfbeta&=\bfbeta^{(0)}+\bfbeta^{(1)}+\bfbeta^{(2)} \in \Omega^{0,\bullet}(\bbC^2,R^\vee)[1], \quad &  \bfgamma&=\bfgamma^{(0)}+\bfgamma^{(1)}+\bfgamma^{(2)} \in \Omega^{0,\bullet}(\bbC^2,R) \,.
    \label{eq:mattercontent}
\end{aligned}
\end{equation}
A chiral multiplet contains to an anti-chiral superfield $\chi$ satisfying $D_{\alpha} \chi = 0$ with equation of motion $\overline D^2 \chi = 0$. Each chiral multiplet gives rise to two semi-chiral superfields. We immediately recognize the bosonic semi-chiral superfield
\begin{equation}
    \bfgamma = \left. \chi \right|_{\theta^\alpha = 0}\,.
\end{equation}
In addition, the complex conjugate chiral superfield $\bar\chi$ contains a fermionic semi-chiral superfield given by
\begin{equation}
    \bfbeta= \left. D_+ \bar\chi \right|_{\theta^\alpha = 0} \,.
\end{equation}
These two superfields form a $\bfbeta\bfgamma$-system where we choose a normalization such that 
\begin{equation}
 \{\bfbeta_{n} \, {}_{\lambda}\, \bfgamma^{m}\} = \delta_m^n\,,
\end{equation}
where $m,n = 1,\dots, \dim R$ are indices for the representation $R$. As a simple modification, one can add a holomorphic superpotential $W(\chi)$ changing the equations of motion to $\overline D^2 \chi = \del_\chi W(\chi)$ which appears unchanged in the BV action \eqref{eq:generalBVaction}.

The holomorphic twist of a $\cN=1$ vector multiplet on the other hand gives rise to a holomorphic $\mathbf{b}\mathbf{c}$ BF theory, where we choose conventions such that 
\begin{equation}
    \left\{\mathbf{b}_A\, {}_{\lambda}\, \mathbf{c}^B \right\} = \delta_A^B \,,
\end{equation}
where $A,B$ are adjoint indices.
The identification of the elementary fields appearing in the BV action with the components of the vector multiplet is rather non-trivial and we refer the reader to the references above for more details.
The vector multiplet BPS letters with respect to the supercharge $\bQ$ are given by the self-dual $(2,0)$ part of the field strength, $F_{++}$, and the gaugino $\bar\lambda_{\dota}$.
Roughly speaking, these fields can be identified with the bottom components of the superfields $\dd^2 z \, \mathbf{b}^{(0)}$ and $\del_i \mathbf{c}^{(0)}$ respectively.
Entirely analogous to the two-dimensional $\beta\gamma$-system, the gauging of a flavor symmetry proceeds by coupling the $\bfbeta\bfgamma$ system to an adjoint valued $\mathbf{b}\mathbf{c}$ system describing the gauge sector.

%%%%%%%%%%%%%%%%%%%%%%%%%%%%%%%%%%%%%%%%%
\section{Symmetries and Anomalies}      %
\label{sec:symmetries}                  %
%%%%%%%%%%%%%%%%%%%%%%%%%%%%%%%%%%%%%%%%%

So far, the brackets introduced above are just a set of formal gadgets allowing us to extract certain pieces of (physical) OPE data. In what follows, we will revisit the interpretation of these brackets as generating symmetries of holomorphic field theories and show how the ternary brackets encode the associated anomalies.

%-------------------
\subsection{Infinite-dimensional symmetry enhancement}
%-------------------
Four-dimensional holomorphic field theories, such as those arising from the minimal twist of $\cN\geq1$ SQFTs, exhibit a rich variety of infinite-dimensional symmetry algebras, analogous in spirit to those familiar from two-dimensional (chiral) conformal field theory. 
However, unlike the two-dimensional case, it is generally insufficient to restrict attention to Lie algebras in the four-dimensional holomorphic setting. 
As we explain below, the natural higher-dimensional analogues of the Virasoro and affine current algebras are most appropriately described by $L_\infty$-algebras; see, for instance, \cite{faonteHigherKacMoodyAlgebras2019, gwilliamHigherKacMoodyAlgebras2018, Saberi:2019fkq, Bomans:2023mkd}. 
This is in large part due to the fact that configuration spaces of points in $\bC^n$ are not affine for $n>1$ and, consequently, the space of holomorphic functions on these spaces (where correlation functions naturally take values) exhibits non-trivial higher Dolbeault cohomology. 

%-------------------
\subsubsection{Spacetime symmetry}
%-------------------

The choice of a twisting supercharge $\bQ$ breaks the $\SO(4)$ Lorentz symmetry to $\SU(2)_+$. 
Together with the holomorphic translations, these generate the group of global holomorphic symplectomorphisms preserving the holomorphic symplectic form 
\begin{equation}
    \omega = \dd z^1 \wedge \dd z^2 \, .
\end{equation}
Infinitesimally, these symmetries are geometric: they are realized as holomorphic vector fields on $\bC^2$. 
More concretely, the $\SU(2)_+$ factor is generated by the traceless holomorphic rotations $z_i \del_j$. 
In the presence of an additional unbroken $\UU(1)$ R-symmetry, the diagonal combination 
\begin{equation}
    \diag(\UU(1)_R\times\UU(1)_{-})\subset \UU(1)_R\times \SU(2)_{-}
\end{equation}
is preserved. This enlarges the $\SU(2)$ above to the full $\UU(2)$ of holomorphic rotations. 
The extra $\UU(1)$ is generated by the trace
\begin{equation}
    E = z^i \del_i \, ,
\end{equation}
often referred to as the holomorphic Euler vector field.%
\footnote{This ``dilatation'' generator counts the total holomorphic degree.}

If the theory preserves $\cN=1$ superconformal symmetry, the spacetime symmetry in the twisted theory is given by the commutant subgroup $\SU(3) \subset \SU(4|1)$. 
The corresponding symmetry algebra is generated by the holomorphic translations $\del_i$, the rotations $z_i \del_j$, and the special conformal transformations $z_i E$.%
\footnote{Analogous to the familiar two-dimensional case, these are precisely the global holomorphic conformal vector fields that extend to $\bC\bP^2$.}

In the twisted theory, the finite-dimensional spacetime symmetry algebras are enhanced to infinite-dimensional ones. 
To see this, let us start by introducing the relevant infinite-dimensional algebras and subsequently show how they arise in holomorphically twisted theories. Consider the sheaf $T^{1,0}$ of holomorphic vector fields on the punctured affine space $\bC^2\backslash\{0\}$. 
Taking the derived global sections gives a natural differential graded (dg) Lie algebra, which we refer to as the 2-Witt algebra, $\lie{witt}(2)$. An explicit model for this algebra is given by the Dolbeault complex of the holomorphic tangent bundle: 
\begin{equation}
    \lie{witt}(2) \simeq \Omega^{0,\bullet}(\bC^2 \backslash \{0\}, T^{1,0}) \,,
\end{equation}
with Lie bracket
\begin{equation}\label{eq:witt2-bracket}
    [V_1,V_2] = \cL_{V_1}V_2\,.
\end{equation}
It will also be useful to introduce the subalgebra of divergence-free holomorphic vector fields,
\begin{equation}
    \lie{svect}(2) \subset \lie{witt}(2) \,,
\end{equation}
consisting of vector fields $V\in\lie{witt}(2)$ such that ${\rm div}\,V = \del_i V^i = 0$.

Consider now a unitary $\cN=1$ SQFT with a conserved stress tensor, sitting in a supercurrent multiplet. 
The most general supercurrent multiplet is the $\cS$-multiplet \cite{Komargodski:2010rb}, which admits improvements to the $\cR$-multiplet when an unbroken $\UU(1)_R$ symmetry is present \cite{Gates:1983nr}, and to the superconformal supercurrent multiplet when the theory is superconformal.%
\footnote{The $\cS$-multiplet satisfies
\begin{equation}
    D^\alpha \cS_{\alpha{\dota}} = \overline{D}_{\dota} \overline X + \overline\chi_{\dota}\,,\qquad D_\alpha \overline X = 0\,,\qquad D_\alpha \overline \chi_{\dota} = D_\alpha \chi^\alpha - \overline D^\dota \overline\chi_{\dota} = 0\,,
\end{equation}
where $\overline X$ is an anti-chiral superfield, and $\chi$ a complex linear superfield. In particular it follows that $D^2 S_{\alpha{\dota}} = 2\ii\del_{\alpha\dota} \overline X$. When the multiplet can be improved so that $\overline X=0$ we recover the $\cR$-multiplet. When the multiplet can be improved so that $\overline \chi=0$ we recover the FZ-multiplet. When the multiplet can be improved so that $\overline X = \overline \chi = 0$ we recover the superconformal supercurrent multiplet.}

Consider the reduced superfield $\bS_i = D^\alpha \cJ_{\alpha i} |_{\theta^{\alpha} = 0}$, where $\cJ$ is the relevant supercurrent superfield. If the theory possesses an unbroken R-symmetry, this is a semi-chiral superfield. In the absence of such symmetry, the relevant semi-chiral superfield is $\bS = \del^i \bS_i$.

Analogous to the two-dimensional case, the modes of the holomorphic stress tensor generate the infinite-dimensional symmetry algebra. Concretely, the action of a vector field $V=V^i(z)\,\del_i \in  \lie{witt}(2)$ on a local operator $\bO$ is given by
\begin{align}
    (-V \cdot \bO)(0) &= \oint_{S^3} \frac{\dd^2 z}{(2\pi \ii)^2} V^i(z) \bS_i(z) \,\bO(0)\,.
\end{align}
When $V$ is holomorphic, i.e. $\bar{\del} V = 0$, it is $\bar{\del}$-cohomologous to a vector field of the form $V^i(z) = \sum_{m,n} a^{mni} \rho_{m,n}$, where $\rho_{m,n}$ are representatives of $H^{0,\bullet}(\bC^2\backslash\{0\})$ as defined in \eqref{eq:dolbeault-representatives}. In this case the action of $V$ becomes
\begin{align}
    (-V \cdot \bO)(0) &= \sum_{m,n,i} a^{mni} \left\{ \bS_i, \bO \right\}_{m,n}(0)\,,
\end{align}
and in particular, for $V = e^{\lambda \cdot z} \del_i$ one recovers the binary $\lambda$-bracket of $\bS_i$ and $\bO$. As shown in \cite{Bomans:2023mkd}, the action as defined above satisfies the 2-bracket of $\lie{witt}(2)$ given in \eqref{eq:witt2-bracket} and is encoded, in part, by the binary $\lambda$-bracket,
\begin{equation}   
    \llbrak{\bS_i}{\bS_j} =\, \del_i\bS_j +  \lambda_i \bS_j +  \lambda_j\bS_i\,. \label{eq:ss}
\end{equation}

If the theory is superconformal, the semi-chiral property of $\bS_i$ ensures that the action of $\bQ$ on the above integral implements the differential on $\lie{witt}(2)$, and is equivalent to the conservation of the surviving components of the stress tensor in the twisted theory. 
In other words, whenever $\overline{\del}V^i = 0$, the codimension-one operators
\begin{equation}
    u_{\alpha}(M_3) = \alpha^{mn,i} \,\oint_{M_3} \f{\dd^2 z}{(2\pi \ii)^2}  \rho_{m,n} \bS_i\,,\qquad V^i = \alpha^{mn,i}\rho_{m,n}\partial_i\,,
\end{equation}
are topological and implement infinitesimal holomorphic spacetime transformations. Finite spacetime symmetry transformations are implemented by the topological codimension-one operators,
\begin{equation}
    U_{\alpha}(M_3) = \exp\left[\ii \,\alpha^{mn,i} \oint_{M_3} \f{\dd^2 z}{(2\pi \ii)^2}  \rho_{m,n} \bS_i \right]\,,
\end{equation}
analogous to the usual symmetry operators in physical theories. Deforming $M_3$ to $M_3^\prime$, one finds
\begin{equation}
    U_{\alpha}(M_3) U_\alpha(M_3^\prime)^{-1} = \exp\left[ \ii\, \alpha^{mn,i}\oint_{M_3\cup \overline{M_3^\prime}} \f{\dd^2 z}{(2\pi\ii)^2}\rho_{m,n} \bS_i \right] = \exp\left[ \ii\, \alpha^{mn,i}\int_{M_4} \f{\dd^2 z}{(2\pi\ii)^2}\rho_{m,n} \bar\del\bS_i \right]\,,
\end{equation}
with $\partial M_4 = M_3\cup \overline{M_3^\prime}$. Since $\bS_i$ is semi-chiral, this vanishes in $\bQ$-cohomology, showing that $U_{\alpha,i}$ is indeed topological.

If the theory does not possess an unbroken $\UU(1)_R$, the enhancement is to $\lie{svect}(2)$ instead. Restricting $V$ to be divergence-free implies $V^i = \epsilon^{ij} \del_j H$, i.e. $V$ is Hamiltonian with respect to $\omega$. Acting on $\bO$ gives
\begin{align}
    (V \cdot \bO)(0) 
    &= \oint_{S^3} \frac{\dd^2 z}{(2\pi \ii)^2} H(z) \, \del^i \bS_i(z) \,\bO(0)\,,
\end{align}
so as desired, $\lie{svect}(2)$ is generated by the modes of the semi-chiral superfield $ \bS = \del^i \bS_i$. 

The dg Lie algebra $\lie{witt}(2)$ admits no ordinary central extensions, but does possess two $L_\infty$ central extensions \cite{williamsHolomorphicSmodelIts,hennionGelfandFuchsCohomologyAlgebraic2022,Williams:2024xrq,Fuks1986}, resulting in the 2-Virasoro algebra $\lie{vir}(2)$. 
These central extensions can be expressed in terms of the Jacobian matrix $(JV)^i{}_j := \del_j V^i$ as
\begin{equation}
\begin{aligned}[]
    [V_1, V_2, V_3] &= C \oint_{S^3} \frac{\Tr (\del J V_1)\,\Tr (\del J V_2)\,\Tr (J V_3)}{(2\pi \ii)^2} \\
    &\quad + A \oint_{S^3} \frac{\Tr (\del J V_1 \,\del J V_2)\,\Tr(J V_3) + \Tr (\del J V_1)\,\Tr(\del J V_2\, J V_3)}{(2\pi \ii)^2} \\
    &\hspace{5cm} + \frac{\Tr (J V_3 \,\del J V_1)\,\Tr (\del J V_2)}{(2\pi \ii)^2} \, ,
\end{aligned}
\end{equation}
where $\del = \dd z^i \del_{z^i}$.
For more details on the derivation of these 3-cocycles, see Section 5.4 of \cite{Saberi:2019fkq}, equations (2.43) and (2.44) of \cite{Williams:2024xrq}, or Appendix D.4 of \cite{Bomans:2023mkd}; see also Section 5 of \cite{Williams:2024mcc} for higher-dimensional analogues.%
\footnote{We note that there is a typo in the cocycles appearing in Eq. (D.38) of \cite{Bomans:2023mkd} that is corrected in the above.} %
When $V\in\lie{svect}(2)$, these cocycles vanish since $\Tr(JV)=\Tr(\del JV)=0$. 

The constants $C$ and $A$ are proportional to the holomorphic Weyl and Euler anomalies:
\begin{equation}
    C = \frac{6}{\pi^2} c^{\rm hol}\,, \qquad A = \frac{1}{\pi^2} a^{\rm hol}\,,
\end{equation}
with \cite{Williams:2024xrq}
\begin{equation}
    c^{\rm hol} = \frac{3c-5a}{27}\,, \qquad a^{\rm hol} = \frac{2(a-c)}{3}\,.
\end{equation}
Indeed, the holomorphic conformal anomalies are simply proportional to the 't Hooft anomalies for $\UU(1)_R$, $c^{\rm hol} = -\tfrac{k_{RRR}}{48}$, $a^{\rm hol} = \tfrac{k_R}{24}$. 
Furthermore, the central extension can also be encoded in a ternary $\lambda$-bracket \cite{Bomans:2023mkd}:
\begin{equation}\label{eq:sss}
    \lllbrak{\bS_i}{\bS_j}{\bS_k} = 16 (a-c) \cI^{\rm vec}_{ijk} + \frac{4}{3}(3c-2a) \cI^{\cN=4}_{ijk} \,,
\end{equation}
where
\begin{equation}
\begin{aligned}
    \cI^{\rm vec}_{ijk}(\lambda_1, \lambda_2) &= -\frac{1}{24\pi^2} [\lambda_1 \lambda_2]^2 \big(\lambda_{1i} \epsilon_{jk} + \lambda_{2j} \epsilon_{ki} - (\lambda_{1k} + \lambda_{2k}) \epsilon_{ij}\big)\,,\\
    \cI^{\cN=4}_{ijk}(\lambda_1, \lambda_2) &= \frac{1}{3 \pi^2}[\lambda_1 \lambda_2]\lambda_{1i} \lambda_{2j}(\lambda_{1k} + \lambda_{2k})\,,
\end{aligned}
\end{equation}
and $[\lambda_1 \lambda_2] = \epsilon^{ij} \lambda_{1i} \lambda_{2j}$.

In the discussion above we considered infinite-dimensional dg Lie algebras at the chain level. For instance, we introduced the 2-Virasoro algebra,
\begin{equation}
    \lie{vir}(2)_{a,c} = \left( \Omega^{0,\bullet}(\bC^2\backslash\{0\},T^{1,0})\oplus A\oplus C,\bar\del,[-,-],[-,-,-] \right)\,.%/{\left(A-\f{2(a-c)}{3\pi^2},C-\f{2(3c-5a)}{27\pi^2}\right)}\,.
\end{equation}
This dg Lie algebra acts naturally on local operators, but not all of its elements should be interpreted as symmetries. The genuine symmetry algebra is instead captured by the Dolbeault cohomology,
\begin{equation}
    H(\lie{vir}(2)_{a,c}) = \left( H^{0,\bullet}(\bC^2\backslash\{0\}) , [-,-],[-,-,-]) \right)\,,
\end{equation}
since, as emphasized in \cite{Budzik:2023xbr,Bomans:2023mkd}, only closed vector fields give rise to topological operators, which is the hallmark of a symmetry.

Passing to cohomology, however, has a notable drawback: the three-bracket disappears, because the cohomology of an $L_\infty$ algebra always forms an ordinary graded Lie algebra. From the perspective of the holomorphic twist this loss is unsatisfactory.
On one hand, since the Jacobi identity for the non-centrally extended algebra of holomorphic vector fields on punctured space is a strict dg Lie algebra, we observe that the cocycles $a,c$ descend to the $\overline\partial$-cohomology thus defining the structure of an $L_\infty$ algebra on $H(\mathfrak{vir}(2)_{a,c})$ with only the $2$ and $3$-ary brackets being nonzero.
A more refined approach is to transfer the full $L_\infty$ structure from the co-chains to cohomology by homotopy transfer. In this way one recovers an infinite hierarchy of higher brackets: the binary bracket survives as usual, while the ternary bracket reappears with the same formula as at the chain level, and in addition one obtains nontrivial $n$-ary brackets for all $n$.

%-------------------
\subsubsection{Flavor symmetries}
%-------------------

If the SQFT enjoys a continuous flavor symmetry $\cF$, it contains a conserved current $J_\mu$ residing in a short multiplet.
From the previous section (see also \cite{Bomans:2023mkd}) it immediately follows that the twisted theory contains a semi-chiral superfield $\bJ$ containing the conserved current:
\begin{equation}
    \bQ \bJ = \overline{\del}\bJ\,.
\end{equation}
As in the case of spacetime symmetries, this implies that $\cF$ is enhanced to an infinite-dimensional symmetry algebra in the twisted theory.

A concrete model\footnote{A model is a (often simpler) chain complex that has the same cohomology as the original one, i.e. they are related by a quasi-isomorphism.} for this (dg) Lie algebra is given by $\ff$-valued Dolbeault forms
\begin{equation}
    \Omega^{0,\bullet}(\bC^2\backslash \{0\}) \otimes \ff\,,
\end{equation}
equipped with the Dolbeault differential, where $\ff = {\rm Lie}(\cF)$. An element $\chi = \chi^a T_a \in \Omega^{0,\bullet}(\bC^2\backslash \{0\}) \otimes \ff$ acts on local operators via
\begin{equation}
    \oint_{S^3} \frac{\dd^2 z}{(2\pi \ii)^2} \chi^a(z) \bJ_a(z)\,.
\end{equation}
The brackets are encoded, in part, in the binary $\lambda$-brackets of two currents
\begin{equation}
\label{eq:biflav}
    \llbrak{\bJ_a}{\bJ_b} = f^c_{ab} \bJ_c\,,
\end{equation}
where $f^c_{ab}$ are the structure constants of $\ff$ defined as $[T_a, T_b] = f^c_{ab} T_c$ (cf. Eq. (3.3) of \cite{Bomans:2023mkd}).
We can decompose the full flavor symmetry algebra as
\begin{equation}
    \ff = \ff_1 \oplus \ldots \oplus \ff_p \oplus \lie{u}(1)_1 \oplus \ldots \oplus \lie{u}(1)_q
\end{equation}
with $\ff_F$ a simple factor.
We denote the corresponding currents $\bJ_{Fa}$ ($F=1,\dots,p$ and  $a=1,\dots,{\rm dim}\ff_F$) and $\bJ_f$ ($f=1,\dots,q$).

Unlike the two-dimensional case, the (dg) Lie algebra $\Omega^{0,\bullet}(\bC^2\backslash \{0\}) \otimes \ff$ does not admit non-trivial strict Lie algebra central extensions.
There are, nevertheless, higher-dimensional versions of the Kac--Moody cocycle if one works in the derived setting.
The higher-dimensional analogues of the Kac--Moody central extension \cite{faonteHigherKacMoodyAlgebras2019, gwilliamHigherKacMoodyAlgebras2018, Saberi:2019ghy} appear instead as $L_\infty$ central extensions.
In the present setting, the central extension takes the form
\begin{equation}
   [\chi_1, \chi_2, \chi_3] = K ~ \oint_{S^3} \frac{1}{(2\pi \ii)^2} \tr\left(\del \chi_1 (\del \chi_2 ~ \chi_3 + \del \chi_3 ~ \chi_2)\right)\,.
\end{equation}
This encodes the possibility of a non-vanishing ternary bracket of three currents $\bJ$.
As we will see in more detail in Section \ref{sec:anomalies}, its non-zero value signals an anomaly of the $\cF$ symmetry in exactly the same way as a non-vanishing level $k$ for an affine current algebra in two dimensions signals the anomalous nature of the symmetry.
The coefficient $K$ is proportional to the corresponding pure flavor 't Hooft anomaly.
This ternary bracket on the centrally extended algebra is encoded in the ternary $\lambda$-bracket of three currents \cite{Bomans:2023mkd}\footnote{We note that the symbol $k_G$ appearing in loc. cit. is often used to denote the mixed flavor-gravitational-gravitational anomaly, also called the flavor level. We choose to use the more common notation $k_{FFF}$ to indicate that it is a pure flavor 't Hooft anomaly.}
\begin{equation}
\label{eq:jjj}
\begin{aligned}
    \lllbrak{\bJ_{F_a}}{\bJ_{F_b}}{\bJ_{F_c}} & = \frac{k_{FFF}}{2\pi^2} \tr_F\left(T_a (T_b T_c + T_c T_b)\right) [\lambda_1 \lambda_2]\,,\\
    \lllbrak{\bJ_{F_a}}{\bJ_{F_b}}{\bJ_{f}} & = \frac{k_{FFf}}{\pi^2} \tr_F\left(T_a T_b\right) [\lambda_1 \lambda_2]\,,\\
    \lllbrak{\bJ_{f}}{\bJ_{g}}{\bJ_{h}} & = \frac{k_{fgh}}{\pi^2} [\lambda_1 \lambda_2]\,.
\end{aligned}
\end{equation}

If the $\cN \geq 1$ theory has both super\emph{conformal} symmetry and a continuous flavor symmetry, there is an action of the semi-direct product $\lie{witt}(2) \ltimes \Omega^{0,\bullet}(\bC^2\backslash \{0\}) \otimes \ff$, where holomorphic vector fields act on the Dolbeault complex in the natural way.
As above, the brackets of this semi-direct product can be expressed between the binary $\lambda$-bracket
\begin{equation}
\label{eq:biflavgrav}
    \llbrak{\bS_i}{\bJ_a} = \del_i \bJ_a + \lambda_i \bJ_a\,,
\end{equation}
which simply states that $\bJ_a$ is a primary for the $\lie{vir}(2)$ symmetry (transforming in the canonical bundle $K$).

There are generally non-trivial ternary brackets involving both $\bS_i$ and $\bJ$.
There is a two-parameter family of brackets with two vector fields and one flavor transformation of the form
\begin{equation}
    [V_1, V_2, \chi] = K_1 \oint_{S^3} \frac{1}{(2\pi \ii)^2} \Tr (\del JV_1) \Tr (\del JV_2) \tr \chi + K_2 \oint_{S^3} \frac{1}{(2\pi \ii)^2} \Tr (\del JV_1 \del JV_2) \tr \chi\,,
\end{equation}
where, in a given theory, the central elements $K_1$ and $K_2$ are identified with mixed $R$-$R$-flavor and gravitational-gravitational-flavor 't Hooft anomaly coefficients, respectively.
These can only be non-zero for the abelian factors in $\lie{f}$ and give rise to ternary $\lambda$-brackets of the form
\begin{equation}
\label{eq:ssj}
    \lllbrak{\bS_i}{\bS_j}{\bJ_f} = \f{k_{RRf}}{4\pi^2}[\lambda_1 \lambda_2] \lambda_{1i} \lambda_{2j} - \frac{k_f}{12 \pi^2} [\lambda_1 \lambda_2]\lambda_{1j} \lambda_{2i}\,.
\end{equation}
There can also be brackets with one vector field and two flavor transformations of the form
\begin{equation}
    [V, \chi_1, \chi_2] = K_3 \oint_{S^3} \frac{1}{(2\pi \ii)^2} \Tr (\del JV) \tr (\del \chi_1 \chi_2 + \del \chi_2 \chi_1)\,,
\end{equation}
where $K_3$ gets identified with a mixed $R$-flavor-flavor 't Hooft anomaly.
This leads to ternary $\lambda$-bracket of the form
\begin{equation}
\label{eq:sjj}
\begin{aligned}
    \lllbrak{\bS_i}{\bJ_{Fa}}{\bJ_{Fb}} & = -\f{k_{RFF}}{2\pi^2} \tr(T_a T_b)[\lambda_1 \lambda_2] \lambda_{1i}\,,\\
    \lllbrak{\bS_i}{\bJ_{f}}{\bJ_{g}} & = -\f{k_{Rfg}}{2\pi^2}[\lambda_1 \lambda_2] \lambda_{1i}\,.
\end{aligned}
\end{equation}
We note that the 't Hooft anomaly $k_{RFF}$ is often replaced by the flavor level $k_F$; these are related by the statement that $N_f$ chiral multiplets with their superconformal $R$-charge $R = \f23$ has flavor $k_{SU(N_f)} = 1$, hence $k_F = 3 k_{RFF}$.
When the $\cN \geq 1$ theory is non-conformal and lacks an unbroken $\UU(1)_R$ symmetry, the 3-cocycles proportional to $K_1$ and $K_3$ become trivial, leaving only the $K_2$ cocycle non-vanishing. This is consistent with the fact that the former two are tied to anomalies of the broken $\UU(1)_R$ symmetry.

As for the spacetime symmetries, the genuine symmetry algebra is obtained only after passing to cohomology of the 2-Kac Moody dg Lie algebra, which isolates the closed elements that correspond to topological operators. Analogous as for the spacetime symmetries, we can form the topological codimension-one operators
\begin{equation}\label{eq:top-infinitesimal}
    u_{\alpha}(M_3) = \alpha^{mn,a} \,\oint_{M_3} \f{\dd^2 z}{(2\pi \ii)^2}  \rho_{m,n} \bJ_a\,,\qquad \chi = \alpha^{mn,a} \rho_{mn} T_a
\end{equation}
and their finite counterparts,
\begin{equation}\label{eq:top-finite}
    U_{\alpha}(M_3) = \exp\left[\ii\, \alpha^{mn,a} \oint_{M_3} \f{\dd^2 z}{(2\pi \ii)^2}  \rho_{m,n} \bJ_a \right]\,,
\end{equation}
analogous to the usual symmetry operators in physical theories.

%-------------------
\subsubsection{Lagrangian examples}
%-------------------

It is straightforward to construct explicit currents realizing the infinite-dimensional symmetries above in the Lagrangian examples of Section \ref{sec:N=1example}.
Consider first the case in which the physical $\cN=1$ theory possesses an unbroken $U(1)_r$ $R$-symmetry. 
In this situation, the classical BV action $S_{\rm BV}$ can be made invariant under infinitesimal holomorphic coordinate changes: the Lie algebra of holomorphic vector fields on $\bC^2$ acts as a symmetry of the holomorphically twisted theory \cite{Saberi:2019fkq}. 

Let $\chi_n$ denote a set of chiral multiplets, with $R$-charges $r_n$,%
\footnote{The $r$-charges must be assigned consistently so that all chiral multiplets in the same irreducible $\fg$-representation carry the same value of $r_n$.}  
chosen so that the superpotential $W(\gamma)$ is quasi-homogeneous of total weight~$2$. 
A general holomorphic vector field $V = V^i(z) \,\del_i$ then acts on the twisted superfields as
\begin{equation}
    \begin{aligned}
        \delta_V \mathbf{c} &= V^{i} \del_{i} \mathbf{c} \,, 
        & \delta_V \mathbf{b} &= V^{i} \del_{i} \mathbf{b} + (\del_{i}V^{i}) \,\mathbf{b}\,,\\
        \delta_V \bfgamma^n &= V^{i} \del_{i} \bfgamma^n + \frac{r_n}{2}(\del_iV^i)\, \bfgamma^n \,, 
        & \delta_V \bfbeta_n &= V^{i} \del_{i} \bfbeta_n + \Big(1-\frac{r_n}{2}\Big)(\del_{i}V^{i}) \,\bfbeta_n\,.
    \end{aligned}
\end{equation}
More invariantly, $\bfgamma^n$ is a Dolbeault $(0,\bullet)$-form valued in the $\f{r_n}{2}$-th power of the canonical bundle $K^{\f{r_n}{2}}$, and the above transformations are simply the corresponding holomorphic Lie derivatives. 
The associated Noether current is the holomorphic stress-tensor superfield,
\begin{equation}
    \mathbf{S}_i = -\, \mathbf{b} \,\del_i \mathbf{c} \,+\, \sum_n \left[ \bfbeta_n \,\del_i \bfgamma^n - \frac{r_n}{2} \,\del_i\!\left(\bfbeta_n \bfgamma^n\right) \right] .
\end{equation}
This symmetry survives at the quantum level, at least perturbatively, precisely when the $U(1)_r$ symmetry of the physical theory is not broken by an ABJ anomaly.

Similarly, one can construct holomorphic flavor currents whenever the physical theory admits a flavor symmetry $\cF$ acting linearly on the chiral multiplets, commuting with the $\fg$-gauge symmetry, and leaving the superpotential $\bm W$ invariant. 
Let $(T_a)^n{}_m$ denote the matrices representing the $\cF$ action in the chiral-multiplet space. 
The corresponding conserved holomorphic currents are 
\begin{equation}
    \bJ_a \,=\, -\bfbeta_n \,(T_a)^n{}_m \, \bfgamma^m \, .
\end{equation}
As in the stress-tensor case, these classical symmetries may fail to persist in the quantum theory due to ABJ-type anomalies. 
We will see in Section~\ref{sec:anomalies} how the presence of such anomalies modifies the symmetry algebra and the associated current conservation equations.

%-------------------
\subsection{Coupling to background fields}
\label{sec:backgrounds}
%-------------------

The conserved currents $\bS_i$ and $\bJ_a$ allow us to couple our holomorphic QFT to background fields for the corresponding symmetries; the current $\bS_i$ allows us to couple to a(n infinitesimal) deformation of complex structure $\mu = \mu^i \del_i \in \Omega^{0,1}(\bC^2, T^{1,0})$, i.e. a background Beltrami differential, and the current $\bJ_a$ allows us to couple to a(n infinitesimal) deformation of the background holomorphic $\cF$ bundle $A = A^a T_a \in \Omega^{0,1}(\bC^2)\otimes \ff$, i.e. a background (partial) connection.
These backgrounds must satisfy the corresponding Maurer-Cartan equations:
\begin{equation}
    \overline{\del} \mu + \tfrac{1}{2}[\mu, \mu] = 0\,, \qquad (\overline{\del} + \mu)A + \tfrac{1}{2}[A,A] = 0\,.
\end{equation}
Minimal coupling to these backgrounds amounts to deforming the action by%
\footnote{The fact that $A$ couples to $J$ but $\mu$ couples to $-\bS$ is tied to the fact that integrals of $V^i \bS_i$ generate the action of $- V^i \partial_i$.}%
\begin{equation}
    S[\mu, A] = S_0 + \int \dd^2 z ~ \left( A^a \bJ_a - \mu^i \bS_i\right)\,,
\end{equation}
where $S_0$ is the action with vanishing background fields.
Such a classical coupling satisfies the classical master equations and is invariant with respect to the background (infinitesimal) gauge transformation $\xi = (V, \chi) \in \Omega^{0,0}(\bC^2, T^{1,0}) \oplus \Omega^{0,0}(\bC^2)\otimes \ff$
\begin{equation}
    \delta_{\xi} \mu = \overline{\del} V + [\mu, V]\,, \qquad \delta_{\xi} A = (\overline{\del} + \mu)\chi + [A, \chi] + V A\,.
\end{equation}
The fact that $S[\mu, A]$ both solves the classical master equation and is gauge invariant can be encapsulated in a equivariant version of the classical master equation; see, for instance, Section 12.2 of \cite{costelloFactorizationAlgebrasQuantum2021} for a detailed discussion of coupling to background gauge fields for more general $L_\infty$ actions in classical field theories.
In brief, we begin by extending the background fields $A$ and $\mu$ to elements of the Chevalley-Eilenberg complex for the (dg) Lie algebra
\begin{equation}
    \cL = \Omega^{0, \bullet}(\bC^2, T^{1,0}) \ltimes \Omega^{0, \bullet}(\bC^2) \otimes \ff
\end{equation}
of holomorphic vector fields and flavor transformations, i.e.
\begin{equation}
    \mu \rightsquigarrow \bm{\mu} \in \Omega^{0,\bullet}(\bC^2, T^{1,0})[1]\,, \qquad A \rightsquigarrow \bA \in \Omega^{0,\bullet}(\bC^2) \otimes \lie{f}[1]
\end{equation}
with differential
\begin{equation}
    \dd_{\rm CE} \bm{\mu} = \overline{\del} \bm{\mu} + \tfrac{1}{2}[\bm{\mu},\bm{\mu}]\,, \qquad \dd_{\rm CE} \bA = (\overline{\del} + \bm{\mu})\bA + \tfrac{1}{2}[\bA, \bA]\,.
\end{equation}
We then consider the action
\begin{equation}
    S^\cL[\bm{\mu}, \bA] = S_0 + I^\cL = S_0 + \int \dd^2 z ~ \overset{\cI^\cL}{\overbrace{\left(\bA^a \bJ_a - \bm{\mu}^i \bS_i\right)}}\,.
\end{equation}
which then satisfies the $\cL$-equivariant classical master equation
\begin{equation}
    \dd_{\rm CE} S^\cL + \frac{1}{2} \{S^\cL, S^\cL\}_{\rm BV} = 0\,.
\end{equation}
The equivariant classical master equation encodes the condition that $\dd_{\rm CE} + \{S^\cL, -\}_{\rm BV}$ squares to zero. A action $S^\cL[\bm{\mu}, \bA]$ with $S^{\cL}[0,0] = S_0$ that satisfies this condition defines a \emph{classical $\cL$ background} for the field theory defined by $S_0$.

%-------------------
\subsection{Anomalies in holomorphic field theories}
\label{sec:anomalies}
%-------------------

We now turn to a description of (perturbative) anomalies for the infinite-dimensional symmetries described above; this is largely a recapitulation of the discussions of anomalies in \cite{williamsHolomorphicSmodelIts, gwilliamHigherKacMoodyAlgebras2018} (for the action of holomorphic vector fields) and \cite {Williams:2018ows, Saberi:2019ghy, Williams:2024xrq} (for the action of holomorphic flavor transformations) using a language more familiar to physicists.
See e.g. Chapter 13 of \cite{costelloFactorizationAlgebrasQuantum2021} for a detailed account of anomalous symmetries in the BV formalism and Chapter 14 thereof, as well as the references therein, for a variety of examples; in particular, \cite{Saberi:2019fkq} describes anomalies to the infinite-dimensional symmetry algebras we study in Section \ref{sec:extended-SUSY} appearing the holomorphic twist of theories with extended supersymmetry.

%-------------------
\subsubsection{Anomalies from Feynman diagrams}
\label{sec:diagrams}
%-------------------

The existence of a quantum anomaly for a classical $\cL$ symmetry can be formulated as a failure to satisfy the $\cL$-equivariant quantum master equation
\begin{equation}
\label{eq:equivQME}
    (\dd_{\rm CE} + \bQ) I^\cL + \f12 \{I^\cL, I^\cL\}_{\rm BV} = 0\,,
\end{equation}
where $\bQ$ is the BV/BRST differential of the QFT with background fields switched off and $\{-,-\}_{\rm BV}$ is the (suitably renormalized) BV bracket.
A local functional $I^\cL$ satisfying \eqref{eq:equivQME} is called an inner quantum $\cL$ background for the original QFT.
In this case, the $\cL$ symmetry survives quantization and is free of anomalies. 

There are two distinct ways in which the failure to solve \eqref{eq:equivQME} can manifest:
\begin{enumerate}
    \item Failure up to background-field terms: It may be impossible to solve \eqref{eq:equivQME} exactly, but the obstruction depends solely on the background fields. In this case, $I^\cL$ is only defined modulo such terms, and is referred to simply as a quantum $\cL$ background.
    \item Failure even modulo background-field terms: Here, an obstruction remains even after ignoring terms depending only on the background fields. This indicates that the classical $\cL$ symmetry is explicitly broken at the quantum level. A standard example is the axial symmetry in gauge theories, which does not persist quantum mechanically due to the ABJ anomaly.
\end{enumerate}
As in ordinary four-dimensional QFTs, 1-loop anomalies of continuous symmetries in four-dimensional holomorphic quantum field theories can be detected by the failure of triangle diagrams to be invariant (cf. Section 4 of \cite{Williams:2018ows}).
For example, consider the coupling to a background $\ff_F$ gauge field as described above, assuming no ABJ anomalies or other background fields are present. One finds:
\begin{equation}
    (\dd_{\rm CE} + \bQ) I^\cL + \f12 \{I^\cL, I^\cL\}_{\rm BV} = -\frac{k_{FFF}}{6\pi^2} \int \tr (\bA_F \del \bA_F \del \bA_F)\,.
\end{equation}
This coupling thus defines an inner quantum $\cL$ background if $k_{FFF} = 0$ and otherwise is simply a quantum $\cL$ background.
Note that the right-hand side is precisely is precisely the gauge variation of the quantum effective action.

%-------------------
\subsubsection{Anomalies from brackets and non-conservation of currents}
\label{sec:nonconservation}
%-------------------

Another perspective on an anomaly is as the failure of the corresponding currents to be (covariantly) conserved in the presence of non-trivial background fields.
In holomorphic quantum field theories, this can equivalently be stated as the failure of the current to be (covariantly) semi-chiral in the presence of such backgrounds.

To make this relation precise, it is useful to write the background-coupled quantum BV/BRST operator $\bQ^\cL$ as a formal perturbative expansion in higher-arity brackets:
\begin{equation}
\bQ^\cL = \dd_{\rm CE} + \bQ + \{ \cI^\cL \,{}_0\, -\} + \frac{\hbar}{2} \{ \cI^\cL \,{}_0\, \cI^\cL \,{}_0\, -\} + \frac{\hbar^2}{6} \{ \cI^\cL \,{}_0\, \cI^\cL \,{}_0\, \cI^\cL \,{}_0\, -\} + \dots ~.
\end{equation}

As a simple example, consider the currents $\bJ_{Fa}$ in the presence of a background $\ff_F$ gauge field $\bA_F$ and a complex structure deformation $\bm{\mu}$.
Their variation takes the form
\begin{equation}
\begin{aligned}
    \bQ^\cL \bJ_{Fa} & = \bar{\partial} \bJ_{Fa} + \del_i(\bm{\mu}^i \bJ_{Fa}) + f^c_{ab} \bA_F^b \bJ_{Fc}\\
    & \quad + \hbar \bigg(-\frac{k_{FFF}}{2\pi^2} \epsilon^{ij}\tr_F (T_a \del_i \bA_F \del_j \bA_F) - \frac{k_{RFF}}{32\pi^2} \epsilon^{ij} \tr_F (T_a \del_i \bA_F) \Tr(\del_j J \bm{\mu})\bigg)\,,
\end{aligned}
\end{equation}
where $J\bm\mu$ denotes the Jacobian of the Beltrami differential $\mu$. The first term is the ordinary $\bQ$-variation expressing the semi-chiral nature of $\bJ_{Fa}$. The second and third terms come from the binary brackets in Eq.~\eqref{eq:biflav} and \eqref{eq:biflavgrav}; together they give the action of the covariant Dolbeault differential on $\bJ_{Fa}$.
The remaining terms on the right-hand side measure the failure of $\bJ_{Fa}$ to be (covariantly) semi-chiral, i.e. covariantly conserved, in the presence of a background and hence measures the anomaly of the flavor symmetry in the ordinary sense.
Since these arise directly from the ternary brackets in Eqs.~\eqref{eq:jjj} and \eqref{eq:sjj}, we see that such higher brackets encode the various anomalies of the $F$ flavor symmetry. 
When a background gauge field $\bA$ is turned on for the full flavor group $\cF$, additional contributions appear from the ternary brackets in Eq.~\eqref{eq:ssj}, which occur only for abelian factors.

Note that the above formula also applies to ABJ anomalies where the failure of a current to be semi-chiral is itself an operator rather than a background field, but its interpretation changes. 
To highlight this distinction, we denote the dynamical semi-chiral operator by $\mathbf{c}$, while the background superfield is written as $\bA$.
As an example, suppose we gauge the $\ff_F$ flavor symmetry.
If the $\lie{u}(1)_f$ flavor symmetry suffers from an ABJ anomaly coming from a non-vanishing the mixed anomaly $k_{FFf} \neq 0$ then find that the current $\bJ_f$ is no longer semi-chiral, instead satisfying
\begin{equation}
    \bQ \bJ_f = \bar{\partial} \bJ_f - \frac{\hbar k_{FFf}}{2\pi^2} \epsilon^{ij}\tr_F (\del_i \bm{c}_F \del_j \bm {c}_F)\,.
\end{equation}
In particular, we see that the classically semi-chiral operator $\bJ_f$ fails to be semi-chiral quantum mechanically.
A similar phenomenon happens when the $U(1)_R$ used to define the twisting homomorphism suffers from an ABJ anomaly from a non-vanishing $k_{RFF}$ where the stress tensor $\bS_i$ fails to be semi-chiral:
\begin{equation}
    \bQ \bS_i = \bar{\partial} \bS_i - \frac{\hbar k_{RFF}}{4\pi^2} \epsilon^{jk} \del_i \big(\tr_F (\del_j \bm{c}_F \del_k \bm {c}_F)\big)\,.
\end{equation}
Although $\bS_i$ is not semi-chiral in the presence of such an ABJ anomaly, the operator $\bS = \del^i \bS_i$ is still semi-chiral and generates an action of $\lie{svect}(2)$.
As we describe in more detail in Section~\ref{sec:anomalousN=2} for situations where $\cN=2$ superconformal symmetry is broken by an ABJ anomaly, we propose an alternative perspective on the operators appearing in the non-conservation equation.
Namely, one should instead include the operator(s) appearing in these non-conservation equations as part of the symmetry algebra and interpret non-conservation equation such as above as a differential on this extended symmetry algebra.
Of course, only the cohomology of this extended symmetry algebra will act on $\bQ$-cohomology but we expect that this chain-level symmetry algebra should nonetheless provides useful constraints.

An alternative view on the ABJ anomaly is to interpret it in terms of non-genuine topological operators. Define the instanton density operator 
\begin{equation}
    \bN = \f{\hbar}{(2\pi \ii)^2}\epsilon^{jk} \tr_F \del_j \mathbf{c}_F \del_k \mathbf{c}_F\,,   
\end{equation}
which is semi-chiral, i.e.  $\bQ \bN = \bar\del \bN$. The ABJ anomaly is the obstruction to the semi-chirality of the current $\bJ$,
\begin{equation}\label{eq:non-cons-non-semi-chiral}
\bQ \bJ = \bar\del \bJ + k_{FFf}\,\bN\,.
\end{equation}
With this property in mind, we can redefine the topological operator introduced in Eq.~\eqref{eq:top-finite} as
\begin{equation}\label{eq:non-genuine-top}
    \hat{U}_\alpha(M_3,M_4) = \exp\left[\ii \,\alpha^{mn} \oint \rho_{mn} \bJ + \ii \, \alpha^{mn} \,k_{FFf}\,\int_{M_4} \rho_{m,n}\bN \right]\,,
\end{equation}
where $\del M_4 = M_3$. By construction, this operator is again topological, precisely because of Eq.~\eqref{eq:non-cons-non-semi-chiral}. However, it cannot be improved to a genuine extended operator since the induced Chern-Simons term $\mathbf{c}_F \del \mathbf{c}_F$ on $M_3$ is not properly quantised, and hence is ill-defined on non-trivial gauge bundles. Thus, the ABJ anomaly can be rephrased as the statement that $\hat U$ is a non-genuine topological operator.

In some situations, however, the story does not end here, since part of the anomalous symmetry can be salvaged. Specializing to $\alpha^{mn} = \alpha\, \delta^{m}_0\delta^{n}_0$, it was shown in \cite{Choi:2022jqy,Cordova:2022ieu} (in an untwisted setting) that a discrete subgroup of the anomalous $\UU(1)$ symmetry survives as a non-invertible symmetry. The key idea is to stack the original symmetry defect with a suitable TQFT that exactly cancels the anomaly.
Appropriate TQFTs are only available for the discrete dense subset $\bQ/\bZ \subset \UU(1)$. This mechanism applies verbatim in the holomorphic context.
Note that in the holomorphic twist there are no non-trivial line defects, so the non-invertible character of these symmetries, emphasized in \cite{Choi:2022jqy,Cordova:2022ieu}, becomes invisible.%
\footnote{In principle, one could imagine extending this construction to an infinite discrete enhancement by stacking the operators associated with different $\rho_{m,n}$ with the appropriate TQFTs. However, the existence of such TQFTs seems unlikely. Indeed, in the analogous, and better understood, setting of 2d chiral CFTs we do not observe an infinite enhancement of discrete symmetries.}

An important restriction arises here: anomalies for TQFTs with discrete symmetry always involve products of at least two characteristic classes, so this cancellation mechanism is only available for abelian gauge groups.
For non-abelian gauge groups the anomaly is proportional to $c_2(F_G)$, which cannot be decomposed in such a way.
A physical way to phrase this distinction is that abelian gauge fields in four dimensions have no instantons on the sphere, whereas non-abelian gauge fields do.
The existence of non-abelian instantons obstructs the possibility of a residual discrete symmetry.\footnote{We thank Andrea Antinucci for discussions on this point.}

So far, this construction has been considered only for internal (flavor) symmetries.%
\footnote{We note that there have been parallel efforts to embed spacetime symmetries into the same framework as internal symmetries \cite{Apruzzi:2025hvs}, and that non-invertible extensions of such symmetries have also been discussed in the literature \cite{Seiberg:2024wgj,Seiberg:2024yig}.}
In our context, however, there is no obstruction to extending it also to spacetime symmetries. In particular, take $V = z \cdot \partial$, the Euler field generating dilations. We can write the analog of \eqref{eq:non-genuine-top},
\begin{equation}
\begin{aligned}
    \hat{U}_\alpha(M_3,M_4) =& \exp\left[\ii \,\alpha \oint z^i \bS_i + \ii \,\alpha\,k_{FFR} \,\int_{M_4} z^i \partial_i\bN \right]\\
    =& \exp\left[\ii \,\alpha \oint z^i \bS_i - \ii \, \alpha \,k_{FFR} \,\int_{M_4} \bN \right]\,.
\end{aligned}
\end{equation}
This has precisely the same form as Eq.~\eqref{eq:non-genuine-top}, and hence can be stacked with the same TQFT to produce genuine symmetry operators for a discrete subgroup of the dilation symmetry. 

%-------------------
\subsubsection{Anomalies from descent}
\label{sec:descent}
%-------------------

As in more familiar QFTs, anomalies of holomorphic field theories can be described via descent from holomorphic Chern classes defined two (real) dimensions higher.
For simplicity, we consider the trivial holomorphic $F$ bundle on $\bC^2$, with a partial connection $\overline{\del} + A$ satisfying $(\overline{\del} + A)^2 = 0$, which can be interpreted as a deformation of the bundle's complex structure.
The full connection on the gauge bundle is then $\nabla = \del + (\overline{\del} + A)$, so its curvature is simply
\begin{equation}
    F = \del A\,.
\end{equation}

The characteristic class $c_3 = \tr F^3$ is gauge-invariant because $\dd_{\rm CE} F = [c, F]$ and is locally exact, with Chern-Simons primitive,
\begin{equation}
    c_3 = \del \tr A \del A \del A = \dd \tr (A \del A \del A) = \dd \tr (A F^2)\,.
\end{equation}
The Chern-Simons form itself is not gauge-invariant, but its variation is exact due to the gauge-invariance of $c_3$:
\begin{equation}
    \dd_{\rm CE} \tr (A \del A \del A) = \dd \bigg(\tr (c \del A \del A)\bigg)\,.
\end{equation}
As in the standard case, the gauge-variation of the quantum effective action computed above is precisely the integral of this primitive.
Thus, the anomalous behavior of symmetries in four-dimensional holomorphic field theories can be captured by degree-3 characteristic classes, just as in ordinary four-dimensional QFTs.

Given a four-dimensional $\cN=1$ SQFT, the anomaly polynomial of its holomorphic twist is obtained as a specialization of the untwisted theory’s anomaly polynomial: The characteristic classes corresponding to flavor symmetries are unchanged under the twist.
For the gravitational and R-symmetry anomalies more care is needed because the twist mixes spacetime and R-symmetries. In particular, the gravitational anomalies of the holomorphic theory are linear combinations of the original gravitational anomalies and mixed anomalies involving 
$\UU(1)_R$.

Following Section 4.1 of \cite{Williams:2024xrq}, the specialization proceeds as follows: First, break $\Spin(4)$ characteristic classes to ${\rm MU}(2)$ classes by sending
\begin{equation}
    p_1 \mapsto 2 {\rm ch}_2 = c_1^2 - 2c_2\,,
\end{equation}
%,
reflecting the breaking of most of the Lorentz group by the choice of complex structure. Then, specialize the R-symmetry class according to 
\begin{equation}
    c_{1,r} = -\f12 c_1\,,
\end{equation}
where $c_{1,r}$ is the first Chern class of the background R-symmetry connection.
This accounts for the twisting homomorphism.

%-------------------
\subsection{Spectral flow and \texorpdfstring{$a$}{a}-maximization}
\label{sec:aext}
%-------------------

When a holomorphic field theory has a semi-chiral stress tensor $\bS_i$ and semi-chiral abelian currents $\bJ_f$, we can define a family of stress tensors $\bS^\epsilon_i$ by way of spectral flow.
Explicitly, given a collection of abelian%
\footnote{More generally, one could include an arbitrary collection of commuting regular elements of the flavor symmetry algebra but this will break the flavor symmetry to the commutant of these elements.} %
currents $\bJ_f$, then the brackets of
\begin{equation}
\label{eq:spectralflow}
    \bS^\epsilon_i = \bS_i + \epsilon^f \del_i \bJ_f
\end{equation}
with itself are still given by Eq. \eqref{eq:ss} for any choice of $\epsilon^f$ and so this also defines an action of $\lie{witt}(2)$.
This can be thought of as choosing a different twisting homomorphism or, equivalently, choosing a different $\UU(1)_R$ $R$-symmetry.
There isn't a preferred member of this family from the perspective of the holomorphic field theory; each one defines a perfectly good action of $\lie{vir}(2)_{a(\epsilon),c(\epsilon)}$, although the value of the central charges $a(\epsilon)$,$c(\epsilon)$ depend on the choice of $\epsilon^f$ as well as the mixed anomaly coefficients.

If the holomorphic field theory arises form the twist of a not-necessarily-unitary $\cN=1$ SCFT then there is a preferred choice: the superconformal stress tensor $\bS^{\rm sc}_i$.
Work of Osborn \cite{Osborn:1998qu} shows that the three-point function of two superconformal stress tensors and an abelian current has a unique tensor structure.
Following \cite{Bomans:2023mkd}, when we translate this three-point function to the corresponding ternary $\lambda$-brackets to see that it too must only contain a single tensor structure.
However, the general form of the $\lambda$-bracket given in Eq. \eqref{eq:ssj} contains two independent tensor structures. Therefore, if $\bS^{\rm sc}_i$ comes from the superconformal stress tensor, i.e. if the twisted spin comes from the superconformal $\UU(1)_R$, the anomaly coefficients $k_f$ and $k_{RRf}$ must be proportional.
Since the form of the three-point function is universal, we can determine the proportionality constant in a free theory, obtaining,
\begin{equation}\label{eq:ssj-superconformal}
    \lllbrak{\bS^{\rm sc}_i}{\bS^{\rm sc}_j}{\bJ_f} = \f{k_f}{36 \pi^2}[\lambda_1 \lambda_2](\lambda_{1i} \lambda_{2j} - 3 \lambda_{1j}\lambda_{2i})\,.
\end{equation}
Comparing to Eq. \eqref{eq:ssj} yields the relation
\begin{equation}
\label{eq:aext}
    k_{f} = 9 k_{RRf}\,,
\end{equation}
between the mixed $R$-$R$-flavor and gravitational-gravitational-flavor 't Hooft anomalies for the superconformal $R$-charge.
If the parent $\cN=1$ SCFT is unitary, then there are additional positivity constraints on the theory's mixed anomalies \cite{Anselmi:1997am}:
\begin{equation}
\label{eq:amax}
    k_{Rfg} < 0\,.
\end{equation}
As explained in \cite{Intriligator:2003jj}, Eq. \eqref{eq:aext} implies that the superconformal $R$-charge corresponds to a choice of $\epsilon^f$ that extremizes the central charge $a(\epsilon)$.
Additionally imposing Eq. \eqref{eq:amax} implies that this extrema must in fact be a local maximum.

It is presently unclear if either Eq. \eqref{eq:aext} or \eqref{eq:amax} can be deduced purely within the holomorphic field theory, i.e. without relying on properties of an underlying physical SCFT.
That said, the constraint in Eq. \eqref{eq:aext} can be phrased in terms of the anomalies of the holomorphic field theory.
In this language, the mixed gravitational-gravitational-flavor anomalies must be proportional to the local functional
\begin{equation}
    \int \tr(\bA_f) \left(\Tr(\del J\bm{\mu}) \Tr(\del J \bm{\mu}) - 3\Tr(\del J\bm{\mu} \del J\bm{\mu}) \right)\,.
\end{equation}
This expression represents the class $c_{1,f}({\rm ch}_1^2 - 3 {\rm ch}_2)$, where $c_{1,f}$ is the first Chern class of the background $\lie{u}(1)_f$ gauge field and ${\rm ch}_1$, ${\rm ch}_2$ are the first and second gravitational Chern classes.

%%%%%%%%%%%%%%%%%%%%%%%%%%%%%%%%%%%%%%%%% 
\section{Extended Supersymmetry}        %
\label{sec:extended-SUSY}               %
%%%%%%%%%%%%%%%%%%%%%%%%%%%%%%%%%%%%%%%%%

In theories with $\cN>1$ supersymmetry, the $\bQ$-cohomology of the supersymmetry algebra is enlarged by the presence of additional supercurrents. 
In what follows we will mainly consider the superconformal case, since for $\cN\geq 3$ the large amount of supersymmetry is widely believed to imply superconformal symmetry.%
\footnote{For $\cN\geq 3$ in four dimensions, the supercurrent multiplet is ultrashort: representation theory admits only the $B_1B_1[0,0]_2$ stress–tensor multiplet, which contains the conformal currents and enforces $T_\mu^\mu = 0$ \cite{Sohnius:1978wk} as it leaves no space for a virial current. There is no long A-type multiplet with the same quantum numbers \cite{Cordova:2016emh}, so any interacting theory with $\cN\geq 3$ supersymmetry and a stress tensor is necessarily superconformal. This was first shown in the supercurrent formalism in \cite{Sohnius:1978wk,Sohnius:1981sn} and is manifest in the modern unitarity classification \cite{Cordova:2016emh}.}
In the $\cN=2$ supersymmetric case we will comment on the anomalous breaking of supersymmetry.

Analogous to the two-dimensional case, in theories with extended $\cN\geq 2$ supersymmetry the 2-Virasoro symmetry algebra of the twisted theory is enlarged to an $\cN-1$ supersymmetric 2-Virasoro algebra.
Equivalently, these algebras arise as the holomorphic twists of $\cN\geq 2$ superconformal algebras \cite{Saberi:2019fkq,Hahner:2024hak}. 
At the classical level, ignoring central extensions, \cite{Hahner:2024hak} showed that the 2-Virasoro algebra already enhances the twist of $\cN\geq 1$ superconformal symmetry, by explicitly characterizing the twist of the full Weyl multiplet.
In the following, we describe the symmetry algebras appearing in twists of superconformal theories with extended supersymmetry; a summary is given in Table~\ref{tab:syms-enhanced}.

\begin{table}[htb!]
    \centering
    \begin{tabular}{c|c|c|c}
        $\cN=k$ &  superconformal algebra  &  $\bQ$-cohomology &  $\infty$-dim. enhancement\\
        \hline
        $\cN=1$ & $\lie{sl}(4|1)$ & $\lie{sl}(3)$ & ${\mathfrak{vir}(2)}_{a,c}$\\
        $\cN=2$ & $\lie{sl}(4|2)$ & $\lie{sl}(3|1)$ & ${\mathfrak{svir}(2|1)}_{a,c}$\\
        $\cN=3$ & $\lie{sl}(4|3)$ & $\lie{sl}(3|2)$ & ${\mathfrak{svir}(2|2)}_{a}$\\
        $\cN=4$ & $\lie{psl}(4|4)$ & $\lie{psl}(3|3)$ & ${\mathfrak{svir}(2|3)}_{a}$ \\
    \end{tabular}
    \caption{The $\bQ$-cohomology of the $\cN=k$ superconformal algebras and their infinite-dimensional enhancement. For $\cN=3,4$ there is only a single central charge $a=c$.}
    \label{tab:syms-enhanced}
\end{table}

%-------------------
\subsection{\texorpdfstring{$\cN=2$}{N=2} supersymmetry}
\label{sec:N=2}
%-------------------

For theories with extended $\cN=2$ superconformal symmetry, the superconformal stress tensor multiplet decomposes into a set of $\cN=1$ multiplets; specifically, the $\cN=2$ stress tensor multiplet splits as
\begin{equation}
    A_2\overline{A}_2[0,0]^{(0;0)}_2 = \underbrace{\rule[-1.2ex]{0pt}{0pt}{A}_1\overline{A}_1[1,1]^{(0)}_3}_{\bS_i} \oplus \underbrace{\rule[-1.2ex]{0pt}{0pt}{A}_1\overline{A}_2[1,0]^{(1/3)}_{5/2}}_{\widetilde\bG} \oplus \underbrace{\rule[-1.2ex]{0pt}{0pt}{A}_2\overline{ A}_1[0,1]^{(-1/3)}_{5/2}}_{\bG_i} \oplus \underbrace{\rule[-1.2ex]{0pt}{0pt}{A}_2\overline{A}_2[0,0]^{(0)}_2}_{\bR}\,.
\end{equation}
This set contains additional supersymmetry-current multiplets, each contributing semi-chiral operators, along with a conserved flavor current multiplet associated with the residual $R$-symmetry. The corresponding semi-chiral superfields are denoted
\begin{equation}
    \widetilde \bG\,,\qquad\bG_i\,,\qquad \bR \,.
\end{equation}
The semi-chiral superfield $\bR$ is fermionic and generates bosonic symmetries, whereas $\widetilde{\bG}$ and $\bG_i$ are bosonic and generate fermionic symmetries. Since our focus in this section is on twisting superconformal theories, we consistently use the superconformal $\UU(1)_R^{\cN=1}$ $R$-symmetry, unless otherwise specified.\footnote{In an $\cN=2$ theory, the $\cN=1$ superconformal $R$-charge is related to the $\cN=2$ one by $r{\cN=1} = \frac13 r_{\cN=2} + \frac43 I_2$, where $I_2$ denotes the Cartan generator of the $\SU(2)_R$ symmetry. See Appendix \ref{app:4dSCA} for more details.} With this choice of twisting homomorphism, the $\lambda$-brackets of the semi-chiral operators in the stress tensor multiplet include those given in Eq.~\eqref{eq:ss}, together with the following:
\begin{align}
     &\llbrak{\bS_i}{\widetilde\bG} = \, \del_i\widetilde\bG + \f43 \lambda_i\widetilde\bG\,, & &\llbrak{\bR}{\widetilde\bG} = \, -\widetilde\bG\,,\\
    &\llbrak{\bS_i}{\bG_j} = \,\del_i \bG_j + \f23\lambda_i \bG_j + \lambda_j \bG_i\,,& &\llbrak{\bR}{\bG_i} = \, \bG_i\,,\\
    &\llbrak{\bS_i}{\bR} = \,\del_i\bR + \lambda_i\bR\,,&&\llbrak{\bG_i}{\widetilde\bG} = \, \bS_i + \f23 \del_i\bR + \lambda_i \bR\,.
\end{align}
The equations in the left column express that $\bG_i$, $\widetilde{\bG}$, and $\bR$ transform as $\lie{vect}(2)$ primaries, while the right column encodes the (holomorphic extension of the) twisted supersymmetry algebra. All other binary brackets vanish; for example, $\llbrak{\widetilde{\bG}}{\widetilde{\bG}} = 0$.

The algebra presented above admits a super-geometric interpretation, cf. \cite{Saberi:2019fkq}; see also Section 6.3.5 of \cite{Hahner:2024hak}.
As noted previously, the modes of~$\bS_i$ generate the dg Lie algebra $\lie{witt}(2)$, defined as the derived global sections of the holomorphic tangent bundle of $\bC^2 \setminus \{0\}$.
Similarly, the modes of $\bS_i$, $\bR$, $\bG_i$, and $\widetilde{\bG}$ generate the dg Lie algebra given by the derived global sections of the following holomorphic super vector bundle:
\begin{equation}
    T_{\bS_i} \oplus \cO_{\bR} \oplus \Pi(T \otimes K^{1/3}_{\bG_i} \oplus K^{-1/3}_{\widetilde{\bG}})
\end{equation}
over $\bC^2\backslash \{0\}$, where $T$ is the (holomorphic) tangent sheaf, $\cO$ the structure sheaf, $K$ the canonical sheaf, and $\Pi$ denotes a shift in Grassmann parity.
The subscript labels the bundle for which the corresponding generator is a section.
We can view this as the dg Lie superalgebra of derived holomorphic super vector fields on the complex supermanifold $\bC^{2|1} \backslash \{0\}$.
By \textit{derived} super vector fields, we mean derived global sections of the (holomorphic) tangent bundle of this super manifold.
Concretely, if we put local coordinates $z^i$ and $\theta$ on this total space, the above currents generate the action of the following super vector fields
\begin{equation}
\begin{aligned}
    & & \bS_i \rightsquigarrow - \omega \del_i - \tfrac{1}{3}(\del_i \omega) \theta \del_\theta & & \\
    \bG_i & \rightsquigarrow \omega \theta \del_i & & & \widetilde{\bG} & \rightsquigarrow \omega \del_\theta\\
    & & \bR \rightsquigarrow \omega \theta \del_\theta \hspace{1.25cm} & & 
\end{aligned}
\end{equation}
via $S^3$ integrals weighted by $\omega \in \Omega^{(0,\bullet)}(\bC^2\backslash \{0\})$.
The above binary $\lambda$-brackets can be recovered by commutators of these vector fields.
Note that the additional factor of $\tfrac{1}{3} (\del_i \omega) \theta \del_\theta$ in the vector field generated by $\bS_i$ can be viewed as declaring $\theta$ transforms non-trivially under coordinate transformations of spacetime $\bC^2$.
In particular, they say that the fermionic coordinate $\theta$ transforms as a section of $K^{1/3}$.
We can more invariantly characterize this symmetry as the dg Lie superalgebra of derived holomorphic super vector fields on the complex supermanifold
\begin{equation}
    \operatorname{Tot} \left(\Pi K^{1/3} \to \bC^2\backslash \{0\}\right) \,.
\end{equation}

We now turn to the central extensions of the above algebra.
Recall that conformal anomalies of the untwisted theory manifest as (central) ternary brackets.
These appear as central extensions of the super version of the Witt algebra.
The non-vanishing brackets in the stress tensor multiplet, beyond the one appearing in Eq. \eqref{eq:sss}, are:
\begin{align}
    &\lllbrak{\bR}{\bS_i}{\bS_j} = \f{2(a-c)}{3\pi^2}[\lambda_1 \lambda_2]\left( 3\lambda_{1i}\lambda_{2j}-\lambda_{2i}\lambda_{1j}+2\lambda_{2i}\lambda_{2j} \right) \,,\\
    &\lllbrak{\bS_i}{\bR}{\bR} = \f{c}{\pi^2} [\lambda_1 \lambda_2] \,\lambda_{1i} \,,\\
     &\lllbrak{\bR}{\bR}{\bR} = \f{3(4a-3c)}{\pi^2}[\lambda_1 \lambda_2]\,,\\
    &\lllbrak{\wti\bG}{\bS_i}{\bG_j} = \f{2}{\pi^2}[\lambda_1 \lambda_2]\left((a-c)\lambda_{1i}\lambda_{2j}-\f a 3\lambda_{1j}\lambda_{2i} - \f c 3\lambda_{2i}\lambda_{2j} \right) \,,\\
    &\lllbrak{\wti\bG}{\bR}{\bG_i} = \f{1}{2\pi^2}[\lambda_1 \lambda_2] \left(2c\,\lambda_{1i}+4(a-c)\lambda_{2i} \right) \,.
\end{align}
To complete the complex geometric description of $\lie{svir}(2|1)_{a,c}$, it would be desirable to find explicit representatives for the 3-cocycles corresponding to the central extensions of this algebra. We leave this for future work.\footnote{A natural first attempt is to mimic the construction of the 3-cocycle for $\lie{svir}(2|1)_{a,c}$ by replacing vector fields, Jacobians, and traces with their super analogues: super vector fields, super-Jacobians, and supertraces on $\bC^{2|1}$. A closer inspection, however, shows that this straightforward generalization fails to reproduce all of the three-brackets introduced above.}

%-----------------------
\subsubsection{Anomalous \texorpdfstring{$\cN=2$}{N=2} superconformal symmetry}
\label{sec:anomalousN=2}
%-----------------------

It can, and often does, happen that the above symmetry is realized classically but there is an anomaly implying it does not survive quantization.%
\footnote{In $\cN=2$ gauge theories the $\beta$-function is one-loop exact so a necessary and sufficient condition for quantum conformality is $2h_\fg^\vee - \sum_{{\rm hypers}\,i}T(R_i) = 0$, where $h^\vee_\fg$ is the dual Coxeter number of the gauge algebra and $T(R_i)$ the Dynkin index of the representation $R_i$ of the $i$th hypermultiplet. If this condition is not satisfied an ABJ anomaly will arise.}
This breaking manifests as an ABJ anomaly for the symmetry generated by $\bR$.
Correspondingly, $\bR$ fails to be semi-chiral and we find
\begin{equation}
    \bQ \bR = \bar{\del} \bR + \frac{k}{2\pi^2} \bN
\end{equation}
for some operator $\bN$ and an anomaly coefficient $k$, cf. Section  \ref{sec:anomalies}.
The would-be superconformal stress tensor $\bS_i$ and the supercurrent $\widetilde{\bG}$ also fail to be semi-chiral:
\begin{equation}
    \bQ \bS_i = \bar{\del} \bS_i + \frac{k}{6\pi^2} \del_i \bN\,, \qquad \bQ \widetilde{\bG} = \bar{\del} \wti{\bG} + \frac{k}{2\pi^2} \epsilon^{ij} \del_i \bL_j\,.
\end{equation}
We see that $\widetilde{\bS}_i = \bS_i - \f{1}{3}\del_i \bR$ is semi-chiral and, moreover, generates an action of $\lie{witt}(2)$ with vanishing central charges; the twisted spin generated by $\widetilde{\bS}_i$ is obtained by a twisting homomorphism with the Cartan subalgebra of ${\rm SU}(2)_R$. From this point of view it is clear that no central charges can appear, since these are proportional to either $\Tr(T^a)$ or $d_{abc}$, both of which vanish for $\SU(2)$.

There is one additional operator $\bM$ appearing in the multiplet with $\bN$ and $\bL_i$.
The action of twisted $\cN=2$ superconformal symmetry on these operators takes the following form: the action on $\bN$ is given by
\begin{equation}
\begin{aligned}
   && \hspace{-1cm} \llbrak{\bS_i}{\bN} = \del_i \bN + \lambda_i \bN\,, \hspace{-1cm}&&\\
   \llbrak{\bG_i}{\bN} & = 0\,, &&& \llbrak{\widetilde{\bG}}{\bN} & = \epsilon^{ij}(\del_i \bL_j + \lambda_i \bL_j)\,,\\
   && \llbrak{\bR}{\bN} = 0\,, &&
\end{aligned}
\end{equation}
the action on $\bM$ is given by
\begin{equation}
\begin{aligned}
   && \hspace{-2cm} \llbrak{\bS_i}{\bM} = \del_i \bM + \f{2}{3} \lambda_i \bM\,, \hspace{-1cm}&&\\
   \llbrak{\bG_i}{\bM} & = -\bL_i\,, &&& \llbrak{\widetilde{\bG}}{\bM} & = 0\,,\\
   && \llbrak{\bR}{\bM} = -2\bM\,, &&
\end{aligned}
\end{equation}
and, finally, the action on $\bL_i$ is given by
\begin{equation}
\begin{aligned}
   && \hspace{-1cm} \llbrak{\bS_i}{\bL_j} = \del_i \bL_j + \f{1}{3} \lambda_i \bL_j + \lambda_j \bL_i\,, \hspace{-1cm}&&\\
   \llbrak{\bG_i}{\bL_j} & = -\epsilon_{ij} \bN\,, &&& \llbrak{\widetilde{\bG}}{\bL_j} & = \del_i \bM + 2 \lambda_i \bM\,,\\
   && \llbrak{\bR}{\bL_i} = -\bL_i\,. &&
\end{aligned}
\end{equation}
We find that the binary and ternary brackets of elements of this multiplet necessarily vanish, as do the mixed ternary brackets with the superconformal symmetry generators.

%-------------------
\subsubsection{Flavor symmetries}
%-------------------

Whenever the theory enjoys an additional global symmetry the spectrum contains a conserved current. 
Suppose the untwisted theory has a global symmetry by the Lie algebra $\lie{f}$.
We assume this Lie algebra is either simple or abelian.
The conserved current multiplet, after twisting, gives rise to a semi-chiral superfield $\bJ$ whose modes generate a 2-Kac-Moody algebra.
Analogous to the discussion above, in the presence of extended superconformal symmetry, this multiplet splits in a collection of $\cN=1$ multiplets. 
Concretely, an $\cN=2$ conserved current multiplet decomposes into three $\cN=1$ multiplets,
\begin{equation}
    B_1\overline{B}_1[0,0]^{(2;0)}_2 = \underbrace{\rule[-1.2ex]{0pt}{0pt}{A}_2\overline{A}_2[1,1]^{(0)}_3}_{\bJ} \oplus \underbrace{\rule[-1.2ex]{0pt}{0pt}{L}\overline{B}_1[0,0]^{(4/3)}_{2}}_{\bm{\mu}} \oplus \underbrace{\rule[-1.2ex]{0pt}{0pt}{B}_1\overline{ L}[0,0]^{(-4/3)}_{2}}_{\emptyset}\,.
\end{equation}
where $\bm\mu$ is the $\cN=2$ moment map operator and the third multiplet does not contribute a semi-chiral superfield since its anti-chiral half does not satisfy a shortening condition. The $\lambda$-brackets involving these fields are given by
\begin{align}
    \llbrak{\bJ_a}{\bJ_b} =& \, f_{ab}^c \bJ_c\,,& \llbrak{\bJ_a}{\bm\mu_b} =& f_{ab}^c\, \bm\mu_c\\
     \llbrak{\bS_i}{\bJ_a} =&\, \del_i\bJ_a +  \lambda_i\bJ_a\,,& \llbrak{\bS_i}{\bm{\mu}_a} =&\,\del_i\bm\mu_a + \f23 \lambda_i\bm\mu_a\,,\\
     \llbrak{\bR}{\bJ_a} =&\, 0\,,&
    \llbrak{\bR}{\bm\mu_a} =&\, {\bm \mu}_a\,,\\
    \llbrak{\bG_i}{\bJ_a} =&\, -(\del_i\bm{\mu}_a + \lambda_i\bm\mu_a)\,,&
    \llbrak{\widetilde\bG}{\bm\mu_a} =&\, \bJ_a\,.
\end{align}
Invariantly, the symmetries generated by the currents $\bJ,\bm\mu$ can be thought of as the dg Lie superalgebra of derived global holomorphic sections of the sheaf
\begin{equation}
    \cO \otimes \lie{f}_{\bJ} \oplus \Pi K^{1/3} \otimes \lie{f}_{\bm\mu}
\end{equation}
over $\bC^2\backslash \{0\}$.
Equivalently, it can be viewed as the dg Lie superalgebra of $\ff$-valued derived global holomorphic function on the supermanifold ${\rm Tot}(\Pi K^{1/3} \to \bC^2\backslash\{0\})$.

In addition, the 3-brackets involving the conserved currents encode the flavor 't Hooft anomalies $k_{FFF}$ and levels $k_F$ of the physical theory. The 3-brackets involving only flavor currents are just those given in Eq. \eqref{eq:jjj}.
The non-vanishing 3-brackets involving two operators in the flavor symmetry multiplet and one in the stress tensor multiplet are
\begin{align}
    &\lllbrak{\bJ_a}{\bJ_b}{\bS_i} = -\f{k_{F}}{6\pi^2}\tr_F(T_a T_b)[\lambda_1 \lambda_2](\lambda_{1i}+\lambda_{2i})\,,\\
    &\lllbrak{\bJ_a}{\bJ_b}{\bR} = -\f{k_{F}}{2\pi^2}\tr_F(T_a T_b)[\lambda_1 \lambda_2]\,, \\
    &\lllbrak{\bm\mu_a}{\bJ_b}{\widetilde\bG} = \f{k_F}{2\pi^2}\tr_F(T_a T_b)[\lambda_1 \lambda_2] \,,
\end{align}
where the first equation is simply Eq. \eqref{eq:sjj} with $k_{RFF}$ replaced by $k_F$.
The 3-brackets with 1 flavor and two superconformal components all vanish. Note that this vanishing only happens when choosing the superconformal twist, this is a consequence of $a$-maximization.

%-------------------
\subsubsection{Example: Lagrangian $\cN=2$ theories}
%-------------------
We now describe a collection of Lagrangian examples furnishing this symmetry algebra; these are the main examples studied in \cite{Saberi:2019fkq}.
A Lagrangian $\cN=2$ theory of $G$ gauge theory coupled to hypermultiplets in the pseudoreal representation $\cR$ can be realized by an $\cN=1$ $G$ gauge theory coupled an $\cR$-valued chiral $\bfgamma^n$, $n = 1, \ldots, \dim \cR$ and an adjoint chiral multiplet $\bfphi^A$, $A=1,\ldots, \dim \fg$, with superpotential $\mathbf{W} = -\tfrac{1}{2}\Omega_{ml} (T_A)^l{}_n \bfphi^A \bfgamma^m \bfgamma^n$, where $\Omega_{mn}$ denotes the components of a non-degenerate invariant pairing for the Lie algebra of $G$.

It is convenient to view the fields as living on the superspace $\bC^{2|1}$, where we denote the odd coordinate by $\theta$.
The twisted $\cN=2$ hypermultiplet becomes
\begin{equation}
    \mathbf{X}^n = \bfgamma^n + \theta (\bfbeta_m\Omega^{mn})\,,
\end{equation}
and the $\cN=2$ gauge field becomes
\begin{equation}
    \mathbf{A}^A = \mathbf{c}^A + \theta \bfphi^A\,, \qquad \mathbf{B}_A = \bflambda_A + \theta \mathbf{b}_A\, .
\end{equation}
The action is then expressed in terms of the Berezinian integral
\begin{equation}
    S_{\rm BV} = \int_{\bC^{2|1}}\dd^2z \dd\theta\bigg(\mathbf{B} (\bar\del \mathbf{A} + \tfrac{1}{2}[\mathbf{A}, \mathbf{A}]) +  \tfrac{1}{2}\Omega_{mn} \mathbf{X}^n \bar{\del}_\mathbf{A} \mathbf{X}^n\bigg).
\end{equation}
This theory is clearly holomorphically parametrization invariant.
The stress tensor and additional supercurrent superfields are the Noether currents for the twisted $N=2$ superconformal symmetries; they take the following form, cf. Eq. (27) of \cite{Saberi:2019fkq}:
\begin{equation}
\begin{aligned}
    & & \hspace{-4cm} \bS_i = - \mathbf{b}_A \del_{i} \mathbf{c}^A + \tfrac{2}{3}\bflambda_A \del_{i} \bfphi^A - \tfrac{1}{3} \bfphi^A \del_i \bflambda_A + \tfrac{2}{3}\bfbeta_n\del_i \bfgamma^n - \tfrac{1}{3}\bfgamma^n \del_i \bfbeta_n\,, \hspace{-3.5cm} & &\\
    \bG_i & = \, \tfrac{1}{2}  \Omega_{mn}\bfgamma^n \del_i \bfgamma^m - \bflambda_A \del_{i} \mathbf{c}^A\,, &&& \widetilde \bG & =\, \tfrac{1}{2} \Omega^{nm} \bfbeta_n \bfbeta_m - \mathbf{b}_A \bfphi^A\,,\\
    & & \bR =\, \tfrac{1}{2} \bfbeta_n \bfgamma^n - \bflambda_A \bfphi^A\,. &&
\end{aligned}
\end{equation}

These currents are classically semi-chiral, but this can fail quantum mechanically due to an ABJ anomaly.
In particular, the quantum-corrected action of $\bQ$ on the current $\bJ$ takes the form
\begin{equation}
    \bQ \bR = \bar{\del} \bR - \f{1}{2\pi^2} \epsilon^{ij}\left(\tr_{\rm ad}(\del_i \mathbf{c} \del_j \mathbf{c}) - \tfrac{1}{2}\tr_{\rm \cR}(\del_i \mathbf{c} \del_j \mathbf{c})\right)\,.
\end{equation}
The fact that the right-hand side is nonzero is due to an ABJ anomaly and vanishes if $2h^\vee - \f{1}{2} T(\cR) = 0$, i.e. when the untwisted theory is $\cN=2$ superconformal; this is a manifestation of the anomaly multiplet found in Section \ref{sec:anomalousN=2} with anomaly coefficient $k = 2h^\vee - T(\cR)$ and precisely the obstruction/anomaly found in Section 7 of \cite{Saberi:2019fkq}.%
\footnote{More precisely, when the gauge group has multiplet simple factors the operator $\bL$ is a sum of $\tr \del \mathbf{c} \del \mathbf{c}$ with weights capturing their individual anomaly coefficients with $U(1)_r$.} %
The operators in the anomaly multiplet are given by
\begin{equation}
    \bN = \epsilon^{ij}\tr (\del_i \mathbf{c} \del_j \mathbf{c})\,, \qquad \bM = \tr(\bfphi^2)\,, \qquad \bL_i = \tr(\bfphi \del_i \mathbf{c})\,.
\end{equation}
where the trace is taken in the representation in which the gauge theory fields transform.

The theory described above enjoys a flavor symmetry $F$ given by the normalizer of $G$ in $\rm{Sp}(\cR)$ modulo the adjoint action of $G$.
For example, when $G = {\rm SL}(N_c)$ and $\cR = ((\bC^{N_c}) + (\bC^{N_c})^*)^{\oplus N_f}$ then there is an ${\rm SL}(N_f)$ flavor symmetry.
If the matrices generating this action on $\cR$ are given by $(\tau_a)^m{}_n$, $a = 1, \dots, \dim F$, then the operators in the $\cN=2$ flavor symmetry multiplet are given by
\begin{equation}
    \bJ_a = - \bfbeta_m (\tau_a)^m{}_n \bfgamma^n\,, \qquad \bm{\mu}_a = \f{1}{2}(\tau_a)_{mn} \bfgamma^m \bfgamma^n\,,
\end{equation}
where $(\tau_a)_{mn} = \Omega_{ml} (\tau_a)^l{}_n = (\tau_a)_{nm}$\,.
As mentioned in Section 4.2 of \cite{Saberi:2019fkq}, there symmetry does not suffer from an anomaly.

%-------------------
\subsubsection{Non-example: Free chiral}
%-------------------

We note that the presence of the twisted $\cN=2$ supersymmetry algebra is not sufficient to declare an enhancement of supersymmetry.
To see that this is the case, we note that a single $\cN=1$ chiral possesses this symmetry; this is a 4d analogue of the 3d example described in e.g. Section 2.4.4 of \cite{Garner:2023zko}.
The corresponding currents are given by
\begin{equation}
\begin{aligned}
    & &\bS_i = \tfrac{1}{3}\bfbeta \del_{i} \bfgamma - \tfrac{2}{3} \bfgamma \del_i \bfbeta\,, & &\\
    \bG_i & = \, -\tfrac{1}{2}  \bfbeta \del_i \bfbeta\,, &&& \widetilde \bG & =\, \tfrac{1}{2} \bfgamma^2\,,\\
    & & \bR =\, -\tfrac{1}{2} \bfbeta \bfgamma\,. \hspace{1cm}&&
\end{aligned}
\end{equation}
This twisted $\cN=2$ structure does not arise from the fact that a single $\cN=1$ chiral can be viewed as a half-hypermultiplet. For instance, the $\wti \bG$ operator originates from a scalar in the physical theory, not from a spinor.
A clear diagnostic that this structure does not result from twisting a genuine $\cN=2$ SCFT is that $\bS_i$ is \emph{not} the superconformal stress tensor: while it extremizes $a$, it does not maximize it.
In fact, the $\bS_i$ above follow from assigning $R$-charge $r=\f43$ to $\bfgamma$.

Although this twisted $\cN=2$ superconformal structure is not inherited from a true $\cN=2$ SCFT, it does admit a natural interpretation in the physical theory. Deforming the twisted theory by $\wti\bG$ corresponds to giving a mass to the chiral multiplet, rendering the physical theory gapped. Since the IR limit of such a theory is topological, its holomorphic twist is likewise topological. The operators $\bG_i$ are precisely the homotopies for holomorphic translations generated by $\bS_i-\f13\del_i\bR$.

%-------------------
\subsection{\texorpdfstring{$\cN=3$}{N=3} supersymmetry}
%-------------------

We now turn to the case of $\cN=3$ supersymmetry. 
The decomposition of the $\cN=3$ stress-tensor multiplet into 
$\cN=1$ multiplets is as follows:
\begin{equation}
\begin{aligned}
    B_1\overline{B}_1[0,0]^{(0,2,0)}_2 =& \underbrace{\rule[-1.2ex]{0pt}{0pt} {A}_1\overline{A}_1[1,1]^{(0)}_3}_{\bS_i} \oplus \underbrace{\rule[-1.2ex]{0pt}{0pt}{A}_1\overline{A}_2[1,0]^{(1/3)}_{5/2}}_{\widetilde\bG} \oplus \underbrace{\rule[-1.2ex]{0pt}{0pt}{A}_2\overline{ A}_1[0,1]^{(-1/3)}_{5/2}}_{\bG_i} \oplus \underbrace{\rule[-1.2ex]{0pt}{0pt}{A}_2\overline{A}_2[0,0]^{(0)}_2}_{\bR}\\
    & \oplus \underbrace{\rule[-1.2ex]{0pt}{0pt}{B}_1\overline{L}[0,0]^{(-4/3)}_{2}}_{\bM} \oplus \underbrace{\rule[-1.2ex]{0pt}{0pt}{B}_1\overline{L}[0,1]^{(-5/3)}_{5/2}}_{\bL_i} \oplus \quad c.c.\,.
\end{aligned}
\end{equation}
where $c.c.$ denotes the conjugate multiplets of the second line. However since these are long semi-long on the anti-chiral side they do not contribute to the semi-chiral ring. Note that the $\cN=1$ $\UU(1)_r$ generator is identified as 
\begin{equation}
    r_{\cN=1} = \f19\left(r_{\cN=3} + 4 I_3 \right)\,,   
\end{equation}
where $I_3$ is the $\UU(1)$ generator appearing in the decomposition $\SU(3)\to \SU(2)\times \UU(1)$ whose charges are integers.%
\footnote{In our conventions the $\mathbf{3}$ of $\SU(3)$ decomposes into $\mathbf{2}_1+\mathbf{1}_{-2}$ where subscripts are $I_3$ charges.}
The corresponding semi-chiral operators will be denoted
\begin{equation}
    \bS_i\,,\quad \widetilde{\bG}_I\,, \quad \bG^I_i\,, \quad \bR^I{}_J\,, \quad \bM_{I}\,, \quad \bL_{i}\,.
\end{equation}
where $I = 1, 2$. These semi-chiral fields transform under the remaining $\UU(2)$ flavor symmetry as $\bf{1}_0\,,\bf{2}_1\,,\bf{2}_{-1}\,,\bf{3}_0+\bf{1}_0\,,\bf{2}_1$ and $\bf{1}_2$ respectively, where the generator of the $\UU(1)\subset \UU(2)$ flavor symmetry is $\f13(I_3-2r_{\cN=3})$.

We now enumerate the binary $\lambda$-brackets involving these generators. 
In particular, the operators $\bR^I{}_J$ transform as 2-Virasoro primaries and generate a 2-Kac–Moody algebra with underlying symmetry algebra $\gl(2)$:
\begin{align}
    &\llbrak{\bS_i}{\bR^I{}_J} = \,\del_i\bR^I{}_J + \lambda_i\bR^I{}_J\,, &&\llbrak{\bR^I{}_J}{\bR^K{}_L} = \delta^K{}_J \bR^I{}_L - \delta^I{}_L \bR^K{}_J\,.
\end{align}
The operators $\bS_i$ and $\bR^I{}_J$ generate the holomorphic enhancement of the bosonic subalgebra of the $\lie{sl}(3|2)$ twisted superconformal algebra.
Their $\lambda$-brackets with the currents $\bL_i$, which generate the remaining bosonic symmetries, are
\begin{align}
    &\llbrak{\bS_i}{\bL_{j}} = \,\del_i \bL_{j} + \f13\lambda_i \bL_{j} + \lambda_j \bL_{i}\,,& &\llbrak{\bR^I{}_J}{\bL_{i}} = \, \delta^I_J \bL_{i}\,.
\end{align}
The remaining bracket among two bosonic symmetries vanishes:
\begin{equation}
    \llbrak{\bL_i}{\bL_j} = 0\,.
\end{equation}
The $\lambda$-brackets of $\bS_i$ and $\bR^I{}_J$ with the fermionic generators take the form
\begin{align}
    &\llbrak{\bS_i}{\widetilde{\bG}_I} = \, \del_i\widetilde\bG_I + \f43 \lambda_i\widetilde{\bG}_I\,, & &\llbrak{\bR^I{}_J}{\widetilde{\bG}_K} = -\delta^I_K \widetilde{\bG}_J\,,\\
    &\llbrak{\bS_i}{\bG^I_i} = \,\del_i \bG^I_j + \f23\lambda_i \bG^I_j + \lambda_j \bG^I_i\,,& &\llbrak{\bR^I{}_J}{\bG^K_i} = \, \delta^K_J \bG^I_i + \lambda_i(\epsilon^{IK}\bM_{J}+\delta^I_J \epsilon^{KL}\bM_L)\,,\\
    &\llbrak{\bS_i}{\bM_{I}} = \,\del_i \bM_{I} + \f23\lambda_i \bM_{I}\,,& &\llbrak{\bR^I{}_J}{\bM_{K}} = \, \delta^I_J \bM_{K} -\delta^I_K \bM_{J}\,,
\end{align}
and the only non-vanishing bracket of these generators with $\bL_i$ is
\begin{equation}
    \llbrak{\bL_{i}}{\widetilde{\bG}_I} = \, \epsilon_{IJ} \bG^J_i + \del_i \bM_{I} + 2 \lambda_i \bM_{I}\,.
\end{equation}
Finally, the non-vanishing brackets among the fermionic generators are
\begin{align}
    &\llbrak{\bG^I_i}{\bG^J_j} = \, -\f12\epsilon^{IJ}(2\del_i \bL_{j} - \del_j \bL_{i} + \lambda_i \bL_{j} + \lambda_j \bL_{i})\,,\\
    &\llbrak{\bG^I_i}{\widetilde{\bG}_J} = \, \delta^I_J \bS_i + \del_i \bR^I{}_J + \f32 \lambda_i \bR^I{}_J - \f13 \delta^I_J\left(\bR^K{}_K + \f32 \lambda_i \bR^K{}_K\right)\,,\\
    &\llbrak{\bG^I_i}{\bM_{J}} = \,\f12 \delta^I_J \bL_{i}\,,\\
    &\llbrak{\widetilde{\bG}_I}{\bM_{J}} = \, -\f12 \epsilon_{IK} \bR^K{}_J\,.
\end{align}
As in the previous cases, the conformal anomalies are captured by ternary brackets among the generators of the twisted superconformal symmetry algebra.
In the $\cN=3$ case, superconformal symmetry is sufficiently restrictive to enforce equality of the anomaly coefficients, $a = c$ \cite{Aharony:2015oyb}.
Besides the ternary bracket in Eq.~\eqref{eq:sss} (with $a = c$), the following additional ternary brackets are non-vanishing:
\begin{align}
    &\lllbrak{\bR^I{}_J}{\bR^K{}_L}{\bR^M{}_N} = \frac{a}{\pi^2} \left(2 \delta^I_L \delta^K_N \delta^M_J + 2 \delta^I_N \delta^K_J \delta^M_L - 4 \delta^I_J \delta^K_L \delta^M_N \right)[\lambda_1 \lambda_2]\,,\\
    &\lllbrak{\bS_i}{\bR^I{}_J}{\bR^K{}_L} = \frac{4a}{6\pi^2} \left(\delta^I_L \delta^K_J + \delta^I_J \delta^K_L\right)[\lambda_1 \lambda_2]\lambda_{1i}\,,\\
    &\lllbrak{\widetilde{\bG}_I}{\bS_i}{\bG^J_j} = -\frac{2 a}{3\pi^2} \delta^J_I [\lambda_1 \lambda_2] \lambda_{2i}(\lambda_{1j}+\lambda_{2j})\,,\\
    &\lllbrak{\bG^I_i}{\widetilde{\bG}_J}{\bR^K{}_L} = \frac{a}{\pi^2} \delta^I_L \delta^K_J [\lambda_1 \lambda_2]\lambda_{1i}\,\\
    &\lllbrak{\widetilde{\bG}_I}{\widetilde{\bG}_J}{\bL_{i}} = \frac{2a}{\pi^2} \epsilon_{IJ} [\lambda_1 \lambda_2](\lambda_{1i}+\lambda_{2i})
\end{align}
This symmetry algebra admits a natural supergeometric interpretation, analogous to the $\cN=2$ case (see Section 4.2 of \cite{Saberi:2019fkq} and Section 6.3.5 of \cite{Hahner:2024hak}).
To see this, let us introduce local holomorphic coordinates $z^i$, $\theta^I$ ($I=1,2$) on $\bC^{2|2} \backslash \{0\}$. In these coordinates, the operators $\bS_i$, $\bR^I_J$, and $\bL_i$ implement the actions of vector fields on this supermanifold,
\begin{equation}
    \bS_i \rightsquigarrow - \omega \del_i - \tfrac{1}{3}(\del_i \omega) \theta^I \del_{\theta^I}\,, \qquad \bR^I{}_J \rightsquigarrow \omega \theta^I \del_{\theta^J}\,, \qquad \bL_{i} \rightsquigarrow \f12 \omega \epsilon_{IJ} \theta^I \theta^J \del_i\,,
\end{equation}
via $S^3$ integrals weighted by $\omega \in\Omega^{(0,\bullet)}(\bC^2\backslash\{0\})$. Similarly, the operators $\wti{\bG}_I$, $\bG^I_i$, and $\bM_I$ implement the action of the vector fields
\begin{equation}
        \widetilde{\bG}_I \rightsquigarrow  \omega \del_{\theta^I}\,, \qquad  \bG^I_i \rightsquigarrow \omega \theta^I \del_i + (\del_i \omega) \theta^I \theta^J \del_{\theta^J}\,, \qquad \bM_{I} \rightsquigarrow \frac{1}{2} \omega \epsilon_{KL} \theta^K \theta^L \del_{\theta^I}\,.
\end{equation}
Equivalently, this structure can be identified with the dg Lie superalgebra of derived super vector fields on the supermanifold
\begin{equation}
    {\rm Tot}(\Pi K^{1/3} \otimes \bC^2 \to \bC^2\backslash\{0\})\,.
\end{equation}
To complete the complex geometric description of $\lie{vir}(2|2)_a$, it would be desirable to find an explicit representative of the 3-cocycle corresponding to the unique central extension of this algebra. We leave this for future work.

The fact that algebra of super vector fields admits only a single non-trivial cocycle, implying that the twisted superconformal algebra must have $a=c$, was established by Fuks \cite{Fuks1986}.

To end this subsection we note that interacting unitary $\cN=3$ SCFT do not admit flavor symmetries apart from the $R$-symmetry.
This follows from the observation that the associated conserved current multiplets necessarily include higher spin fields \cite{Cordova:2016emh}.
We also note that there is an $\cN=3$ analogue of the anomaly multiplet described in Section \ref{sec:anomalousN=2} but there does not exist an $\cN=3$ analogue of the differential describing the breaking of superconformal symmetry.
This observation nicely agrees the fact that there are no $\cN=3$ preserving deformations that break conformal symmetry \cite{Cordova:2016xhm}; this is due to the fact that the $\cN=3$ superconformal stress tensor multiplet sits in a $B$-type multiplet and there is no way for it to recombine into a long multiplet.

%-------------------
\subsection{\texorpdfstring{$\cN=4$}{N=4} supersymmetry}
%-------------------

Finally, let us discuss $\cN=4$ supersymmetry.
The decomposition of the $\cN=4$ superconformal stress tensor multiplet into $\cN=1$ multiplets is given by:
\begin{equation}
\begin{aligned}
    B_1\overline{B}_1[0,0]^{(0,2,0)}_2 =& \underbrace{\rule[-1.2ex]{0pt}{0pt}{A}_1\overline{A}_1[1,1]^{(0)}_3}_{\bS_i} \oplus \underbrace{\rule[-1.2ex]{0pt}{0pt}{A}_1\overline{A}_2[1,0]^{(1/3)}_{5/2}}_{\widetilde\bG} \oplus \underbrace{\rule[-1.2ex]{0pt}{0pt}{A}_2\overline{ A}_1[0,1]^{(-1/3)}_{5/2}}_{\bG_i} \oplus \underbrace{\rule[-1.2ex]{0pt}{0pt}{A}_2\overline{A}_2[0,0]^{(0)}_2}_{\bR}\\
    & \oplus \underbrace{\rule[-1.2ex]{0pt}{0pt}{B}_1\overline{L}[0,0]^{(-4/3)}_{2}}_{\bM} \oplus \underbrace{\rule[-1.2ex]{0pt}{0pt}{B}_1\overline{L}[0,1]^{(-5/3)}_{5/2}}_{\bL_i} \oplus \underbrace{\rule[-1.2ex]{0pt}{0pt}{B}_1\overline{L}[0,0]^{(-2)}_{3}}_{\bN} \oplus \quad c.c.\,.
\end{aligned}
\end{equation}
where $c.c.$ denotes the conjugate multiplets of the second line. However since these semi-long multiplet are long on the anti-chiral side they do not contribute to the semi-chiral ring. The generator of the $\cN=1$ superconformal R-symmetry is given by $r_{\cN=1} = -\f13 I_4$, where $I_4$ is the generator of the $\UU(1)$ in the decomposition $\SU(4) \to \SU(3)\times \UU(1)$.%
\footnote{In our conventions, the $\bf 4$ of $\SU(4)$ decomposes into ${\bf 3}_{1}+{\bf 1}_{-3}$ where subscripts denote $I_4$ charges.}
The corresponding semi-chiral operators will be denoted
\begin{equation}
    \bS_i\,,\quad \widetilde{\bG}_I\,, \quad \bG^I_i\,, \quad \bR^I{}_J\,, \quad \bM_{(IJ)}\,, \quad \bL_{I i}\,, \quad \bN\,.
\end{equation}
where $I = 1, 2, 3$ and $\bR^I{}_I = 0$. The semi-chiral fields transform in the following $\SU(3)$ representation, respectively: $\bf{1}$, $\bf{3}$, $\bar{\bf{3}}$, $\bf{8}$, $\bf{6}$, $\bar{\bf{3}}$ and $\bf{1}$.

The operators $\bR^I{}_J$ are 2-Virasoro primaries generating a 2-Kac-Moody algebra based on the Lie algebra $\lie{sl}(3)$:
\begin{align}
    &\llbrak{\bS_i}{\bR^I{}_J} = \,\del_i\bR^I{}_J + \lambda_i\bR^I{}_J\,, &&\llbrak{\bR^I{}_J}{\bR^K{}_L} = \delta^K{}_J \bR^I{}_L - \delta^I{}_L \bR^K{}_J\,.
\end{align}
Together with the $\bS_i$, they generate the (holomorphic enhancement of the) bosonic subalgebra of $\lie{psl}(3|3)$.
The remaining bosonic symmetries are generated by the (fermionic) operators $\bL_{Ii}$. Their brackets with $\bS_i$ and $\bR^I{}_J$ takes the form
\begin{align}
    &\llbrak{\bS_i}{\bL_{Ij}} = \,\del_i \bL_{Ij} + \f13\lambda_i \bL_{Ij} + \lambda_j \bL_{Ii}\,,& &\llbrak{\bR^I{}_J}{\bL_{Ki}} = \, \f13 \delta^I_J \bL_{Ki} -\delta^I_K \bL_{Ji}\,,
\end{align}
and they bracket trivially with themselves.

The brackets of the generators of the fermionic symmetries with $\bS_i$ and $\bR^I{}_J$ take the form
\begin{align}
    &\llbrak{\bS_i}{\widetilde{\bG}_I} = \, \del_i\widetilde\bG_I + \f43 \lambda_i\widetilde{\bG}_I\,, & &\llbrak{\bR^I{}_J}{\widetilde{\bG}_K} = \, \f13 \delta^I_J \widetilde{\bG}_K -\delta^I_K \widetilde{\bG}_J\,,\\
    &\llbrak{\bS_i}{\bG^I_j} = \,\del_i \bG^I_j + \f23\lambda_i \bG^I_j + \lambda_j \bG^I_i\,,\hspace{-3.3mm}& &\llbrak{\bR^I{}_J}{\bG^K_i} = \, -\f13 \delta^I_J \bG^K_i + \delta^K_J \bG^I_i + \lambda_i \epsilon^{IKL}\bM_{JL}\,,\\
    &\llbrak{\bS_i}{\bM_{IJ}} = \,\del_i \bM_{IJ} + \f23\lambda_i \bM_{IJ}\,,& &\llbrak{\bR^I{}_J}{\bM_{KL}} = \, \f23 \delta^I_J \bM_{KL} -\delta^I_K \bM_{JL}-\delta^I_L \bM_{KJ}\,,\\
    &\llbrak{\bS_i}{\bN} = \,\del_i \bN + \lambda_i \bN\,,& &\llbrak{\bR^I{}_J}{\bN} = \, 0\,,
\end{align}
while their brackets with $\bL_{Ii}$ take the form
\begin{align}
    &\llbrak{\bL_{Ii}}{\widetilde{\bG}_J} = \, \epsilon_{IJK} \bG^K_i + \del_i \bM_{IJ} + 2 \lambda_i \bM_{IJ}\,, & &\llbrak{\bL_{Ii}}{\bG^J_j} = \, \delta^J_I \epsilon_{ij} \bN\,,\\
    &\llbrak{\bL_{Ii}}{\bM_{JK}} = \,0\,,& &\llbrak{\bL_{Ii}}{\bN} = \, 0\,.
\end{align}
Finally, the non-vanishing brackets among the fermionic symmetries are:
\begin{gather}
    \llbrak{\bG^I_i}{\bG^J_j} = \, -\f12\epsilon^{IJK}(2\del_i \bL_{Kj} - \del_j \bL_{Ki} + \lambda_i \bL_{K j} + \lambda_j \bL_{K i})\,,\\
    \begin{aligned}
        &\llbrak{\bG^I_i}{\widetilde{\bG}_J} = \, \delta^I_J \bS_i + \del_i \bR^I{}_J + \f32 \lambda_i \bR^I{}_J\,, & &\llbrak{\bG^I_i}{\bM_{JK}} = \,\f12(\delta^I_J \bL_{Ki} + \delta^I_K \bL_{Ji})\,,\\
        &\llbrak{\widetilde{\bG}_I}{\bN} = \,\epsilon^{ij}(\del_i \bL_{Ij} + \lambda_i \bL_{Ij})\,,& &\llbrak{\widetilde{\bG}_I}{\bM_{JK}} = \, \f12 (\epsilon_{IJL} \bR^L{}_K + \epsilon_{IKL} \bR^L{}_J)\,.
    \end{aligned}
\end{gather}
As in the previous two cases, this algebra also admits a super-geometric interpretation. It coincides with the one described in \cite{Saberi:2019fkq}.
\footnote{Utilizing the theory of pure spinors, there is a different characterization of the residual superconformal symmetry present in the holomorphic twist of $4d$ $\cN=4$ supersymmetry found in section 6.3.6 of \cite{Hahner:2024hak}. 
At the time of writing, we have not been able to relate these two descriptions.}

Concretely, introducing local holomorphic coordinates $z^i$ and $\theta^I$ on $\bC^{2|3}\backslash\{0\}$, the fermionic currents $\bS_i$, $\bR^I{}_J$, and $\bL_{Ii}$ generate the action of the vector fields
\begin{equation}
    \begin{aligned}
        \bS_i \rightsquigarrow - \omega \del_i - \tfrac{1}{3}(\del_i \omega) \theta^I \del_{\theta^I}\,, & & & & \bR^I{}_J \rightsquigarrow \omega \left(\theta^I \del_{\theta^J} - \f13 \delta^I_J \theta^K\,, \del_{\theta^K}\right)\,,\\
        & & \hspace{-2cm} \bL_{I i} \rightsquigarrow \f12 \omega \epsilon_{IJK} \theta^J \theta^K \del_i + \f16 (\del_i \omega) \epsilon_{JKL} \theta^J \theta^K \theta^L \del_{\theta^I}\,, \hspace{-2cm} & &
    \end{aligned}
\end{equation}
and the bosonic currents $\widetilde{\bG}_I$, $\bG^I_i$, $\bM_{IJ}$, and $\bN$ generate the action of the vector fields
\begin{equation}
    \begin{aligned}
        \widetilde{\bG}_I & \rightsquigarrow  \omega \del_{\theta^I}\,, & \bG^I_i & \rightsquigarrow \omega \theta^I \del_i + (\del_i \omega) \theta^I \theta^J \del_{\theta^J}\,,\\
        \bM_{IJ} & \rightsquigarrow \frac{1}{2} \omega(\epsilon_{IKL} \theta^K \theta^L \del_{\theta^J} + \epsilon_{JKL} \theta^K \theta^L \del_{\theta^I})\,, &  \bN & \rightsquigarrow \f16 \epsilon_{IJK} \theta^I \theta^J \theta^K \epsilon^{ij} (\del_i \omega) \del_j\,,
    \end{aligned}
\end{equation}
where, as above, $\omega \in \Omega^{(0,\bullet)}(\bC^2\backslash \{0\})$.
Invariantly, this is the dg Lie superalgebra of derived super vector fields on the supermanifold
\begin{equation}
    {\rm Tot}(\Pi K^{1/3} \otimes \bC^3 \to \bC^2\backslash \{0\})
\end{equation}
that are divergence-free with respect to the super holomorphic volume form $\dd^2 z \dd^3 \theta$.

The ternary $\lambda$-brackets of these currents encode the $a$ and $c$ conformal anomalies of the underlying $\cN=4$ theory.
As in $\cN=3$ case, $\cN=4$ superconformal symmetry enforces $a = c$.
Accordingly, the non-vanishing ternary brackets consist of Eq. \eqref{eq:sss} (with $a = c$) together with
\begin{align}
    &\lllbrak{\bR^I{}_J}{\bR^K{}_L}{\bR^M{}_N} = \frac{a}{\pi^2} d^{IKM}_{JLN}[\lambda_1 \lambda_2]\,,\\
    &\lllbrak{\bS_i}{\bR^I{}_J}{\bR^K{}_L} = \frac{4a}{6\pi^2} \kappa^{IK}_{JL}[\lambda_1 \lambda_2]\lambda_{1i}\,,\\
    &\lllbrak{\widetilde{\bG}_I}{\bS_i}{\bG^J_j} = -\frac{2 a}{3\pi^2} \delta^J_I [\lambda_1 \lambda_2] \lambda_{2i}(\lambda_{1j}+\lambda_{2j})\,,\\
    &\lllbrak{\bG^I_i}{\widetilde{\bG}_J}{\bR^K{}_L} = \frac{a}{\pi^2} \kappa^{IK}_{JL} [\lambda_1 \lambda_2]\lambda_{1i}\,\\
    &\lllbrak{\widetilde{\bG}_I}{\widetilde{\bG}_J}{\bL_{Ki}} = \frac{2a}{\pi^2} \epsilon_{IJK} [\lambda_1 \lambda_2](\lambda_{1i}+\lambda_{2i})
\end{align}
where
\begin{align}
    \kappa^{IK}_{JL} & = \delta^K_J \delta^I_L - \f13 \delta^I_J \delta^K_L\,,\\
    d^{IKM}_{JLN} & = 2\left(\delta^K_J \delta^M_L \delta^I_N + \delta^K_N \delta^M_J \delta^I_L - \f23 \delta^I_J \delta^K_N \delta^M_L - \f23 \delta^K_L \delta^M_J \delta^I_N - \f23 \delta^M_N \delta^K_J \delta^I_L + \f49\delta^I_J \delta^K_L \delta^M_N\right)
\end{align}
are the metric and $d$-symbol for $\lie{sl}(3)$. As in the previous cases discussed above, we leave the determination of a representative of the 3-cocycle reproducing these brackets for future work.

As for $\cN=3$ SCFTS, conserved flavor currents reside in multiplets containing higher spin currents.
Therefore interacting unitary $\cN=4$ SCFTs do not admit flavor symmetries apart for the R-symmetry \cite{Cordova:2016emh}.
We also note, as with $\cN=3$, there is no $\cN=4$ analogue of the superconformal symmetry breaking described in Section \ref{sec:anomalousN=2}, which is compatible with the fact \cite{Cordova:2016xhm} that there are no $\cN=4$ preserving deformations that break conformal symmetry due to the fact that the $\cN=4$ stress tensor multiplet belongs to a $B$-type multiplet.

%-------------------
\subsubsection{Example: \texorpdfstring{$\cN=4$}{N=4} SYM}
%-------------------

With $\cN=4$ supersymmetry, it is widely believed there exists only a single interacting example: maximally supersymmetric Yang-Mills theory.
Its holomorphic twist can be realized as a special case of an $\cN=2$ theory with $\cR = \fg \oplus \bar \fg$. Equivalently, it can be viewed as BF theory on the superspace $\bC^{2|2}$ or, upon choosing a non-degenerate invariant bilinear form $\Tr$ on $\fg$, as a Chern-Simons theory on $\bC^{2|3}$.
Its action on these superspaces takes the form
\begin{equation}
    S_{BV} = \int_{\bC^{2|3}} \dd^2 z \dd^3 \theta \Tr(\tfrac{1}{2} \cA \bar \del \cA + \tfrac{1}{3} \cA^3) = \int_{\bC^{2|2}} \dd^2 z \dd^2 \theta \mathbf{B} (\bar \del \mathbf{C} + \tfrac{1}{2}[\mathbf{C}, \mathbf{C}])
\end{equation}
where
\begin{equation}
    \cA = \mathbf{C} + \theta^3 \mathbf{B} = \mathbf{c} + \theta^I \bfphi_I + \tfrac{1}{2} \epsilon_{IJK} \theta^{I} \theta^J \bflambda^K + \theta^1 \theta^2 \theta^3 \mathbf{b}\,.
\end{equation}
In the latter description, the manifest symmetry is the algebra of holomorphic vector fields on $\bC^{2|2}$, corresponding to the symmetry algebra of a holomorphically twisted $\cN=3$ theory.
In contrast, the theory on $\bC^{2|3}$ makes the full (twisted) $\cN=4$ supersymmetry algebra, given by divergence-free vector fields on $\bC^{2|3}$, manifest, provided the odd coordinates $\theta^1,\theta^2,\theta^3$ transform as sections of $K_{\bC^2}^{1/3}$.

The currents realizing this symmetry are constructed in the proof of Proposition 4.9 of \cite{Saberi:2019fkq}.%
\footnote{We note that the currents used in this proof are not all physical. In particular, they include operators not counted by the superconformal index, such as $\tr(\bm{c}^3)$, which generates a degree-$1$ symmetry proportional to $\theta^1 \theta^2 \theta^3 \in \Omega^{0,\bullet}(\bC^{2|3})$.
By contrast, the currents generating divergence-free vector fields, which we present below, are all counted by the superconformal index.}
The stress tensor $\bS_i$ and $\rm{SU}(3)$ currents $\bR^I{}_J$ take the usual form
\begin{equation}
    \bS_i = -\tr (\bm{b} \del_i \bm{c}) + \f{2}{3} \tr(\bflambda^I \del_i \bm{\phi}_I) - \f{1}{3} \tr(\bm{\phi}_I \del_i \bflambda^I)\,, \qquad \bR^I{}_J = \f{1}{3} \delta^I_J \tr(\bflambda^K \bm{\phi}_K) - \tr(\bflambda^I \bm{\phi}_J)\,.
\end{equation}
The currents $\bL_{Ii}$ generating the remaining bosonic symmetries are given by
\begin{equation}
    \bL_{Ii} = \tr(\bm{\phi}_I \del_i \bm{c})\,.
\end{equation}
The currents generating the fermionic symmetries are as follows.
The $\bG^I_i$ and $\widetilde{\bG}_I$ currents are direct analogues of those appearing in $\cN=2$ and are given by
\begin{equation}
    \bG^I_i = \f{1}{2} \epsilon^{IJK} \tr(\bm{\phi}_J \del_i \bm{\phi}_K) - \tr(\bflambda^I \del_i \bm{c})\,, \qquad \widetilde{\bG}_I = \f{1}{2} \epsilon_{IJK} \tr(\bflambda^J \bflambda^K) + \tr(\bm{b} \bm{\phi}_I)\,.
\end{equation}
The remaining currents $\bM_{(IJ)}$ and $\bN$ are given by
\begin{equation}
    \bM_{IJ} = -\tr(\bm{\phi}_I \bm{\phi}_J)\,, \qquad \bN = \epsilon^{ij}\tr(\del_i \bm{c} \del_j \bm{c})\,.
\end{equation}
%

%%%%%%%%%%%%%%%%%%%%%%%%%%%%%%%%%%%%%%%%%
\section{Superconformal Deformations}   %
\label{sec:deformations}                %
%%%%%%%%%%%%%%%%%%%%%%%%%%%%%%%%%%%%%%%%%

With extended superconformal symmetry the range of possible twists expands significantly. Beyond the familiar supersymmetry (Poincar\'e) twists classified in \cite{Eager:2018dsx, Elliott:2020ecf}, one finds superconformal twists and even more exotic variants, see \cite{Elliott:2024jvw} for a catalogue of these twists. Many of these can be obtained as deformations of the holomorphic twist, a fact first observed in \cite{Elliott:2015rja} in the context of $\cN=4$ super Yang-Mills theory. As described in \cite{Saberi:2019fkq}, such deformations correspond precisely to Maurer-Cartan elements of the twisted superconformal symmetry algebra of the holomorphically twisted theory. In our framework, they arise uniformly by turning on a background coupling for the odd elements of the twisted superconformal symmetry algebra (as described in Section \ref{sec:backgrounds}); see also \cite{Borsten:2025hrn} for a related perspective. Constant odd vector fields correspond to supersymmetry twists, linear ones implement superconformal twists, and more general odd vector fields yield a variety of further twists, which we will not discuss further.

For concreteness, consider $\cN=2$ theories.
Deforming the action by the integrated current $\bG_{2}$ yields a four-dimensional theory that remains holomorphic in the $z^1$ plane but becomes topological in the $z^2$ plane (and vice versa for $\bG_1$).
This results in the holomorphic-topological or Kapustin twist \cite{Kapustin:2006hi}.
Similarly, deforming by $\wti\bG$ produces a fully topological theory, realizing the Donaldson-Witten twist \cite{Witten:1988ze}.
These two deformations amount to adding to the holomorphic supercharge another Poincar\'{e} supercharge.
More generally, one can twist by any nilpotent element of the relevant superconformal algebra; many of those are given by a deformation of the holomorphic twist and can therefore be accessed in the same way as described above.
Through these deformations, algebraic properties of local operators in these more supersymmetric twists can often be deduced from those in the holomorphic twist, as illustrated for the $A$- and $B$-twists of three-dimensional $\cN=4$ theories in Section 2.4 of \cite{Garner:2022vds}.

In this section we will focus on deforming the holomorphic twist by the integral of $z^2 \widetilde{\bG}$. This corresponds to turning on a non-constant background field for the twisted superconformal symmetry and has the effect of adding a superconformal supercharge $\{\wti \bG, -\}_{0,1} \sim \wti{S}^{22} = \wti{S}^{2 \dot{-}}$ to the holomorphic supercharge $\bQ = Q^1_-$, thereby realizing the twist of \cite{Beem:2013sza}.
As shown in Section 6 of \cite{Saberi:2019fkq}, the resulting theory is two-dimensional holomorphic theory localized in the $z^2=0$ plane.
Its local operators form a vertex operator algebra (VOA) whose stress tensor (with central charge $c_{2\dd} = -12 c$) corresponds to the supercurrent $\bG_1$.
Our goal in this section is to explain how the OPEs of this VOA can be extracted directly from those of the holomorphic twist.

%-------------------
\subsection{Example: hypermultiplet}
\label{sec:hyper}
%-------------------

Before turning to the general procedure for extracting VOA data from holomorphic field theories, let us first examine the case of the free hypermultiplet. This example is particularly instructive because the superconformal deformation can be made fully explicit, allowing us to compute the singular part of the VOA OPE directly from the four-dimensional holomorphic OPE. In more intricate theories such a direct calculation is no longer feasible, and one must instead rely on more abstract methods as discussed below.

As discussed in Section \ref{sec:N=2}, the holomorphic twist of a single free hypermultiplet is captured by the Lagrangian
\begin{equation}
    S_{\rm holo} = \frac{1}{2} \int_{\bC^{2|1}}\dd^2z \dd\theta\bigg(\Omega_{mn} \mathbf{X}^n \bar{\del} \mathbf{X}^m\bigg) = \int_{\bC^2} \dd^2z \bfbeta_m \bar \del \bfgamma^m\,,
\end{equation}
where the equations of motion are encoded in the differential
\begin{equation}
    \bQ \bfgamma^m = \bar \del \bfgamma^m\,, \qquad \bQ \bfbeta_m = \bar \del \bfbeta_m\,.
\end{equation}
The superconformal deformation is implemented by adding the interaction $\int z^2 \widetilde{\bG}$ to the action. 
We denote by $\mathbb{Q}$ the differential obtained after turning on this deformation, whose action on the fundamental fields is given by
\begin{equation}
\begin{aligned}
    \mathbb{Q} \bfgamma^m &= \bQ\bfgamma^m + \{\wti\bG, \bfgamma^m\}_{0,1} + z^2 \{\wti\bG, \bfgamma^m\}_{0,0}  = \bar \del \bfgamma^m  - z^2 \bfbeta_n \Omega^{nm}\,,\\
    \mathbb{Q} \bfbeta_m  &= \bQ \bfbeta_m + \{\wti\bG, \bfbeta_m\}_{0,1} + z^2 \{\wti\bG, \bfbeta_m\}_{0,0} = \bar \del \bfbeta_m\,.
\end{aligned}
\end{equation}
where the OPE is readily obtained by performing Wick contractions.

\begin{figure}[ht]
	\centering
	\begin{tikzpicture}
		\node (int) at (0,1.5) {$\otimes$};
		\node (vert) at (0,2) {$-\tfrac{1}{2} \Omega^{nm}\displaystyle{\int} \dd^2z z^2 \bfbeta_m \bfbeta_n$};
		\node (gamma1) at (-1.5,-0.25) {$\gamma^n$};
		\node (b1) at (-1,-0.25) {$\bullet$};
		\node (gamma2) at (1.5,0.25) {$\gamma^m$};
		\node (b2) at (1,0.25) {$\bullet$};
		\draw[middlearrow={latex}, thick] (b1.center) to [in=180, out=90] (int.center);
		\draw[middlearrow={latex}, thick] (b2.center) to [in=0, out=90] (int.center);
		\draw (-2.5,-0.5) -- (-1.5,0.5) -- (2.5,0.5) -- (1.5,-0.5) -- cycle;
	\end{tikzpicture}
	\caption{An illustration of the single Feynman diagram contributing to the singular part of the OPE of $\gamma^n$, $\gamma^m$ after turning on the superconformal deformation.}
	\label{fig:sympbosonFeynman}
\end{figure}
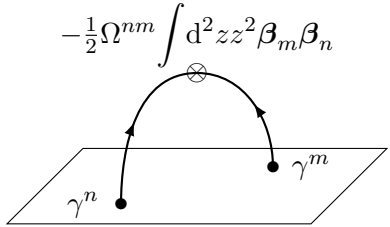

The cohomology of $\mathbb{Q}$ can be computed using a spectral sequence. On the zeroth page, we take cohomology with respect to $\bar{\del}$, which leaves operators constructed from the zero-form components $\beta_m$, $\gamma^m$, and their holomorphic derivatives. On the next page, we compute the cohomology with respect to the remaining part of the differential, denoted $Q_{\rm SC}$, which is given by
\begin{equation}
    Q_{\rm SC} \gamma^m = - z^2 \beta_n \Omega^{nm}\,, \qquad Q_{\rm SC} \beta_m = 0\,.
\end{equation}
The resulting cohomology vanishes away from the locus $z^2 = 0$. Restricting to $z^2 = 0$, it is generated by the fields $\gamma^m(z^1,0)$ together with their holomorphic derivatives $\del_1^k \gamma^m(z^1,0)$. For instance, we have
\begin{equation}
	Q_{\rm SC}~ \del_2 \gamma^m(z^1, 0) = -\beta_n(z^1, 0) \Omega^{nm}\,.
\end{equation}
Although the fields $\bfgamma^m$ have regular OPEs amongst themselves in the holomorphic twist, the deformation by $z^2 \widetilde{\bG}$ generates a singular contribution. The singular part of the OPE is captured by the Feynman diagram shown in Figure~\ref{fig:sympbosonFeynman}.

The corresponding Feynman integral is straightforward to evaluate:
\begin{equation}
	\Omega^{mn}\int_{\bC^2}\dd^2w w^2 P\big((z^1,0);(w^1, w^2)\big)P\big((0,0);(w^1, w^2)\big) = \frac{\Omega^{nm}}{4\pi^2 z^1}
\end{equation}
After absorbing the overall normalization factor $1/4\pi^2$ into a rescaling of the fields $\gamma^a$, we obtain the OPE
\begin{equation}
	\gamma^n(z^1,0) \gamma^m(0,0) \sim \frac{\Omega^{nm}}{z^1}.
\end{equation}
Since the $\mathbb{Q}$-cohomology is generated by polynomials in $\gamma^m(z^1,0)$ and their holomorphic derivatives, we conclude that the algebra of local operators obtained by deforming the holomorphic twist with the integrated current $z^2 \widetilde{\bG}$ is precisely the symplectic boson vertex algebra supported on the $z^2=0$ plane, thereby reproducing the results of \cite{Beem:2013sza} as in \cite{Saberi:2019fkq}.

%-------------------
\subsection{Vertex algebras from four-dimensional descent}
%-------------------
The explicit computation above illustrates how the superconformal deformation of the holomorphic twist localizes the four-dimensional theory to the plane $z^2 = 0$, yielding a two-dimensional holomorphic field theory. The local operators of this superconformal deformation naturally assemble into a vertex operator algebra (VOA). 
In a VOA, the operator product expansion (OPE) of two local operators $\cO_1(z)$ and $\cO_2(0)$ can be extracted by integrating $\cO_1(z)$ along a contour linking $\cO_2(0)$, weighted by an appropriate holomorphic function.
The singular part of the OPE is encoded in the two-dimensional $\lambda$-bracket:
\begin{equation}\label{eq:lambda2d}
    \{\cO_1 {}_\lambda \cO_2\} = \oint_{S^1} \frac{\dd z}{2\pi \ii} e^{\lambda z} [\cO_1(z) \cO_2(0)]\,,
\end{equation}
where as in the four-dimensional case, $[\cO_1(z)\cO_2(0)]$ denotes the OPE of two local operators. The regular part of the OPE is captured by the normal-ordered product
\begin{equation}\label{eq:normal2d}
    :\!\cO_1 \cO_2\!: = \oint_{S^1} \frac{\dd z}{2\pi \ii} z^{-1} [\cO_1(z) \cO_2(0)]\,,
\end{equation}
together with normal-ordered products involving derivatives of $\cO_1$ that account for the remaining regular terms.
For the hypermultiplet, this can be demonstrated explicitly by constructing the spectral sequence associated with the superconformal deformation and evaluating the relevant Feynman diagrams. 

In more general settings, particularly  when the (extended) superconformal symmetry is not realized classically or in computing the OPEs of composite operators, such analytic control is typically unavailable. In what follows we outline a more systematic approach. In this subsection we demonstrate that the two-dimensional $\lambda$-bracket and normal-ordered products admit equivalent, or more precisely cohomologous, realizations as three-dimensional contour integrals extending into the full four-dimensional space. We then explain how these $S^3$ integrals can be directly related to the four-dimensional $\lambda$-brackets in the next subsection, thereby providing a concrete prescription for extracting VOA $\lambda$-brackets from holomorphically twisted data.

We begin by noting that the twisted $\cN=2$ superconformal algebra allows us extend ordinary reduced superfields.
Given a reduced superfield $\bO$, we define its extension
\begin{equation}
	\mathbb{O} = \bO + \dd z^2 \mathbf{\Psi}_\bO\,, \qquad \mathbf{\Psi}_\bO = \{\widetilde{\bG} {}_{(0,0)} \bO\} = \{\widetilde{\bG} , \bO\}_{0,0}\,.
\end{equation}
The extension $\mathbb{O}$ should be thought of as extending $\bO$ to a reduced superfield on $\bC^{2|1}$ and then identifying the fermionic coordinate $\theta$ with the holomorphic $1$-form $\dd z^2$.
If instead of the superconformal deformation we were to deform the holomorphic twist to the holomorphic–topological (Kapustin) twist, i.e. by turning on the interaction $\int \bG_2$, then the components of $\mathbb{O}$ could be identified with the holomorphic–topological descendants of $\bO^{(0)}$; as this is the Hamiltonian for $\theta \partial_2 \sim \dd z^2 \partial_2$, can be thought of a deforming the Dolbeault differential to the THF differential $\dd' = \bar\partial_{\bC^2} + \dd z^2 \partial_2 = \bar \partial_{\bC} + \dd_{\bR^2}$.

For the superconformal deformation, we instead deform by the interaction $\int z^2 \wti{\bG}$, which is the Hamiltonian for $z^2 \partial_\theta \sim z^2 \iota_{\partial_2}$ and so the Dolbeault differential gets replaced by $\bar \partial + \iota_{z^2 \partial_2}$.
The analogue of a semi-chiral superfield in the superconformal deformation then an extended superfield $\mathbb{O}$ that satisfies
\begin{equation}
	\mathbb{Q} \mathbb{O} =  (\bar \del + \iota_{z^2\del_2}\big) \mathbb{O}
\end{equation}
If $\mathbb{O}$ satisfies this condition then its zero-form component defines a local operator in the $\mathbb{Q}$ cohomology, provided it is placed somewhere on the $z^2 = 0$ plane.
In terms of the component (reduced) superfields this condition reads
\begin{equation}
	\mathbb{Q} \bO = \bar\del \bO + z^2 \mathbf{\Psi}_\bO = 0\,, \qquad \mathbb{Q}\mathbf{\Psi}_\bO = \bar\del \mathbf{\Psi}_\bO\,.
\end{equation}
We refer to any extended superfield $\mathbb{O}$ or, equivaletly, a pair of reduced superfields $(\bO, \mathbf{\Psi}_\bO)$ obeying this condition as a \emph{Schur superfield}.
If $\mathbb{O}$ is Schur, then its derivative $\del_1 \mathbb{O}$ is also Schur but its derivative $\del_2 \mathbb{O}$ is not.

Having defined Schur superfields, we now introduce their $\lambda$-bracket and normal-ordered product.
Explicitly, for Schur superfields $\mathbb{O}_1$ and $\mathbb{O}_2$, the $\lambda$-bracket is given by
\begin{equation}
\begin{aligned}
	\{\!\{\mathbb{O}_1 {}_\lambda \mathbb{O}_2\}\!\}(w^1, w^2) & = \oint_{S^3} \frac{\dd^2 z}{(2\pi \ii)^2} ~ e^{\lambda \cdot (z-w)} \iota_{\del_2} \mathbb{O}_1(z^1, z^2) \mathbb{O}_2(w^1,w^2)\\
	& = \oint_{S^3} \frac{\dd^2 z}{(2\pi \ii)^2} ~ e^{\lambda \cdot (z-w)} \mathbf{\Psi}_{\bO_1}(z^1, z^2) \bO_2(w^1,w^2) + \dd w^2 (\dots)\,,
\end{aligned}
\label{eq:2dbracketasS3integral}
\end{equation}
where we have suppressed the terms proportional to $\dd w^2$ in the second line.
The $\mathbb{Q}$-closedness of this operator follows directly from the fact that $\mathbb{O}_1$ and $\mathbb{O}_2$ are Schur superfields.

To demonstrate how this expression relates to the usual two-dimensional $\lambda$-bracket, as defined above in \eqref{eq:lambda2d}, we construct an explicit homotopy via descent.
As a first attempt, consider the 2-chain
\begin{equation}
	D = \left\{(z^1,z^2) \in \bC^2 \bigg| \begin{aligned}
		|z^1|^2 + |z^2|^2 = 1\\
		\Im\, z^2 = 0, ~\Re\, z^2 \leq 0
	\end{aligned}\right\}\,.
\end{equation}
This chain enjoys two important properties: 1) its boundary is the equatorial $S^1 \subset S^3$ preserved by rotations in the $z^2$ plane and 2) under the action of these rotations, its orbit fills the entire $S^3$.
Using the second property, we may rewrite the $S^3$ integral (centered at $w^1 = w^2 = 0$) as one over $D$:
\begin{equation}
	\oint_{S^3} \frac{\dd^2 z}{(2\pi \ii)^2}e^{\lambda \cdot z} \mathbf{\Psi}_{\bO_1} = \int_{D} \iota_{z^2\del_2} \bigg(\frac{\dd^2 z}{(2\pi \ii)^2}e^{\lambda \cdot z} \mathbf{\Psi}_{\bO_1}\bigg) = -\int_D \frac{\dd z^1}{(2\pi \ii)^2}e^{\lambda z^1} z^2 \mathbf{\Psi}_{\bO_1}
\end{equation}
Since $\mathbb{O}_1$ is a Schur superfield, we can further simplify:
\begin{equation}
	-\int_D \frac{\dd z^1}{(2\pi \ii)^2} e^{\lambda \cdot z} \big(\mathbb{Q} - \bar \del\big) \bO_1 = \mathbb{Q}\bigg(\int_D \frac{\dd z^1}{(2\pi \ii)^2} e^{\lambda \cdot z} \bO_1\bigg) - \int_D \bar \del \bigg(\frac{\dd z^1}{(2\pi \ii)^2} e^{\lambda \cdot z} \bO_1\bigg)\,.
\end{equation}
We would be done if the last term were a total derivative, but unfortunately it is not.
To fix this, we introduce a corrected homotopy.

Define the 3-chain
\begin{equation}
    \Sigma = \left\{(z^1,z^2) \in \bC^2 \bigg| 
        \begin{aligned}
            |z^1|^2 &+ |z^2|^2 \geq 1\\
		\Im z^2 = 0 &\text{  and  } 
            \Re z^2 \leq 0
	\end{aligned}\right\}\,.
\end{equation}
Its boundary consists of the 2-chain $D$ (with reversed orientation) together with the locus $z^2 = 0$, $|z^1|^2 \geq 1$.
The orbit of $\Sigma$ under $z^2$-rotations is the region $(z^1,z^2) \in \bC^2$ with $|z^1|^2 + |z^2|^2 \geq 1$.

With these properties, we compute
\begin{equation}
\begin{aligned}
    \mathbb{Q}\left[\int_{\Sigma} \del\bigg(\frac{\dd z^1}{(2\pi \ii)^2}e^{\lambda \cdot z} \bO_1\bigg)\right] & =  -\int_\Sigma \del\bigg( \frac{\dd z^1}{(2\pi \ii)^2}e^{\lambda \cdot z} \big(\bar \del \bO + z^2 \mathbf{\Psi}_{\bO_1}\big) \bigg)\\
    & = -\int_{\del \Sigma} \del \bigg(\frac{\dd z^1}{(2\pi \ii)^2}e^{\lambda \cdot z} \bO_1\bigg)\\
    & \qquad + \int_{\Sigma} \cL_{z^2 \del_2} \bigg(\frac{\dd^2 z}{(2\pi \ii)^2}e^{\lambda \cdot z} \mathbf{\Psi}_{\bO_1}\bigg)\\
    & = \int_{D} \del \bigg(\frac{\dd z^1}{(2\pi \ii)^2}e^{\lambda \cdot z} \bO_1\bigg) - 2 \oint_{S^3} \frac{\dd^2 z}{(2\pi \ii)^2}e^{\lambda \cdot z} \mathbf{\Psi}_{\bO_1}\\
\end{aligned}
\end{equation} 
The second equality follows from the Cartan homotopy formula and Stokes' theorem; the third from the fact that the integrals over $\del \Sigma$ only receive contributions from $D$ as both $\dd^2 z$ and $z^2 \del_2$ vanish on the $z^2 = 0$ plane.
Putting these together, we arrive at the desired result:
\begin{equation}
\begin{aligned}
	\mathbb{Q} & \left[\int_D \frac{\dd z^1}{(2\pi \ii)^2} e^{\lambda \cdot z} \bO_1 + \int_{\Sigma} \del\bigg(\frac{\dd z^1}{(2\pi \ii)^2}e^{\lambda \cdot z} \bO_1\bigg)\right]\\
	& =\oint_{S^1} \frac{\dd z^1}{(2\pi \ii)^2} e^{\lambda_1 z^1} \bO_1 - \oint_{S^3} \frac{\dd^2 z}{(2\pi \ii)^2} e^{\lambda \cdot z} \mathbf{\Psi}_{\bO_1}
\end{aligned}
\label{eq:cohomologous}
\end{equation}
We note that the $\lambda_2$ dependence of the secondary product is totally absent in its formulation as an $S^1$-integral, i.e. this dependence is $\mathbb{Q}$-exact; see Appendix \ref{app:bracketaxiom} for a direct proof of this $\mathbb{Q}$-exactness.
We will set $\lambda_2 = 0$ and, by an abuse of notation, $\lambda_1 = \lambda$ in the following.

The construction of the four-dimensional analogue of the normal-ordered product \eqref{eq:normal2d} is more subtle, since $(z^1)^{-1}$ appearing in the two-dimensional normal-ordered product is not defined on $\bC^2 \setminus {0}$.
Instead, we define the following:
\begin{equation}
	\begin{aligned}
		\noo{\mathbb{O}_1\mathbb{O}_2}\noo(0,0) & = \oint_{S^3} \frac{\dd^2 z}{(2\pi \ii)^2} (\omega_{\rm BM} - \zeta \iota_{\del_2})\mathbb{O}_1(z^1,z^2) \mathbb{O}_2(0,0)\\
		& = \oint_{S^3} \frac{\dd^2 z}{(2\pi \ii)^2} \bigg(\omega_{\rm BM} \bO_1(z^1,z^2) - \zeta \mathbf{\Psi}_{\bO_1}(z^1,z^2)\bigg) \bO_2(0,0)
	\end{aligned}
\end{equation}
where $\zeta$ is the function on $\bC^2\backslash \{0\}$ defined by
\begin{equation}
	\zeta ={\frac{1}{(2\pi i)^2}} \frac{\bar{z}^1}{|z^1|^2 + |z^2|^2}\,.
\end{equation}
That this operator is $\mathbb{Q}$-closed follows immediately from the fact that both $\mathbb{O}_1$ and $\mathbb{O}_2$ are Schur superfields, together with the identity $\bar \del \zeta = z^2 \,\omega_{\rm BM}$.
Repeating the homotopy analysis from above, one can relate this four-dimensional normal-ordered product to the standard two-dimensional definition.
The essential observation is that the restriction of $\zeta$ to the equatorial $S^1$ at $z^2 = 0$ reduces precisely to $(z^1)^{-1}$.
Similarly, more general regularized products involving derivatives of $\mathbb{O}_1$ can be defined in the same way.

%-------------------
\subsection{Two-dimensional \texorpdfstring{$\lambda$}{lambda}-brackets from the holomorphic twist}
\label{sec:2dbracketfrom4d}
%-------------------

Having defined the two-dimensional $\lambda$-bracket and normal-ordered product in terms of $S^3$ integrals, the next step is to relate these constructions to data in the holomorphic twist. This connection provides a direct way to extract VOA data from the holomorphic twist.

Indeed, we can relate the integrals corresponding to the two-dimensional $\lambda$-bracket introduced above to $\lambda$-brackets in the four-dimensional holomorphically twisted theory. Concretely, we identify $z^2 \wti{\bG}$ as an additional interaction term $\cI$ in the BV action and expand perturbatively following Eq. \eqref{eq:pert-int}. To account for the factor of $z^2$ in the definition of the superconformal deformation, we furthermore take a derivative of the $\lambda$ parameters, arriving at the following expression
\begin{equation}
\begin{aligned}
	\{\!\{\mathbb{O}_1 {}_\lambda \mathbb{O}_2 \}\!\} & = \{\{\widetilde{\bG}{}_{(0,0)} \bO_1\} {}_{(\lambda,0)} \bO_2\}\\
	& \qquad + \sum_{n > 0} \frac{(-\pi^2)^n}{n!} \frac{\dd \hfill}{\dd \mu^2_1} \dots \frac{\dd \hfill}{\dd \mu^2_n} \{\widetilde{\bG}{}_{(\mu^1_1, \mu^2_1)} \dots \widetilde{\bG} {}_{(\mu^1_n, \mu^2_n)} \{\widetilde{\bG}{}_{(0,0)} \bO_1\} {}_{(\lambda,0)} \bO_2\}\bigg|_{\mu^1_i = 0, \mu^2_i = 0}\\
	& \qquad + \dd z^2 \left(\dots\right)\,.
\end{aligned}
\label{eq:2dbracketsfrom4d}
\end{equation}
The final term completes this expression into a Schur superfield and is obtained by acting with $\{\widetilde{\bG}{}_{(0,0)} -\}$ on the terms in the first two lines.
We note that the first term of the right-hand side reproduces the holomorphic-topological descent bracket, cf. the three-dimensional analysis of \cite{Garner:2022vds} where the topological descent bracket in the $A$- and $B$-twists were related to the holomorphic-topological descent bracket.
The additional terms can be thought of as the quantum corrections to this quasi-classical vertex algebra, i.e. a (2-shifted) vertex Poisson algebra.

Similarly, the two-dimensional normal ordered product of the VOA is related to the four-dimensional brackets via
\begin{equation}
\begin{aligned}
	\noo\mathbb{O}_1 {} \mathbb{O}_2 \noo & = \bO_1 \bO_2\\
	& \qquad + \sum_{n > 0} \frac{(-\pi^2)^n}{n!} \frac{\dd \hfill}{\dd \mu^2_1} \dots \frac{\dd \hfill}{\dd \mu^2_n} \{\widetilde{\bG}{}_{(\mu^1_1, \mu^2_1)} \dots \widetilde{\bG} {}_{(\mu^1_n, \mu^2_n)}  \bO_1 {}_{(0,0)} \bO_2\}^\gamma\bigg|_{\mu^1_i = 0, \mu^2_i = 0}\\
	& \qquad + \dd z^2 \left(\dots\right)\,.
\end{aligned}
\label{eq:2dproductfrom4d}
\end{equation}
where we have used the quantum corrected regularized products defined in \cite{Gaiotto:2024gii} with $\gamma$ denoting the insertion of an auxiliary propagator between the last two arguments; see \ref{app:bracketaxiom} for a brief review. Although not manifest, the associativity relations of the $\lambda$-brackets and regularized products ensure \eqref{eq:2dbracketsfrom4d} and \eqref{eq:2dproductfrom4d} together define a VOA. More explicitly, in Appendix \ref{app:check2didentities}, we verify that the definition \eqref{eq:2dbracketsfrom4d} satisfies the all the desired properties of two-dimensional $\lambda$-brackets, i.e. those of a Lie conformal algebra. In Appendix \ref{app:check2didentitiesprod}, we further check that \eqref{eq:2dbracketsfrom4d} acts on  \eqref{eq:2dproductfrom4d} as a derivation at vanishing $\lambda$.
 
Returning to the example of the free hypermultiplet, we can now use the general expression to verify the $\lambda$-brackets of its generating fields.
The relevant Schur superfield takes the form
\begin{equation}
    \mathbb{Z}^n = \bfgamma^n + {\rm d}z^2 \bfbeta_m \Omega^{mn}\,.
\end{equation}
Substituting this into the above formula, we obtain
\begin{equation}
\begin{aligned}
    \{\!\{ \mathbb{Z}^n {}_\lambda \mathbb{Z}^m \} \!\} & = \{\{\widetilde{\bG}{}_{(0,0)} \bfgamma^n\} {}_{(\lambda,0)} \bfgamma^m\} + {\rm d}z^2 (\dots)\\
    & = \Omega^{ln} \{\bfbeta_l {}_{(\lambda,0)} \bfgamma^m\} = \Omega^{nm}\,,
\end{aligned}
\end{equation}
where we have omitted the terms proportional to ${\rm d}z^2$.
This precisely reproduces the $\lambda$-bracket of the generating fields of the symplectic boson VOA, in agreement with the computation of Section~\ref{sec:hyper}.
It is worth noting that only binary brackets contribute in this case, due to the fact that the symplectic bosons are linear in the fundamental fields.

Having verified our general expression in the case of the hypermultiplet, we now apply it to recover the $\lambda$-bracket of the two-dimensional stress tensor with itself.
In four dimensions, the Schur superfield corresponding to the two-dimensional stress tensor is
\begin{equation}
    \mathbb{T} = \bG_1 + {\rm d}z^2(\bS_1 - \tfrac{1}{3} \del_1\bR)\,.
\end{equation}
By charge considerations, the quasi-classical $\lambda$-bracket of $\mathbb{T}$ with itself can only receive at first order, i.e. we only need to include the $n=1$ term in Eq. \eqref{eq:2dbracketsfrom4d}.
Applying the formulae from Section~\ref{sec:extended-SUSY} together with the properties of the four-dimensional $\lambda$-brackets, we find that this bracket is given by % (suppressing the ${\rm d}z^2$ terms):
\begin{equation}
\begin{aligned}
    \{\!\{\mathbb{T} {}_\lambda \mathbb{T} \}\!\} &  = \{\{\widetilde{\bG}{}_{(0,0)} \bG_1\} {}_{(\lambda,0)} \bG_1\}\\
    & \quad + (-\pi^2) \frac{\dd \hfill}{\dd \mu^2}\bigg(\{\widetilde{\bG}{}_{(\mu^1,\mu^2)} \{\widetilde{\bG}{}_{(0,0)} \bG_1\} {}_{(\lambda,0)} \bG_1\}\bigg)\bigg|_{\mu^1 = 0, \mu^2 = 0} + {\rm d}z^2(\dots)\\
    & = \{ \bS_1  {}_{(\lambda,0)} \bG_1\} + \frac{\lambda}{3}\{\bR {}_{(\lambda,0)} \bG_1\}\\
    & \quad + (-\pi^2) \frac{\dd \hfill}{\dd \mu^2}\bigg(\{\widetilde{\bG}{}_{(\mu^1,\mu^2)} \bS_1 {}_{(\lambda,0)} \bG_1\} + \frac{\lambda}{3} \{\widetilde{\bG}{}_{(\mu^1,\mu^2)} \bR {}_{(\lambda,0)} \bG_1\}\bigg)\bigg|_{\mu^1 = 0, \mu^2 = 0} + {\rm d}z^2(\dots)\\
    & = \del_1\mathbb{T} + 2 \lambda \mathbb{T} + \lambda^3 (-c)
\end{aligned}
\end{equation}
where we have used the $\lambda$-brackets relations given in section \ref{sec:N=2} and the properties of $\lambda$-brackets in Appendix \ref{app:bracketaxiom}.

Identifying $c_{2d} = -12c$ \cite{Beem:2013sza}, we recognize this as precisely the $\lambda$-bracket of a two-dimensional stress tensor of central charge $c_{2d}$ with itself.

Analogously, the OPE of two flavor symmetry currents, realized in four dimensions by the Schur superfield
\begin{equation}
    \mathbb{J}_a = {\bm \mu}_a + {\rm d}z^2 \bJ_a\,,
\end{equation}
takes the form
\begin{equation}
    \{\!\{\mathbb{J}_a {}_\lambda \mathbb{J}_b\}\!\} = f_{ab}^c ~ \mathbb{J}_c + \lambda (-\tfrac{1}{2} k_F \tr_F(T_a T_b))\,.
\end{equation}
Upon identifying $k_F = -2k_{2d}$ \cite{Beem:2013sza}, we recover the $\lambda$-bracket of affine currents at level $k_{2d}$.

Similar analyses extend to theories with $\cN=3$ and $\cN=4$ supersymmetry, allowing us to identify two-dimensional superconformal subalgebras.
For an $\cN=3$ theory deformed by $\int z^2 \widetilde{\bG}_2$, the additional generators, beyond $\mathbb{T}$, are realized by the Schur superfields
\begin{equation}
    \mathbb{J} = -2 \bM_1 + {\rm d}z^2 (-\bR^1{}_1),
    \qquad
    \mathbb{G} = -\bL_1 + {\rm d}z^2 \big(\bG^1_1 + \del_1 \bM_2\big),
    \qquad
    \widetilde{\mathbb{G}} = \bR^2{}_1 + {\rm d}z^2 (-\widetilde{\bG}_1).
\end{equation}
Together with $\mathbb{T}$, these generate the two-dimensional $\cN=2$ superconformal algebra.
For an $\cN=4$ theory, deforming instead by $\int z^2 \widetilde{\bG}_3$, the additional generators are realized by
\begin{equation}
\begin{aligned}
    \mathbb{J}_{ab} &= -\bM_{ab} + {\rm d}z^2\left(-\tfrac{1}{2}\big(\epsilon_{ac}\bR^{c}{}_{b}+\epsilon_{bc}\bR^{c}{}_{a}\big)\right)\,,\\
    \mathbb{G}_a &= -\bL_{a1} + {\rm d}z^2 \big(\epsilon_{ac}\bG^{c}_1 + \del_1 \bM_{a3}\big)\,,\\
    \widetilde{\mathbb{G}}_a &= \bR^3{}_{a} + {\rm d}z^2 (-\widetilde{\bG}_{a})\,,
\end{aligned}
\end{equation}
with $a,b,c = 1,2$.
Together with $\mathbb{T}$, these Schur superfields generate the small $\cN=4$ superconformal algebra.

%%%%%%%%%%%%%%%%%%%%%%%%%%%%%%%
%	ACKNOWLEDGEMENTS          %
%%%%%%%%%%%%%%%%%%%%%%%%%%%%%%%

 \bigskip
 \bigskip
 \bigskip
% \newpage

 \leftline{\bf Acknowledgments}
 \smallskip
 \noindent It is with pleasure that we thank Andrea Antinucci, Chris Beem, Ingmar Saberi and Sakura Sch\"afer-Nameki for useful and inspiring discussions. 
 The contributions of PB were made possible through the support of grant No. 494786 from the Simons Foundation (Simons Collaboration on the Non-perturbative Bootstrap) and the ERC Consolidator Grant No. 864828, titled “Algebraic Foundations of Supersymmetric Quantum Field Theory'' (SCFTAlg). 
 The work of NG is supported by the ERC Consolidator Grant No. 864828, titled “Algebraic Foundations of Supersymmetric Quantum Field Theory'' (SCFTAlg). 
 The work of JW is supported by the UKRI Frontier Research Grant, underwriting the ERC Advanced Grant ``Generalized Symmetries in Quantum Field Theory and Quantum Gravity”. 
 PB thanks the Aspen Center for Physics (ACP), where part of this work was performed. The ACP is supported by National Science Foundation grant PHY-2210452 and the Simons Foundation (1161654, Troyer).
 BW thanks the hospitality of the Institute Mittag--Leffler as part of the program "Cohomological Aspects of Quantum Field Theory" where some of the work for this paper was performed.

%%%%%%%%%%%%%%%%%%%%%%%%%%%%%%%%%%%%%%%%%%%%%%%%%%%%%%%%%%%%%
% APPENDICES                                                %                                                   %
\appendix                                                   %
%%%%%%%%%%%%%%%%%%%%%%%%%%%%%%%%%%%%%%%%%%%%%%%%%%%%%%%%%%%%%
\newpage 

%%%%%%%%%%%%%%%%%%%%%%%%%%%%%%%%%%%%%%%%%
\section{Conventions}                   %
\label{app:conventions}                 %
%%%%%%%%%%%%%%%%%%%%%%%%%%%%%%%%%%%%%%%%%

With numerous indices at play and various contractions appearing throughout, we outline our conventions below to keep the notation clear and consistent. In the following table we summarize all the indices appearing in this paper, together with their meaning:
\begin{table}[!htb]
    \centering
    \begin{tabular}{c|c|c}
        Index & Range & Meaning \\
        \hline
        $\alpha,\,\beta$ & $\pm$ & "Holomorphic" spinor indices \\
        $\dota,\,\dotb$ & $\dot{\pm}$ & "Anti-holomorphic" spinor indices (before twist) \\  
        $i,\,j$ & $1,2$ & "Anti-holomorphic" spinor indices (after twist) \\
        $\mu,\,\nu$ & $1,\cdots,4$ & Spacetime vector indices\\
        $v,w$ & $1,\cdots, k+1$ & Labels for $k+1$ points/vertices in $\bR^4$\\
        $m,n$ & $\bZ$ & Mode labels \\
        $\cI,\cJ$ & $1,\cdots, \cN$ & Fundamental $\fg_R$ labels\\
        $f,g$ & $1,\cdots, q$ & Label for $\uu(1)$ factors of flavor symmetry algebra\\
        $F,G$ & $1,\cdots, p$ & Label for simple factors of flavor symmetry algebra\\
        $a,b$ & $1,\cdots, {\rm dim} \ff_F$ & Adjoint $\ff_F$ labels\\
    \end{tabular}
    \caption{Collection of indices used throughout this paper.}
    \label{tab:my_label}
\end{table}

We fix a complex structure and split the Lorentz indices $\mu = 1,\cdots,4$ on $\bR^4$ in holomorphic and anti-holomorphic parts. More explicitly, we write $x^{\alpha\dota} = \sigma^{\alpha\dota}_\mu x^\mu$ and define $(z^{\dota},\bz^{\dota}) = (x^{+\dota},x^{-\dota})$. Similarly, we define the (anti-)holomorphic momenta $(\lambda_{\dota},\bar{\lambda}_{\dota}) = (p_{+\dota},p_{-\dota})$. With these definitions, we have for example $x^\mu p^\mu = \f12 (z^\dota\lambda_\dota+\bz^\dota\bar\lambda_\dota)$. The intertwining matrices $\sigma_\mu$ are defined as
\begin{equation}
    (\sigma_\mu)^{\alpha\dota} = \left( 1_{2} \,,\, \ii\,\sigma^{1}\,,\, \ii\,\sigma^{2}\,,\, \ii\,\sigma^{3}\right)^{\alpha\dota}\,,
\end{equation}
where $\sigma^i$ are the standard Pauli matrices. In the main text the dotted indices are replaced with $i,j$ indices. Here we will not do so, hoping this will not cause confusion.

Spinor indices are raised and lowered with the Levi-Civita tensor $\epsilon^{\alpha\beta}$ and $\epsilon^{\dota\dotb}$. We employ conventions where dotted indices are raised and lowered according to the NW-SE convention, i.e. $\lambda^\dota = \epsilon^{\dota\dotb}\lambda_\dotb$ while undotted indices are contracted using SW-NE conventions, i.e. $\eta^\alpha = \eta_\beta\epsilon^{\beta\alpha}$. In these conventions we have, $\epsilon^{12} = \epsilon_{12} = 1$ for both dotted and undotted spinor indices.

The $\lambda_\dota$ (or equivalently the $\bar\lambda_\dota$) are closely related to the anti-holomorphic spinors appearing in the spinor-helicity formalism where we decompose null momenta as $p_{\alpha\dota} = \tilde\lambda_\alpha \lambda_\dota$.\footnote{Strictly speaking the spinor helicity formalism only makes sense in Lorentzian signature as null vectors in Euclidean signature are necessarily trivial. However, if we relax our assumptions and allow for (non-physical) complex momenta $p_\mu$, we can generalize this formalism to our Euclidean setting.} Following the notation used in that context we define the Lorentz invariant contractions of $\lambda$'s as
\begin{equation}
    [ \lambda_k\lambda_l ] = \lambda_{k \dota }\lambda_{l\dotb}\epsilon^{\dota\dotb}\,.
\end{equation}
In addition we define the contraction between (anti-)holomorphic momenta and positions as follows:
\begin{equation}
    \lambda \cdot z = z^\dota\lambda_{\dota}\,,\qquad\qquad \bar\lambda\cdot \bz = \bz^{\dota}\bar\lambda_{\dota}\,.
\end{equation}

\FloatBarrier

%%%%%%%%%%%%%%%%%%%%%%%%%%%%%%%%%%%%%%%%%	
\section{Four-dimensional superconformal algebras}
\label{app:4dSCA}
%%%%%%%%%%%%%%%%%%%%%%%%%%%%%%%%%%%%%%%%%

In this appendix, we collect some facts about four-dimensional superconformal algebras. The spacetime symmetry algebra for $\cN=1,2,3$ superconformal field theories is the superalgebra $\sl(4|\cN)$. When $\cN=4$, the superconformal algebra is given by $\psl(4|4)$. The maximal bosonic subalgebra is $\so(6,\mathbf{C}) \times \gl(\cN,\mathbf{C})_R$ for $\cN=1,2,3$ and $\so(6,\mathbf{C}) \times \sl(4,\mathbf{C})_R$ for $\cN=4$.
	
The four-dimensional conformal algebra $\so(6,\mathbf{C})$ is generated by translations, special conformal transformations, rotations, and dilatations. The generators for these transformations are given by
\begin{equation}
\cP_{\alpha\dota}\,,\qquad \cK^{\dota\alpha}\,,\qquad {\cM_\alpha}^\beta\,,\qquad {\cM^{\dota}}_{\dotb}\,,\qquad \cH\,,
\end{equation}
where we use bi-spinor notation. By adding 4$\cN$ Poincar\'e supercharges $\cQ_\alpha^\cI$, $\widetilde{\cQ}_{\cI\dota}$ and $4\cN$ conformal supercharges $\cS_\cI^\alpha$, $\widetilde{\cS}^{\cI\dota}$ to this algebra we obtain the full four-dimensional superalgebra $\sl(4|2)$. These supercharges are acted upon by the R-symmetry with generators ${\cR^\cI}_\cJ$ where $\cI,\cJ = 1,\dots,\cN$ are indices in the fundamental of $\fg_R$.
	
The commutation relations for the $\so(6,\mathbf{C})$ conformal algebra are
\begin{equation}
    \begin{aligned}
    \left[ {{\cM}_{\alpha}}^{\beta} , {{\cM}_{\gamma}}^{\delta} \right] &= \delta_{\gamma}^{\beta} {{\cM}_{\alpha}}^{\delta}-\delta_\alpha^\delta {{\cM}_{\gamma}}^\beta \,,&
    \left[{\cM^{\dot{\alpha}}}_{\dot{\beta}} , {\cM^{\dot{\gamma}}}_{\dot{\delta}}\right] &= \delta^{\dot{\alpha}}_{\dot{\delta}} {\cM^{\dot{\gamma}}}_{\dot{\beta}}-\delta^{\dot{\gamma}}_{\dot{\beta}} {\cM^{\dot{\alpha}}}_{\dot{\delta}} \,,\\
    \left[ {{\cM}_{\alpha}}^{\beta} , {{\cP}_{\gamma\dot{\gamma}}} \right] &= \delta_\gamma^\beta {\cP}_{\alpha\dot{\gamma}} - \f12 \delta_\alpha^\beta {\cP}_{\gamma\dot{\gamma}}\,,&
    \left[{\cM^{\dot{\alpha}}}_{\dot{\beta}} , {{\cP}_{\gamma\dot{\gamma}}} \right] &= \delta^{\dot{\alpha}}_{\dot{\gamma}}\cP_{\gamma\dot{\beta}}-\f12 \delta^{\dot{\alpha}}_{\dot{\beta}} {\cP}_{\gamma\dot{\gamma}}\,,\\
	\left[ {{\cM}_{\alpha}}^{\beta} , {{\cK}^{\dot{\gamma}\gamma}} \right] &= -\delta_\alpha^\gamma {\cK}^{\dot{\gamma}\beta} + \f12 \delta_\alpha^\beta {\cK}^{\dot{\gamma}\gamma}\,,&
    \left[{\cM^{\dot{\alpha}}}_{\dot{\beta}} , {{\cK}^{\dot{\gamma}\gamma}} \right] &= -\delta^{\dot{\gamma}}_{\dot{\beta}}\cK^{\dot{\alpha}\gamma}+\f12 \delta^{\dot{\alpha}}_{\dot{\beta}} {\cK}^{\dot{\gamma}\gamma}\,,\\
    \left[ \cH , \cP_{\alpha\dot{\alpha}} \right] &= \cP_{\alpha\dot{\alpha}}\,,&
	\left[ \cH , \cK^{\dot{\alpha}\alpha} \right] &= -\cK^{\dot{\alpha}\alpha}\,,\\
	\left[ \cK^{\dot{\alpha}\alpha} , \cP_{\beta\dot{\beta}} \right] &= \delta_\beta^\alpha \delta^{\dot{\alpha}}_{\dot{\beta}}\cH + \delta_\beta^\alpha {\cM^{\dot{\alpha}}}_{\dot{\beta}} + \delta^{\dot{\alpha}}_{\dot{\beta}} {{\cM}_{\beta}}^{\alpha} \,.&&
    \end{aligned}
\end{equation}
The commutation relations obeyed by the R-charges are
\begin{equation}
    \left[ {\cR^\cI}_\cJ ,{\cR^\cK}_\cL \right] = \delta^\cK_\cJ \cR^\cI_\cL - \delta^\cI_\cL \cR^\cK_\cJ\,.
\end{equation}
The non-vanishing commutators between the supercharges are
\begin{equation}
\begin{aligned}
    \left\{ \cQ_\alpha^\cI,\widetilde{\cQ}_{\cJ\dota} \right\} &= \delta^\cI_\cJ \cP_{\alpha\dota}\,,\\
    \left\{ \widetilde{\cS}^{\cI\dota} , \cS_\cJ^\alpha\right\} &= \delta^\cI_\cJ \cK^{\dota\alpha}\,,\\
    \left\{ \cQ_\alpha^\cI,\cS_{\cJ}^\beta \right\} &= \f12\delta^\cI_\cJ\delta_\alpha^\beta\cH + \delta^\cI_\cJ {\cM_\alpha}^\beta - \delta_\alpha^\beta {\cR^\cI}_\cJ\,,\\
    \left\{ \widetilde{\cS}^{\cI\dota},\widetilde{\cQ}_{\cJ\dotb} \right\} &= \f12\delta^\cI_\cJ\delta^\dota_\dotb\cH + \delta^\cI_\cJ {\cM^\dota}_\dotb + \delta^\dota\dotb {\cR^\cI}_\cJ\,,
    \end{aligned}
\end{equation}
Finally, the bosonic generators act on the supercharges as
\begin{equation}
    \begin{aligned}
    \left[{\cM_{\alpha}}^{\beta} , \cQ_{\gamma}^{\cI} \right] &= \delta_{\gamma}^{\beta} \cQ_{\alpha}^{\cI} - \f12 \delta_{\alpha}^{\beta} \cQ_{\gamma}^{\cI}\,, &
    \left[ {\cM^{\dota}}_{\dotb} , \widetilde{\cQ}_{\cI\dot{\gamma}} \right] &= \delta_{\dot{\gamma}}^{\dota} \widetilde{\cQ}_{\cI\dotb} - \f12 \delta^{\dota}_{\dotb}  \widetilde{\cQ}_{\cI\dot{\gamma}}\,,\\
    \left[{\cM_{\alpha}}^{\beta} , \cS_{\cI}^{\gamma} \right] &= -\delta_{\alpha}^{\gamma} \cS_{\cI}^{\beta} + \f12 \delta_{\alpha}^{\beta} \cS_{\cI}^{\gamma}\,,\quad &
    \left[ {\cM^{\dota}}_{\dotb} , \widetilde{\cS}^{\cI\dot{\gamma}} \right] &= -\delta^{\dot{\gamma}}_{\dotb} \widetilde{\cS}^{\cI\dota} + \f12 \delta^{\dota}_{\dotb}  \widetilde{\cS}^{\cI\dot{\gamma}}\,,\\
    \left[ \cH , \cQ_\alpha^\cI \right] &= \f12 \cQ_\alpha^\cI \,,&
    \left[ \cH , \widetilde{\cQ}_{\cI\dota} \right] &= \f12 \widetilde{\cQ}_{\cI\dota} \,.\\
    \left[ \cH , \cS_\cI^\alpha \right] &= -\f12 \cS_\cI^\alpha \,,&
    \left[ \cH , \widetilde{\cS}^{\cI\dota} \right] &= -\f12 \widetilde{\cS}^{\cI\dota} \,.\\
    \left[{\cK^{\dota\alpha}} , \cQ_{\beta}^{\cI} \right] &= \delta_{\beta}^{\alpha}\widetilde{\cS}^{\cI\dota}\,,&
    \left[ {\cK^{\dota\alpha}} , \widetilde{\cQ}_{\cI\dot{\beta}} \right] &= \delta_{\dot{\beta}}^{\dota} \cS_{\cI}^\alpha\,,\\
    \left[{\cP_{\alpha\dota}} , \cS_{\cI}^{\beta} \right] &= -\delta_{\alpha}^{\beta} \widetilde{\cQ}_{\cI\dota}\,,& 
    \left[ {\cP_{\alpha\dota}} , \widetilde{\cS}^{\cI\dot{\beta}} \right] &= -\delta^{\dot{\beta}}_{\dota} \cQ^{\cI}_\alpha\,,\\
    \left[ {\cR^\cI}_\cJ , \cQ_\alpha^\cK \right] &= \delta_\cJ^\cK \cQ_\alpha^\cI - \f14 \delta_\cJ^\cI \cQ_\alpha^\cK\,,&
    \left[ {\cR^\cI}_\cJ , \widetilde{\cQ}_{\cK\dota} \right] &= -\delta_\cK^\cI \widetilde{\cQ}_{\cJ\dota} + \f14 \delta_\cJ^\cI \widetilde{\cQ}_{\cK\dota}\,.
    \end{aligned}
\end{equation}	
All other commutators vanish. In radial quantization, the various generators satisfy the following hermiticity conditions
\begin{equation}
    \begin{aligned}
    \cH^\dagger = \cH\,,& \qquad (\cP_{\alpha\dota})^\dagger = \cK^{\dota\alpha}\,,\qquad ({{\cM}_\alpha}^\beta)^\dagger = {{\cM}_\beta}^\alpha\,,\qquad ({{\cM}^{\dota}}_{\dotb})^\dagger = {{\cM}^{\dotb}}_{\dota}\,,\\
    &({{\cR}^{\cI}}_{\cJ})^\dagger = {{\cR}^{\cJ}}_{\cI}\,,\qquad ({\cQ_{\alpha}^{\cI}})^\dagger = {\cS^{\alpha}_{\cI}}\,,\qquad ({{\widetilde{\cQ}}_{\cI\dota}})^\dagger = {{\widetilde{\cS}}^{\cI\dota}}\,.
    \end{aligned}
\end{equation}

The supercharges transform in the following representations:
\begin{align}
    \cN=&1:\quad & \cQ &\in [1,0]_{\f12}^{(-1)} & \wti{\cQ} &\in [0,1]_{\f12}^{(1)}\,,\\
    \cN=&2: & \cQ &\in [1,0]_{\f12}^{(1;-1)} & \wti{\cQ} &\in [0,1]_{\f12}^{(1;1)}\,,\\
    \cN=&3: & \cQ &\in [1,0]_{\f12}^{(1,0;-1)} & \wti{\cQ} &\in [0,1]_{\f12}^{(0,1;1)}\,,\\
    \cN=&4: & \cQ &\in [1,0]_{\f12}^{(1,0,0)} & \wti{\cQ} &\in [0,1]_{\f12}^{(0,0,1)}\,.
\end{align}
where the representation $[L]^{(R)}_\Delta$ has Lorentz representation $L$ specified by the Dynkin labels of the two $\su(2)$s, $\Delta$ is the scaling dimension and $(R)$ are the Dynkin labels for the R-symmetry representation. For $\cN=1,2,3$, the last entry, separated by a semicolon is the $\UU(1)_r$ charge.

For $\cN=2$ it is sometimes useful to redefine the R-symmetry generators as 
\begin{equation}\label{eq:rR4ddef}
    {\cR^1}_{2} = \cR^+\,,\qquad {\cR^2}_{1} = \cR^-\,, \qquad {\cR^1}_{1} =  \f r2+\cR \,,\qquad {\cR^2}_{2} = \f r2 - \cR\,, 
\end{equation}
where $\cR^\pm$ and $\cR$ form a Chevalley basis of generators for $\sl(2,\mathbf{C})_R$. 

The relation between the Cartan of $\SU(2)^2\simeq \SO(4)$ and $\SO(2)_1\times \SO(2)_2 \subset \SO(4)$ is 
\begin{equation}
    \cM_1{}^1+\cM^{\dot{1}}_{\dot{1}} = \cM_1\,, \qquad     \cM_1{}^1-\cM^{\dot{1}}_{\dot{1}} = \cM_2\,. 
\end{equation}
The Cartan generators of $\SO(6)\simeq \SU(4)$, expressed as the $\SO(2)^3$ block diagonal $\SO(6)$ are given by
\begin{equation}
\begin{aligned}
    \cR^1{}_1+\cR^2{}_2-\cR^3{}_3-\cR^4{}_4 &= R_1\,,\\
    \cR^1{}_1-\cR^2{}_2+\cR^3{}_3-\cR^4{}_4 &= R_2\,,\\
    \cR^1{}_1-\cR^2{}_2-\cR^3{}_3+\cR^4{}_4 &= R_3\,. 
\end{aligned}
\end{equation}
In terms of these generators, the Cartan of the $\cN=2$ R-symmetry embedded in $\cN=4$ is given by
\begin{equation}
    \cR = \f{R_2+R_3}{4}\,,\qquad r = \f{R_1}{2}\,. 
\end{equation}
Finally, the generator of the $\cN=1$ R-symmetry embedded in the $\cN=2$ superconformal algebra is given by
\begin{equation}
    r_{\cN=1} = -\f43 \cR^1{}_1 = -\f23\left(r + 2 \cR\right)\,.
\end{equation}
%

%-------------------
\subsection{\texorpdfstring{$\cN=1$}{N=1} superspace}
\label{subsec:superspace}
%-------------------

In some computations throughout this paper, we employ $\mathcal{N}=1$ superspace techniques. To align with the conventions for the superconformal algebra introduced above, our superspace conventions differ slightly from the standard choices of Wess and Bagger \cite{Wess:1992cp}.

In particular, we define the supercovariant derivatives as
\begin{equation}
\begin{aligned}
    D_\alpha =&\, \f{\del}{\del \theta^\alpha} + \f \ii 2 \sigma^\mu_{\alpha\dota}\bar\theta^\dota\f{\del}{\del x^\mu}\,,\\
    \overline D_\dota =&\, -\f{\del}{\del \bar\theta^\dota} - \f \ii 2 \theta^\alpha \sigma^\mu_{\alpha\dota}\f{\del}{\del x^\mu}\,.
\end{aligned}
\end{equation}
while the differential operators corresponding to the supercharges are given by
\begin{equation}
    \begin{aligned}
        Q_\alpha =&\, \f{\del}{\del \theta^\alpha} - \f \ii 2 \sigma^\mu_{\alpha\dota}\bar\theta^\dota\f{\del}{\del x^\mu}\,,\\
        \wti Q_\dota =&\, -\f{\del}{\del \bar\theta^\dota} + \f \ii 2 \theta^\alpha \sigma^\mu_{\alpha\dota}\f{\del}{\del x^\mu}\,.
    \end{aligned}
\end{equation}
These operators satisfy the following anti-commutation relations:
\begin{equation}
\begin{aligned}
    \acomm{D_\alpha}{\overline D_\dota} =&\, \ii \sigma^\mu_{\alpha\dota}\f{\del}{\del x^\mu} = P_{\alpha\dota}\,,\\
    \acomm{Q_\alpha}{\overline Q_\dota} =&\, - \ii \sigma^\mu_{\alpha\dota}\f{\del}{\del x^\mu} = - P_{\alpha\dota}\,,\\
    \acomm{D_\alpha}{D_\beta} =& \acomm{\overline D_\dota}{\overline D_\dotb} = \acomm{Q_\alpha}{Q_\beta} = \acomm{\overline Q_\dota}{\overline Q_\dotb} = 0 \,,\\
    \acomm{D_\alpha}{Q_\beta} =& \acomm{D_\alpha}{\overline Q_\dotb} = \acomm{\overline D_\dota}{Q_\beta} = \acomm{\overline D_\dota}{\overline Q_\dotb} = 0\,.
\end{aligned}
\end{equation}
Note that in these formulae, $P_{\alpha\dota}$ are the differential operators implementing translations. At first glance, the sign in the first two anti-commutation relations might seem unusual compared to the superconformal algebras introduced above. However, this is consistent given our definitions of the differential operators,
\begin{equation}
    P_\mu = -\ii\del_\mu
\end{equation}
which include an additional minus sign. For a detailed discussion of this point, see for example Appendix A of \cite{Fortin:2011nq}.

%%%%%%%%%%%%%%%%%%%%%%%%%%%%%%%%%%%%%%%%%	
\section{Properties of brackets}        %
\label{app:bracketaxiom}                %
%%%%%%%%%%%%%%%%%%%%%%%%%%%%%%%%%%%%%%%%%

As argued for in \cite{Budzik:2022mpd, Gaiotto:2024gii}, the $n$-ary $\lambda$-brackets satisfy a number of properties which give the space of semi-chiral superfields, equiped which such brackets, the structure of a homotopical generalization of a Lie conformal (super-)algebra. In this appendix we briefly summarize the relevant properties and refer the reader to the references above for more details.

An $n$-ary bracket, 
\begin{equation}
    \{ \bO_1 \, {}_{\lambda_1} \, \cdots \, \bO_{n-1} \, {}_{\lambda_{n-1}} \, \bO_{n} \}\,,
\end{equation}
takes as input $n-1$ pairs of $(\bO_i, \lambda_i)$ and an additional operator $\bO_n$. To give a more uniform description of the properties below, it will be convenient to introduce a formal parameter $\lambda_n$ to the last slot, where
\begin{equation}
    \lambda_n = - \partial - \sum_{i<n} \lambda_i\,,
\end{equation}
where $\partial$ acts on the whole bracket. 

The first property is the sesquilinearity, or shift, property which can be phrased as
\begin{equation}
   \label{eq:shift}
   \{ \bO_1 \, {}_{\lambda_1} \, \cdots \, \partial \bO_{i} \, {}_{\lambda_{i}} \, \dots \, \bO_{n} \} = -\lambda_i\{ \bO_1 \, {}_{\lambda_1} \, \cdots \, \bO_{i} \, {}_{\lambda_{i}} \, \dots \, \bO_{n} \} \,,\qquad \forall i = 1, \dots, n\,.
\end{equation}
Summing the sesquilinearity property over all $n$ gives rise to the translation property,
\begin{equation}
\label{eq:translation}
   \partial \{ \bO_1 \, {}_{\lambda_1} \, \cdots \, \bO_{n-1} \, {}_{\lambda_{n-1}} \, \bO_{n} \} = \sum_{i=1}^n \{ \bO_1 \, {}_{\lambda_1} \, \cdots \, \partial \bO_{i} \, {}_{\lambda_{i}} \, \cdots \, \bO_{n} \}\,. 
\end{equation}
The second property is the graded skew-symmetry of the $\lambda$-bracket, which asserts that for homogeneous elements $\bO_i$, one has
\begin{equation}
    \label{eq:graded-sym}
    \left\{ \bO_{\sigma(1)}\, {}_{\lambda_{\sigma(1)}} \, \cdots\, \bO_n \,{}_{\lambda_{\sigma(n)}} \right\} = (-1)^{%\sgn \,\sigma + 
    \epsilon(\sigma;|\bO_i|)}\left\{ \bO_1 \,{}_{\lambda_1} \,\cdots\,  \bO_n\,{}_{\lambda_n} \right\}\,,
\end{equation}
where $\sigma\in S_n$ and the ($s$-shifted) Koszul sign is defined as,
\begin{equation}\label{eq:shifted-Koszul-sign}
    \epsilon(\sigma;|\bO_i|) = \sum_{i<j \,,\, \sigma(j)>\sigma(j)}(|\bO_i|+s)(|\bO_j|+s) \mod 2\,,
\end{equation}
More generally, in holomorphic-topological theories with $H$ holomorphic and $T$ topological directions, the parity shift is $s=H+T \mod 2$, which in our case simply reduces to $s=0$. 

The bracket is totally (graded) symmetric with respect to the parities $|\bO|$. The $\lambda$-bracket in a four-dimensional holomorphic theory has cohomological degree $-1$ so that we have $|\{ \bO_1 \,{}_\lambda \, \bO_2\}| = |\bO_1|+|\bO_2|-1$. The graded symmetry together with the translation property imply the sesquilinearity of the $\lambda$-bracket.

For $n=2$, applying this definition one recovers the familiar (graded) symmetry relation:
\begin{equation}
    \{ \bO_1 \, {}_{\lambda_1} \, \bO_2 \} = (-1)^{|\bO_1||\bO_2|} \{ \bO_2 \, {}_{\lambda_2} \, \bO_1 \} 
\end{equation}
Finally, the third class of identities are referred to as associativity relations and take the form
\begin{equation}
\label{eq:quadraticIdentity}
\begin{aligned}
    0=&\sum_{k=1}^{n-1}\sum_{\sigma\in {\rm Unsh(k,n-k)}} (-1)^{
    \epsilon(\sigma;|\bO_i|)}\times \\
    &\quad\times
    \{ \{
    \bO_{\sigma(1)} \,{}_{\lambda_{\sigma(1)}}\, \dots \, 
    \bO_{\sigma(k)} \, {}_{\lambda_{\sigma(k)}}\}_{\Lambda(\sigma)}     
        \bO_{\sigma(k+1)} \,{}_{\lambda_{\sigma(k+1)}}\, \dots \, \bO_{\sigma(n)}\,{}_{\lambda_{\sigma(n)}}
    \} \,,
\end{aligned}
\end{equation}
where we define $\Lambda(\sigma) = \sum_{i=1}^k \lambda_{\sigma(i)}$ to preserve the sesquilinearity of the brackets. The permutations in this sum are restricted to the $(k,n-k)$-unshuffle permutations, which are those permutations that satisfy,
\begin{equation}
    \sigma \in {\rm Unsh}(k,n-k) \Leftrightarrow \sigma(1)<\sigma(2)<\cdots<\sigma(k) \quad {\rm and} \quad \sigma(k+1)<\cdots<\sigma(n)\,.
\end{equation}
As discussed in \cite{Budzik:2022mpd}, these associativity relations can be though of as generating the quadratic relations between Feynman integrals. To remove explicit dependence on $\lambda_n$ we will always reorder the resulting bracket so that $\bO_n$ appears in the last slot, using the graded symmetry relations.

For $n=3$, this reduces to the Jacobi-like identity,
\begin{equation}\label{eq:lambda-associativity}
\begin{aligned}
    0 &=\left\{\bO_1\,{}_{\lambda_1}\,\left\{\bO_2\,{}_{\lambda_2}\,\bO_3\right\}\right\}-(-1)^{(|\bO_1|+1)(|\bO_2|+1)} \left\{\bO_2\,{}_{\lambda_2}\,\left\{\bO_1\,{}_{\lambda_1}\,\bO_3\right\}\right\}\\
    &\qquad + (-1)^{|\bO_1|}\left\{\left\{\bO_1\,{}_{\lambda_1}\,\bO_2\right\}_{\lambda_1+\lambda_2}\,\bO_3\,\right\}\,.
\end{aligned}
\end{equation}
Note that for current superfields $\bJ_a$ this relation reduces precisely to the Jacobi identity for the relevant Lie algebra. For future convenience, the associativity relation for $n=4$ is given by
\begin{equation}\label{eq:associativity-n=4}
\begin{aligned}
    0 =& \{ \{\bO_1\,{}_{\lambda_1}\,\bO_2\}{}_{\lambda_1+\lambda_2} \bO_3\,{}_{\lambda_3}\bO_4\} 
    + (-1)^{|\bO_2||\bO_3|} \{ \{\bO_1\,{}_{\lambda_1}\,\bO_3\}{}_{\lambda_1+\lambda_3} \bO_2\,{}_{\lambda_2}\bO_4\}\\ 
    &+ (-1)^{(|\bO_1|+1)(|\bO_2|+|\bO_3|)}\{ \bO_2\,{}_{\lambda_2}\bO_3\,{}_{\lambda_3}\,\{\bO_1\,{}_{\lambda_1}\,\bO_4\}\}\\ 
    &+ (-1)^{|\bO_1|(|\bO_2|+|\bO_3|)}\{ \{\bO_2\,{}_{\lambda_2}\,\bO_3\}{}_{\lambda_2+\lambda_3} \bO_1\,{}_{\lambda_1}\bO_4\}\\
    &+ (-1)^{|\bO_1|+|\bO_3|+|\bO_2||\bO_3|}\{ \bO_1\,{}_{\lambda_1}\bO_3\,{}_{\lambda_3}\,\{\bO_2\,{}_{\lambda_2}\,\bO_4\}\}\\
    &+ (-1)^{|\bO_1|+|\bO_2|}\{ \bO_1\,{}_{\lambda_1}\bO_2\,{}_{\lambda_2}\, \{\bO_3\,{}_{\lambda_3}\,\bO_4\}\}\\
    &+ \{\{\bO_1\,{}_{\lambda_1}\bO_2\,{}_{\lambda_2}\bO_3\}{}_{\lambda_1+\lambda_2+\lambda_3}\bO_4\}
    + (-1)^{|\bO_3|(|\bO_1|+|\bO_2|+1)}\{\bO_3\,{}_{\lambda_3} \,\{\bO_1\,{}_{\lambda_1}\bO_2\,{}_{\lambda_2}\bO_4\}\}\\
    &+ (-1)^{|\bO_2|(|\bO_1|+1)}\{\bO_2\,{}_{\lambda_2} \,\{\bO_1\,{}_{\lambda_1}\bO_3\,{}_{\lambda_3}\bO_4\}\}
    + (-1)^{|\bO_1|}\{\bO_1\,{}_{\lambda_1} \,\{\bO_2\,{}_{\lambda_2}\bO_3\,{}_{\lambda_3}\bO_4\}\}\,.
\end{aligned}
\end{equation}
The space of local operators equipped with this collection of $\lambda$-brackets forms a homotopic analog of a Lie conformal algebra. The holomorphic twist is even richer, and is endowed with the structure of a four-dimensional higher VOA. The $n$-ary $\lambda$-brackets are believed to encode the singular OPE data of this higher VOA \cite{Gaiotto:2024gii}. To access the regular parts of the OPE, we can extend the homotopic Lie conformal algebra with a collection of decorated brackets (see \cite{Gaiotto:2024gii} for more details), 
\begin{equation}
    \big\{\mathbf{O}_1\,{}_{\lambda_1} \cdots \mathbf{O}_{n-1}\,{}_{\lambda_{n-1}} \mathbf{O}_n\big\}^{\gamma}\,,
\end{equation}
where $\gamma$ is a sub-graph of the complete graph with $n$ vertices, one for each of the $n$ inputs of the $\lambda$-bracket. For trivial $\gamma$ one recovers the ordinary $\lambda$-brackets. Physically, these modified brackets correspond to coupling to an auxiliary theory whose propagators are inserted along the edges of $\gamma$, or equivalently regularizing by hand with fixed propagators along $\gamma$ and summing only over diagrams that contain $\gamma$. In the definition \eqref{eq:n-bracket-np} this corresponds to inserting a Bochner-Martinelli kernel for each edge contained in $\gamma$. Roughly speaking, this operation therefore picks out the regular terms in the (multi-)OPE when operators joined by such an edge approach each other.

These decorated brackets satisfy the same symmetry properties as the $\lambda$-brackets, i.e. they are graded symmetric in all entries. Note however that when considering symmetry relations it only makes sense to consider permutations that leave the subgraph $\gamma$ invariant. They also satisfy the same associativity relations, with the difference that the inner and outer bracket in \eqref{eq:quadraticIdentity} have to be dressed with $\gamma_{\rm in}$ and $\gamma_{\rm out}$ such that $\gamma = \gamma_{\rm in} \cup \gamma_{\rm out}$.

\begin{figure}[!htb]
    \centering
    \begin{tikzpicture}
    [
	baseline={(current bounding box.center)},
	line join=round
	]
    \def\gS{1.5};
	% Coordinates of the vertices of the graph
	\coordinate (pd1) at (-0.866*\gS,-0.5*\gS);
	\coordinate (pd2) at (0.*\gS,1.*\gS);
	\coordinate (pd3) at (0.866*\gS,-0.5*\gS);
        % \coordinate (label) at (0*\gS,2*\gS);
        % \draw (label) node {};
	% Draw the edges
	\draw[GraphEdge] (pd1) -- (pd2) node[midway, left] {};
	\draw[GraphEdge,red] (pd1) -- (pd3) node[midway, above] {};
	\draw[GraphEdge] (pd2) -- (pd3) node[midway, right] {};
         % Label the vertices
	\draw (pd1) node[GraphNode] {} node[left] {$\lambda_{2}$};
	\draw (pd2) node[GraphNode] {} node[above] {$\lambda_{1}$};
	\draw (pd3) node[GraphNode] {} node[right] {$\lambda_{3}$};
   \end{tikzpicture}
   \caption{Example of a decorated ternary bracket.}
   \label{fig:gamma-3-bracket}
\end{figure}
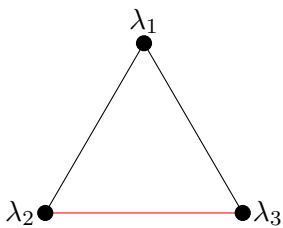

In concrete examples, we specify $\gamma$ by the set of edges, e.g. the red segment in Figure \ref{fig:gamma-3-bracket} will be denoted by $\gamma = (\{2,3\})$. The simplest example is $\{\bO_1 \,{}_{\lambda_1}\,  \bO_2 \}^{(\{1,2\})}$ with a single auxiliary propagator between $\bO_1$ and $\bO_2$. This bracket does not depend on $\lambda_1$ and corresponds to the four-dimensional generalization of the normal ordered product, $\{\bO_1 \,{}_{\lambda_1}\,  \bO_2 \}^{(\{1,2\})} = :\bO_1 \bO_2:$.

As a ternary example, let us consider $\{\bO_1 \,{}_{\lambda_1}\,  \bO_2 \,{}_{\lambda_2}\, \bO_3 \}^{(\{2,3\})}$ with an auxiliary propagator between $\bO_2$ and $\bO_3$, as illustrated in Figure \ref{fig:gamma-3-bracket} below. The corresponding associativity relation is given by
\begin{equation}\label{eq:lambda-associativity-gamma}
\begin{aligned}
    0 &=\left\{\bO_1\,{}_{\lambda_1}\,:\bO_2\,\bO_3:\,\right\}-(-1)^{(|\bO_1|+1)(|\bO_2|+1)} :\bO_2\,\left\{\bO_1\,{}_{\lambda_1}\,\bO_3\right\}:\\
    &\qquad+ (-1)^{|\bO_1|}:\left\{\bO_1\,{}_{\lambda_1}\,\bO_2\right\}\,\bO_3:\,.
\end{aligned}
\end{equation}
%

%-------------------
\subsection{Checking identities of two-dimensional brackets}
\label{app:check2didentities}
%-------------------

In this section we verify that our definitions \eqref{eq:2dbracketsfrom4d} and \eqref{eq:2dproductfrom4d} defined in terms of the four-dimensional brackets, indeed satisfy all the desired identities of standard Lie conformal algebra.
\begin{equation}
\begin{aligned}
    \llb{\partial\mathbb{A}}{\lambda}{\mathbb{B}} &=-\lambda\,\llb{\mathbb{A}}{\lambda}{\mathbb{B}}\\
    \llb{\mathbb{A}}{\lambda}{\partial\mathbb{B}}&=(\lambda+\partial)\,\llb{\mathbb{A}}{\lambda}{\mathbb{B}},\\
     \llb{\mathbb{A}}{\lambda}{\mathbb{B}}&=-\llb{\mathbb{B}}{-\lambda-\partial}{\mathbb{A}},\\
    \llb*{\mathbb{A}}{\lambda}{\llb{\mathbb{B}}{\mu}{\mathbb{C}}} &= \llb*{\mathbb{B}}{\mu}{\llb{\mathbb{A}}{\lambda}{\mathbb{C}}}
    + \llb*{\llb{\mathbb{A}}{\lambda}{\mathbb{B}}}{\lambda+\mu}{\mathbb{C}}.
\end{aligned}
\end{equation}
The first two identities can be checked directly from the definitions, term by term. The first follows immediately from the translation invariance identity \eqref{eq:translation}. The second follows from translation invariance \eqref{eq:translation} together with the shift property \eqref{eq:shift}.

Consider the third identity. The leading term in $\llb{\mathbb{A}}{\lambda}{\mathbb{B}}$ is
\begin{equation}
        \begin{aligned}
        \lb*{\lb{\tilde{\bG}}{(0,0)}{\bA}}{(\lambda,0)}{\bB} & = \lb*{\bB}{(-\partial - \lambda,-\partial)}{\lb{\tilde{\bG}}{(0,0)}{ \bA}}\\
        & = \lb*{\tilde{\bG}}{(0,0)}{\lb{\bA}{(\lambda, 0)}{\bB} } - \lb*{\lb{\tilde{\bG}}{(0,0)}{\bB}}{(-\partial - \lambda,-\partial)}{\bA}\\
        & = - \lb*{\lb{\tilde{\bG}}{(0,0)}{\bB}}{(-\partial - \lambda,-\partial)}{\bA}\\
        & = - \lb*{\lb{\tilde{\bG}}{(0,0)}{\bB}}{(-\partial - \lambda, 0)}{\bA} 
        \end{aligned}
\end{equation}
This is precisely the leading term of $-\llb{\mathbb{B}}{-\lambda-\partial}{\mathbb{A}}$. The remaining terms follow in the same way.
The first two equalities follow from skew-symmetry and the Jacobi identity. The first term on the second line vanishes since $\lb{\bA}{(\lambda, 0)}{\bB}=0$ if both $\bA$ and $\bB$ are Schur operators. It is straightforward to see this in Lagrangian theories. It would be instructive to prove this more generally. The last equality follows from an observation that the expression
\begin{equation*}
    \lb*{\lb{\tilde{\bG}}{(0,0)}{\bB}}{(\lambda_1, \lambda_2)}{\bA} 
\end{equation*}
is independent of $\lambda_2$ if both $\bA$ and $\bB$ are Schur operators, as we remarked earlier in \eqref{eq:cohomologous}. This can also be seen from the definition \eqref{eq:2dbracketasS3integral}. We notice that
\begin{equation}
     \begin{aligned}
         \int_{S^3} \frac{\dd^2 z}{(2\pi \ii)^2} ~ (z^2)^{n+1}\mathbf{\Psi}_{\bO_1}(z^1, z^2) \bO_2(0, 0)  &=   \int_{S^3} \frac{\dd^2 z}{(2\pi \ii)^2} ~ (z^2)^{n} \left(\mathbb{Q} - \bar{\partial}\right) {\bO_1}(z^1, z^2) \bO_2(0, 0) \\
         &=  \mathbb{Q} \int_{S^3} \frac{\dd^2 z}{(2\pi \ii)^2} ~ (z^2)^{n}  {\bO_1}(z^1, z^2) \bO_2(0, 0)
     \end{aligned}
\end{equation}
which leads to
\begin{equation}
     \int_{S^3} \frac{\dd^2 z}{(2\pi \ii)^2} ~ e^{(\lambda_1 z^1 + \lambda_2 z^2)} \mathbf{\Psi}_{\bO_1}(z^1, z^2) \bO_2(0, 0)  =   \int_{S^3} \frac{\dd^2 z}{(2\pi \ii)^2} ~ e^{\lambda_1 z^1 } \mathbf{\Psi}_{\bO_1}(z^1, z^2) \bO_2(0, 0)   + \mathbb{Q}(\dots)
\end{equation}
Lastly, consider the fourth identity. The leading term of $\llb*{\llb{\mathbb{A}}{\lambda}{\mathbb{B}}}{\lambda+\mu}{\mathbb{C}}$ reads
\begin{equation}
    \begin{aligned}
        &\lb*{ \lb*{\tilde{\bG}}{(0,0)}{\lb{\lb{\tilde{\bG}}{(0,0)}{\bA}}{(\lambda, 0)}{{\bB}}}} {(\lambda + \mu, 0)}{{\bC}}\\
        = &\lb*{ \lb{\lb{\tilde{\bG}}{(0,0)}{\bA}}{(\lambda, 0)}{\lb{\tilde{\bG}}{(0,0)}{\bB}}} {(\lambda + \mu, 0)}{{\bC}} - \lb*{ \lb{\lb{\tilde{\bG}}{(0,0)}{\lb{\tilde{\bG}}{(0,0)}{\bA}}}{(\lambda, 0)}{{\bB}}} {(\lambda + \mu, 0)}{{\bC}} \\
        = & \lb*{\lb{\tilde{\bG}}{(0,0)}{\bA}} {(\lambda, 0)} {\lb{ \lb{\tilde{\bG}}{(0,0)}{\bB} } {(\mu, 0)}{\bC}} - \lb*{\lb{\tilde{\bG}}{(0,0)}{\bB}} {(\mu, 0)} {\lb{ \lb{\tilde{\bG}}{(0,0)}{\bA} } {(\lambda, 0)}{\bC}}
    \end{aligned}
\end{equation}
where the first equality uses the Jacobi identity in the first slot. The second equality uses the fact that $\lb{\tilde{\bG}}{(0,0)}{\lb{\tilde{\bG}}{(0,0)}{\bA}} \sim \lb{\lb{\tilde{\bG}}{(0,0)}{\tilde{\bG}}}{(0,0)}{\bA} = 0$. Hence we recover the leading terms of the left-hand side. The same argument applies to the other terms in the definition.

%-------------------
\subsection{Checking that two-dimensional bracket acts as a derivation}
\label{app:check2didentitiesprod}
%-------------------

Setting $\lambda=0$, the $\lambda$-bracket of a Lie conformal algebra in two (real) dimension acts as a derivation with respect to the normally-ordered product,
\begin{equation}
    \llb{\mathbb{A}}{\,0\,}{\noo{\mathbb{B}\mathbb{C}}\noo}
    = \noo{\mathbb{B}\,\llb{\mathbb{A}}{\,0\,}{\mathbb{C}}}\noo
    + \noo{\llb{\mathbb{A}}{\,0\,}{\mathbb{B}}\,\mathbb{C}}\noo .
    \label{eq:actsasderivation}
\end{equation}
Defined in terms of four-dimensional brackets, \eqref{eq:2dbracketsfrom4d} and \eqref{eq:2dproductfrom4d}, this should then follow from their properties as introduced above. In this appendix, we provide evidence for this.

We proceed as above by checking this term by term in the perturbative expansion. The leading term $\lb*{\lb{\tilde{\bG}}{(0,0)}{\bA}}{(0,0)}{:\bB \bC:}$  follows immediately from \eqref{eq:lambda-associativity-gamma} with $\lambda=0$.
For the next terms, consider $\Psi_{\bA} = \lb{\tilde{\bG}}{(0,0)}{\bA}$. From \eqref{eq:associativity-n=4}, we have the following associativity relation
\begin{align}  
\label{eq:associativity-ABCG}
    0 =& -(-1)^{|\bA|}\lb*{\Psi_{\bA}}{(\lambda_{2d}, 0)}{ \{\tilde{\bG} \,{}_{\lambda_2}\,  \bB \,{}_{0}\, \bC \}^{(\{3,4\})} }
    + \{ \,  \tilde{\bG} \,{}_{\lambda_2} \, \lb{\Psi_{\bA}}{(\lambda_{2d}, 0)}{\bB} \,{}_{(\lambda_{2d}, 0) + 0} \,  \bC \} ^{(\{3,4\})}\nn\\
    &+ (-1)^{|\bA||\bB|}\{\tilde{\bG} \,{}_{\lambda_2}\,  \bB \,{}_{0}\, \lb{\Psi_{\bA}}{(\lambda_{2d}, 0)}{\bC}  \} ^{(\{3,4\})}
    - (-1)^{|\bA|} \{\Psi_{\bA} \,{}_{(\lambda_{2d}, 0)}\,  \tilde{\bG} \,{}_{\lambda_2}\, \lb{\bB}{0}{\bC} ^{(\{3,4\})} \} \nn\\
    &+ (-1)^{|\bA||\bB|} \lb*{\bB}{0}{ \{\Psi_{\bA} \,{}_{(\lambda_{2d}, 0)}\,  \tilde{\bG} \,{}_{\lambda_2}\, \bC \}} ^{(\{3,4\})} 
    + \lb*{{ \{\Psi_{\bA} \,{}_{(\lambda_{2d}, 0)}\,  \tilde{\bG} \,{}_{\lambda_2}\, \bB \}}}{(\lambda_{2d}, 0)+\lambda_2}{\bC}^{(\{3,4\})} \nn\\
    &- (-1)^{|\bA|+|\bB|}\{\Psi_{\bA} \,{}_{(\lambda_{2d}, 0)}\,  \bB \,{}_{0}\, \lb{\tilde{\bG}}{\lambda_2}{\bC}  \} ^{(\{3,4\})} 
    - (-1)^{|\bA|} \{ \,  \Psi_{\bA} \,{}_{(\lambda_{2d}, 0)} \, \lb{\tilde{\bG}}{\lambda_2}{\bB} \,{}_{\lambda_2} \,  \bC \} ^{(\{3,4\})}\nn\\
    &+ \lb*{\tilde{\bG}}{\lambda_2}{ \{\Psi_{\bA} \,{}_{(\lambda_{2d}, 0)}\,  \bB \,{}_{0}\, \bC \}^{(\{3,4\})}} 
    + \{\lb{\Psi_{\bA}}{(\lambda_{2d}, 0)}{\tilde{\bG}} \,{}_{(\lambda_{2d}, 0) + \lambda_2}\,  \bB \,{}_{0}\,  \bC \} ^{(\{3,4\})}
\end{align}
Expanding out \eqref{eq:actsasderivation}, we need to show $\lb*{\Psi_A}{\, (\lambda, 0) \,}{ \bullet}$ acts as a derivation on
\begin{equation}
    \frac{\mathrm{d}}{\mathrm{d}\mu_1^2} \{\tilde{\bG} {}_{(\mu_1, \mu_1^2)} \bB {}_{(0,0)} \bC \} ^{(\{2,3\})} \bigg|_{\mu^1_1 = 0, \mu^2_1 = 0}
\end{equation}
and additionally 
\begin{equation}
    \frac{\mathrm{d}}{\mathrm{d}\mu_1^2} \{\tilde{\bG} {}_{(\mu_1, \mu_1^2)} \bA {}_{(0,0)} \bullet \}  \bigg|_{\mu^1_1 = 0, \mu^2_1 = 0}
\end{equation}
acts as a derivation on $\{\bB_0\bC\}^{(\{1,2\})}$. They correspond precisely to the first and the second line respectively upon taking derivative of $\lambda_2^2$ and setting all lambda parameters to zero. In all examples we have checked in this paper, we find this is indeed the case, and the remaining terms in \eqref{eq:associativity-ABCG} vanish. It would be interesting to derive this more generally through a non-perturbative argument. We leave this for future work.

%%%%%%%%%%%%%%%%%%%%%%%%%%%%%%%%%%%%%%%%%%%%%%%%%%%%%%%%%%%%%%%%%%%%%%%%
\newpage
\begingroup

\let\oldthebibliography\thebibliography
\renewcommand{\thebibliography}[1]{%
  \oldthebibliography{#1}%
  \setlength{\itemsep}{4pt}  % define space between items
  \setlength{\parskip}{0pt}  % define paragraph spacing
}
\bibliographystyle{JHEP}
\bibliography{theBib}	
\endgroup	
%%%%%%%%%%%%%%%%%%%%%%%%%%%%%%%%%%%%%%%%%%%%%%%%%%%%%%%%%%%%%%%%%%%%%%%%
\end{document}